\pdfoutput=1   
\documentclass[preprint,12pt]{elsarticle}

\usepackage{amsmath,amssymb}
\usepackage{booktabs}
\usepackage{array}
\usepackage{enumitem}
\usepackage[table]{xcolor}
\usepackage{graphicx}
\usepackage[expansion=false]{microtype}
\usepackage{threeparttable}
\usepackage{pifont}
\usepackage{placeins}
\usepackage{tikz}
\usetikzlibrary{positioning, calc, shapes.geometric, arrows.meta, fit, backgrounds}
\usepackage{pgfplots}
\pgfplotsset{compat=1.17}
\usepgfplotslibrary{groupplots}
\usetikzlibrary{patterns, decorations.pathreplacing, matrix}
\usepackage{tabularx}
\usepackage{makecell}
\usepackage{multirow}
\usepackage{ragged2e}
\usepackage{rotating}   
\usepackage{mathtools}  

\definecolor{navyblue}{RGB}{24,43,73}
\definecolor{slatebg}{RGB}{245,247,250}
\definecolor{accentblue}{RGB}{41,128,185}
\definecolor{alertred}{RGB}{192,57,43}
\definecolor{lightgray}{RGB}{220,224,230}
\definecolor{okgreen}{RGB}{39,124,86}
\definecolor{warnamber}{RGB}{191,139,32}

\colorlet{svInk}{navyblue}               
\colorlet{svBlue}{accentblue}            
\colorlet{svRed}{alertred}               
\definecolor{svGrey}{HTML}{6B7280}       
\definecolor{svRule}{HTML}{C9CED6}       
\definecolor{svPanel}{HTML}{F3F5F8}      
\definecolor{svFamI}{HTML}{0072B2}       
\definecolor{svFamIb}{HTML}{56B4E9}      
\definecolor{svFamII}{HTML}{009E73}      
\definecolor{svFamIII}{HTML}{D55E00}     
\definecolor{svFamIV}{HTML}{CC79A7}      
\definecolor{svGold}{HTML}{E69F00}       
\colorlet{svfMRI}{svFamI}
\colorlet{svEEG}{svFamII}
\colorlet{svMEG}{svFamIII}
\colorlet{sviEEG}{svFamIV}

\newcommand{\svfont}{\sffamily\fontsize{8}{9.6}\selectfont}
\newcommand{\svsmall}{\sffamily\fontsize{7}{8.4}\selectfont}

\tikzset{
  sv/.style      ={font=\svfont, text=svInk, line width=.6pt},
  svbox/.style   ={draw=svInk, rounded corners=2pt, fill=white,
                   inner sep=3pt, align=center, font=\svfont},
  svarrow/.style ={-{Stealth[length=5pt,width=4pt]}, line width=.7pt, draw=svInk},
  svdash/.style  ={dashed, dash pattern=on 3pt off 2pt},
  svnote/.style  ={font=\svsmall, text=svGrey, align=center},
  svtag/.style   ={font=\svsmall\bfseries, text=white, rounded corners=1.5pt,
                   inner sep=1.6pt},
}

\DeclareMathOperator*{\argmin}{arg\,min}
\DeclareMathOperator*{\argmax}{arg\,max}
\DeclarePairedDelimiter{\norm}{\lVert}{\rVert}

\newcommand{\neural}{\mathbf{x}}      
\newcommand{\imgcontent}{s}           
\newcommand{\cue}{c}                  
\newcommand{\exemplar}{I_{c}}         
\newcommand{\est}[1]{\hat{#1}}        
\newcommand{\decrep}{z}               
\newcommand{\encmodel}{g_{\phi}}      
\newcommand{\decmap}{h_{\psi}}        
\newcommand{\gen}{G}                  
\newcommand{\transform}{T}            
\newcommand{\Feat}{\varphi}           
\newcommand{\Lam}{\Lambda}            

\newcolumntype{L}[1]{>{\RaggedRight\arraybackslash\hspace{0pt}}p{#1}}
\newcolumntype{Y}{>{\RaggedRight\arraybackslash\hspace{0pt}}X}   
\newcolumntype{M}[1]{>{\centering\arraybackslash}m{#1}}          
\newcommand{\vsep}{\,\textperiodcentered\,\allowbreak}

\newcommand{\cmk}{\textcolor{svFamII}{\ding{51}}}        
\newcommand{\pmk}{\textcolor{svGold}{(\ding{51})}}       
\newcommand{\xmk}{\textcolor{svRed}{\ding{55}}}          
\newcommand{\nmk}{\textcolor{svGrey}{---}}               
\newcommand{\umk}{\textcolor{svGrey}{?}}                 

\newcommand{\sepyes}{\cmk}   
\newcommand{\sepno}{\xmk}    
\newcommand{\sepunc}{\umk}   
\newcommand{\sepna}{\nmk}    

\newcommand{\lone}{\textsf{L1}}
\newcommand{\ltwo}{\textsf{L2}}
\newcommand{\lthree}{\textsf{L3}}

\newcommand{\adm}[2]{\mbox{#1\,\textcolor{svGrey}{\textperiodcentered}\,#2}}

\newcommand{\hasY}{\cmk}       
\newcommand{\hasP}{\pmk}       
\newcommand{\hasN}{\xmk}       

\newcommand{\cyes}{\cmk}
\newcommand{\cpart}{\pmk}
\newcommand{\cno}{\xmk}
\newcommand{\cna}{\nmk}
\newcommand{\Cy}{\cellcolor{svFamII!16}\cyes}
\newcommand{\Cp}{\cellcolor{svGold!20}\cpart}
\newcommand{\Cn}{\cellcolor{svRed!10}\cno}
\newcommand{\Ca}{\cna}

\newcommand{\famrow}[3]{%
  \multicolumn{#1}{@{}l}{\textcolor{#2}{\rule[-0.25ex]{2.2pt}{1.9ex}}\;#3}\\
  \addlinespace[1pt]}

\newcommand{\pre}{\textsuperscript{\dag}}

\newcommand{\searchdate}{30 September 2026}
\newcommand{\searchdateshort}{Sept.\ 2026}

\begin{document}

\begin{frontmatter}

\title{Seeing Without Seeing: A Review of Visual Mental Imagery Reconstruction from Brain Activity}

\author{Shiyi Zhang}
\ead{sy.zhang4@siat.ac.cn}

\author{Dong Liang}
\ead{dong.liang@siat.ac.cn}

\author{Yihang Zhou\corref{cor1}}
\ead{yh.zhou2@siat.ac.cn}

\cortext[cor1]{Corresponding author.}

\affiliation{organization={Shenzhen Institutes of Advanced Technology,
            Chinese Academy of Sciences},
            city={Shenzhen},
            country={China}}

\begin{abstract}
Reconstructing \emph{visual mental imagery} from brain activity differs from
reconstructing \emph{perceived} images. Imagery is generated top-down, its
signal is weak, and the imagined content is never observed, so decoders must
be trained and scored against a proxy such as a cue or a previously seen
exemplar. This criterion-guided methodological review of mainly
non-invasive studies (fMRI, EEG, MEG) is narrative, with explicit inclusion
criteria, rather than systematic. We organise the field into a two-stage
pipeline, two design axes and four method families, and formalise
\emph{target-free evaluation}, in which the proxy gap is a systematic bias
rather than noise. A two-dimensional admissibility criterion crosses
trial-level separability from working-memory maintenance with the memory
source of the imagined content. Among the studies reviewed, every
image-scored reconstruction concerns a recalled exemplar, none builds a
scorable target for concept-generated content, and all come from fMRI,
consistent with missing image-level targets in electrophysiological
datasets. Published values for five decoders, compared under a common human
forced-choice framing, show seen-to-imagined drops of 5.9 to 26.1 percentage
points and a change in method ordering. We treat this as a diagnostic
signal, not a controlled benchmark result, and specify a within-subject test
for an altered neural code. Beyond this synthesis, we make two proposals.
First, because frozen generative priors can produce photorealistic images
weakly coupled to the neural signal, we combine existing null-signal
controls into a normalised neural dependence with a within-category null, a
diagnostic quantity rather than a validated metric. Second, we propose a
candidate checklist of ten reporting items with illustrative pass criteria.
In an exploratory application to ten studies spanning Families~I--III, none
reports all items in a fully checkable form; this describes reporting
practice, not study quality.
\end{abstract}


\end{frontmatter}

\section{Introduction}
\label{sec:intro}

\subsection{Motivation}
\label{subsec:motivation}

For two decades, decoding visual content from brain activity has mostly
meant reconstructing what a person \emph{sees}. This review addresses what
a person \emph{imagines} without an external stimulus. Imagery re-enacts
perceptual representations \emph{top-down}, so its signal is coarser and
weaker than in perception, and its overlap with perception is strong for
semantic content but weak for retinotopic detail. Methods carried over from
perceptual reconstruction therefore do not transfer directly, and the
content they aim to recover is never observed.

\subsection{Scope and Relation to Prior Reviews}
\label{subsec:scope-intro}

We review methods that decode or reconstruct voluntary visual mental imagery
from human fMRI, EEG and MEG, admitting intracranial recordings where they
bear on the same question (Sections~\ref{subsec:terminology}
and~\ref{subsec:methodology}). Electrophysiology shows imagined content to
be decodable, most robustly from parieto-occipital alpha-band EEG
activity~\cite{Xie2020alpha,StecherKaiser2024alpha}, reinstated in single
neurons~\cite{Kreiman2000imagery} and usable for closed-loop
control~\cite{Cerf2010online}, with null results also
reported~\cite{Shatek2019decoding}; our search identified no image-level
reconstruction from it scored against an image-level target
(Section~\ref{subsec:eegmeg}). Seen-image reconstruction has been
reviewed~\cite{Tanaka2025visual,Guo2025fmri2image}, and its pipelines and
corpora~\cite{Allen2022massive,Gifford2022large} enter here only as an upper
bound and as reusable components.

The closest prior review~\cite{Tanaka2025visual} raises zero-shot and
spurious-prediction concerns for presented stimuli, where the scoring target
exists by construction. Without such a target, further elements are needed:
transfer across a change of state (Sec.~\ref{subsec:famIII}), an
admissibility criterion (Sec.~\ref{subsec:terminology}), the proxy gap
(Eq.~\eqref{eq:eval-bias}) and checkable reporting standards
(Sec.~\ref{subsec:guidelines}); Table~\ref{tab:reviews} compares the other
reviews, including the survey of Guo et al.~\cite{Guo2025fmri2image}\pre{}.
From Shirakawa et al.~\cite{Shirakawa2025spurious} we adopt the results on
output dimension collapse, identification inflation and the recovery check
and ask which carry over when no ground-truth image exists.

\begin{table*}[!htbp]
\centering
\setlength{\tabcolsep}{3pt}
\renewcommand{\arraystretch}{1.04}
\begin{threeparttable}
\caption{Prior reviews compared with this review. Columns 4--7 record
whether a review gives each criterion a dedicated section (note~a), and the
last column states its non-overlap with this review. The table backs delta
rather than priority claims, with per-row evidence in \ref{app:reviews}.}
\label{tab:reviews}
\scriptsize
\begin{tabularx}{\linewidth}{@{}L{1.95cm}L{1.6cm}L{1.42cm}*{4}{M{0.64cm}}Y@{}}
\toprule
\textbf{Review} & \textbf{State covered} & \mbox{\textbf{Modality}} &
\rotatebox[origin=l]{90}{\textbf{Unif.\ taxon.}} &
\rotatebox[origin=l]{90}{\textbf{Target-free eval.}} &
\rotatebox[origin=l]{90}{\textbf{Report.\ stds.}} &
\rotatebox[origin=l]{90}{\textbf{Ethics}} &
\textbf{Non-overlap with this review} \\
\midrule
Kosslyn et al.\ 2001~\cite{Kosslyn2001neural} &
Imagery & fMRI \vsep PET \vsep TMS &
\hasN & \hasN & \hasN & \hasN &
Depictive substrate assumed here, predates decoding \\
\addlinespace[1.5pt]
Pearson et al.\ 2015~\cite{Pearson2015mental} &
Imagery & fMRI \vsep behav. &
\hasN & \hasN & \hasN & \hasN &
Mechanisms and clinical use, decoding only as evidence of depictive format \\
\addlinespace[1.5pt]
Pearson 2019~\cite{Pearson2019human} &
Imagery & fMRI \vsep behav.\ \vsep TMS &
\hasN & \hasP\textsuperscript{b} & \hasN & \hasN &
Vividness/aphantasia continuum, here a mandatory covariate \\
\addlinespace[1.5pt]
Dijkstra et al.\ 2019~\cite{DijkstraBosch2019shared} &
Both & fMRI \vsep MEG &
\hasN & \hasN & \hasN & \hasN &
Neuroscience of the overlap, here a transfer problem with a conditional shift \\
\addlinespace[1.5pt]
Vigotsky et al.\ 2024~\cite{Vigotsky2024mental} &
Mental states (not imagery-specific) & fMRI &
\hasN & \hasY & \hasP\textsuperscript{c} & \hasN &
Generic decoder-validity critique, here localised to imagery with pass
criteria \\
\addlinespace[1.5pt]
Guo et al.\ 2025~\cite{Guo2025fmri2image}\pre{} &
Perception & fMRI only &
\hasY & \hasN & \hasN & \hasN &
Closest in form (pipeline taxonomy), but all methods scored against
presented images, so no proxy gap \\
\addlinespace[1.5pt]
Kamitani et al.\ 2025~\cite{Tanaka2025visual} &
Perception (imagery as application) & fMRI &
\hasP\textsuperscript{d} & \hasP\textsuperscript{d} & \hasN & \hasN &
Closest in substance (latent lineage, zero-shot, spurious prediction), but
for \emph{seen} images \\
\midrule
\rowcolor{svBlue!7}
\textbf{This review} &
\textbf{Imagery} (perception as baseline only) &
fMRI \vsep EEG \vsep MEG \vsep iEEG &
\hasY & \hasY & \hasY & \hasY &
\emph{Delta}: four families with bounded composability, WM separability
$\times$ memory source, proxy gap (Eq.~\eqref{eq:eval-bias}) and $\Lam$
(Eq.~\eqref{eq:lambda}), ten reporting standards applied to ten studies
(Table~\ref{tab:compliance}), modality-restricted search (Sec.~\ref{subsec:eegmeg}) \\
\bottomrule
\end{tabularx}
\begin{tablenotes}[flushleft]
\scriptsize
\item[a] \hasY{} dedicated (sub)section, \hasP{} substantive but not
dedicated, \hasN{} neither. \emph{Unified taxonomy} is one scheme placing
the reviewed methods, \emph{target-free evaluation} an explicit treatment of
scoring against something other than the target, and \emph{reporting
standards} prescriptions a later study can be checked against.
\item[b] Objective measurement of imagery itself, not evaluation of
reconstructions.
\item[c] Whether its decoder-validity cautions form an auditable protocol is
a judgement call.
\item[d] A scheme by latent representation, compositionality and
modularity, with zero-shot and spurious prediction raised for perception,
where ground truth exists.
\item[\dag] Preprint (rule~P3, Sec.~\ref{subsec:methodology}). Guo et al.\
claim the first systematic review of fMRI-to-image learning, which we do not
adjudicate, hence a delta rather than a priority claim.
\end{tablenotes}
\end{threeparttable}
\end{table*}

\subsection{A Unified Pipeline and Taxonomy}
\label{subsec:taxonomy-intro}

\begin{figure*}[!htbp]
\centering
\begin{tikzpicture}[
  font=\svfont,
  block/.style={rectangle, rounded corners=2pt, draw=navyblue, thick,
                fill=slatebg, minimum height=0.95cm, minimum width=1.5cm,
                align=center, inner sep=3pt},
  core/.style={rectangle, rounded corners=2pt, draw=navyblue, very thick,
               fill=navyblue!8, minimum height=1.0cm, minimum width=1.5cm,
               align=center, inner sep=4pt},
  frozen/.style={rectangle, rounded corners=2pt, draw=accentblue, thick,
                 fill=accentblue!12, minimum height=0.95cm, minimum width=2.0cm,
                 align=center, inner sep=3pt},
  meta/.style={rectangle, rounded corners=3pt, draw=navyblue, thick, dashed,
               fill=lightgray!35, minimum height=1.0cm, minimum width=1.5cm,
               align=center, inner sep=4pt},
  flow/.style={arrows={-Stealth[scale=1.1]}, accentblue, thick},
  back/.style={arrows={-Stealth[scale=1.1]}, alertred, thick, dashed},
  aux/.style={arrows={-Stealth[scale=1.1]}, navyblue, semithick},
  link/.style={accentblue, thick, dotted},
  famlabel/.style={font=\svfont\bfseries, navyblue},
]
\node[core] (x)    at (0,0)   {$\neural$\\[1pt]\svsmall neural signal};
\node[core] (z)    at (3.1,0) {$\decrep$\\[1pt]\svsmall latent};
\node[core] (Ihat) at (6.3,0) {$\est{I}$\\[1pt]\svsmall reconstruction};
\draw[flow] (x) -- node[above]{$\decmap$} (z);
\draw[flow] (z) -- node[above]{$\gen$} coordinate[pos=0.5] (gmid) (Ihat);

\node[meta] (T) at (-3.6,0) {$\transform$\\[1pt]\svsmall align / adapt};
\node[block, draw=lightgray!90!black] (raw) at (-3.6,2.35)
      {\svsmall cross-subject /\\\svsmall -session / -state\\[1pt]$\neural$};
\draw[aux]  (raw) -- (T);
\draw[flow] (T) -- node[above, pos=0.42]{\svsmall $\neural\!\mapsto\!\transform(\neural)$} (x);
\node[famlabel, below=0.18cm of T, align=center]
      {Family IV\\[-1pt]\svsmall\mdseries (meta-level)};

\node[block] (disc) at (1.95,2.35)
      {\svsmall classifier / regressor\\[1pt]\svsmall $\to$ label or $\decrep_{\mathrm{target}}$};
\node[block] (rsa)  at (5.75,2.35)
      {\svsmall RSA (model-free)\\[1pt]\svsmall $\mathrm{RDM}(\neural)$ vs.\ $\mathrm{RDM}(\Feat(I))$};
\node[font=\svsmall\itshape, navyblue] at ($(disc.east)!0.5!(rsa.west)$) {or};
\begin{scope}[on background layer]
  \node[fit=(disc)(rsa), rounded corners=3pt, draw=navyblue!35, dashed,
        fill=white, inner sep=4pt] {};
\end{scope}
\draw[flow] (x.north) |- (disc.west);
\node[famlabel] at ($(disc.north)!0.5!(rsa.north)+(0,0.42)$)
      {Family II \ \mdseries\svsmall($\gen=\mathrm{id}$)};

\node[frozen] (Gfroz) at (4.7,-1.95)
      {\svsmall frozen pretrained $\gen$\\\svsmall (diffusion / AR),\\\svsmall conditioned on $\decrep$};
\draw[link] (gmid) -- (Gfroz.north);
\fill[accentblue] (gmid) circle (1.6pt);
\node[famlabel, anchor=east] at ($(Gfroz.west)+(-0.15,0)$) {Family III};

\node[block] (enc) at (6.3,-3.45) {$\encmodel$\\[1pt]\svsmall encoding model};
\draw[back] (Ihat.south) -- node[right, font=\sffamily\svsmall, alertred]{predict} (enc.north);
\draw[back] (enc.west) -| (x.south);
\node[font=\svsmall, alertred, fill=white, inner sep=1.5pt] at (2.65,-3.45)
      {$\arg\min_{I}\,\lVert\neural-\encmodel(\Feat(I))\rVert^{2}$};
\node[famlabel, below=0.15cm of enc] {Family I};

\begin{scope}[on background layer]
  \node[fit=(x)(z)(Ihat), rounded corners, draw=navyblue!30,
        fill=navyblue!3, inner sep=8pt] (bbmain) {};
  \node[anchor=south east, inner sep=2pt, font=\svsmall\itshape, navyblue!70] at (bbmain.north east)
        {Unified two-stage pipeline\ \ $\neural\to\decrep\to\est{I}$};
\end{scope}
\end{tikzpicture}
\caption{Unified decoding pipeline of Eq.~\eqref{eq:pipeline},
$\neural\xrightarrow{\decmap}\decrep\xrightarrow{\gen}\est{I}$, and the four
method families. \textbf{Family~I} inverts an encoding model $\encmodel$,
\textbf{Family~II} decodes discriminatively with $\gen=\mathrm{id}$, and
\textbf{Family~III} conditions a \emph{frozen} pretrained $\gen$ on
$\neural$. \textbf{Family~IV} is a meta-level alignment transform on the
input, composable with Families~I--III under the conditions of
Sec.~\ref{subsec:famIV}.}
\label{fig:pipeline}
\end{figure*}
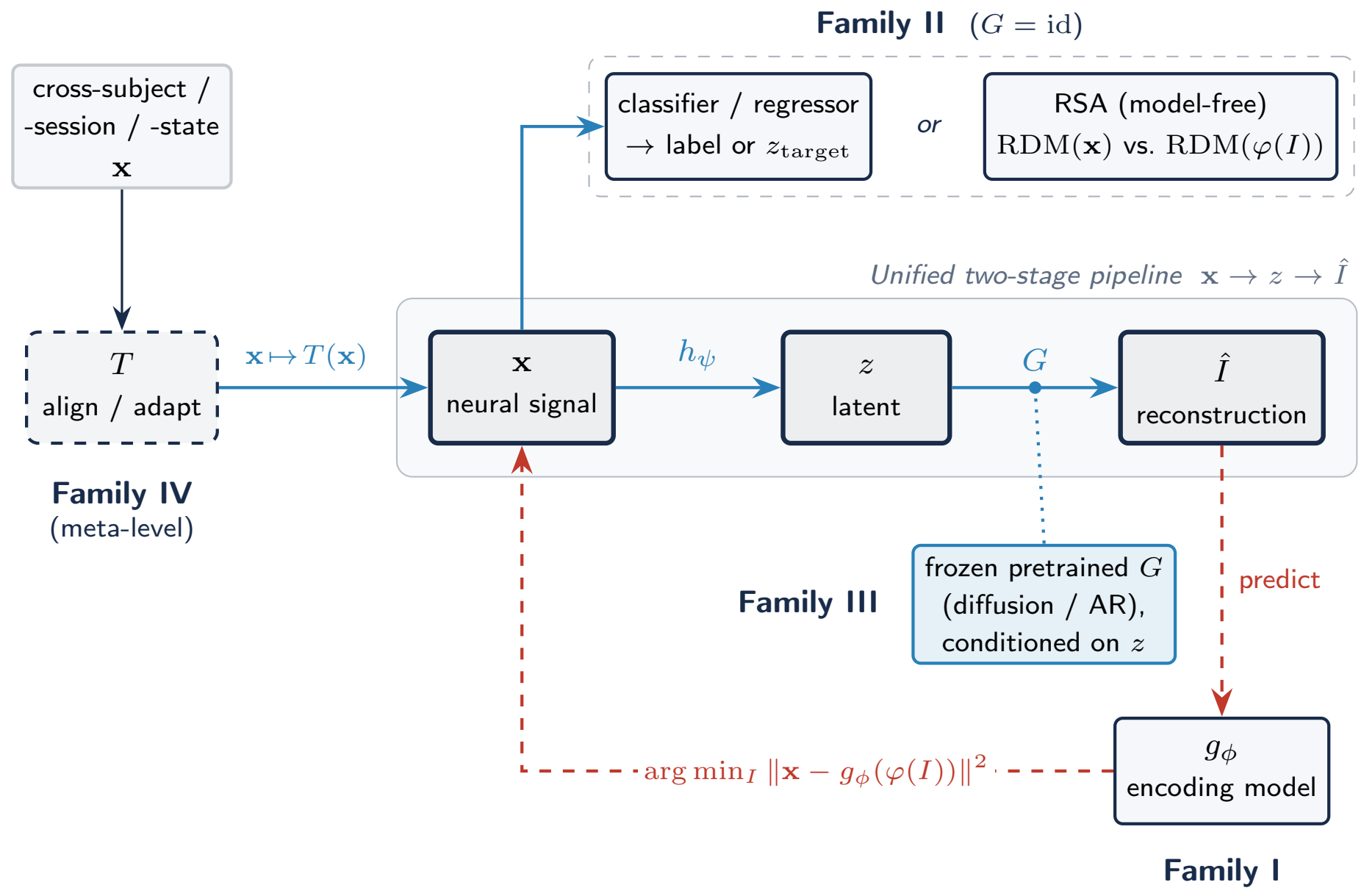

Almost every method instantiates a two-stage map in which a decoder
$\decmap$ transforms the neural response $\neural$ into a representation
$\decrep$ and a generator $\gen$ renders it into a reconstruction $\est{I}$
(Eq.~\eqref{eq:pipeline}, Fig.~\ref{fig:pipeline}). Methods differ along two
design axes (Section~\ref{subsec:pipeline}) and fall into four families
(Section~\ref{subsec:families}).

\subsection{Main Challenges}
\label{subsec:challenges-intro}

\begin{figure*}[!htbp]
\centering
\begin{tikzpicture}[
  font=\svfont,
  sig/.style={rectangle, rounded corners=2pt, draw=navyblue, thick,
              fill=slatebg, minimum height=1.0cm, minimum width=1.7cm,
              align=center, inner sep=4pt},
  lat/.style={rectangle, rounded corners=2pt, draw=navyblue, thick,
              fill=navyblue!8, minimum height=1.0cm, minimum width=1.9cm,
              align=center, inner sep=4pt},
  genG/.style={rectangle, rounded corners=2pt, draw=accentblue, very thick,
               fill=accentblue!12, minimum height=1.0cm, minimum width=1.9cm,
               align=center, inner sep=4pt},
  img/.style={rectangle, rounded corners=2pt, thick, align=center,
              minimum height=1.0cm, minimum width=1.9cm, inner sep=4pt},
  okflow/.style={arrows={-Stealth[scale=1.1]}, accentblue, thick},
  badflow/.style={arrows={-Stealth[scale=1.1]}, alertred, thick},
  warn/.style={rectangle, rounded corners=3pt, draw=alertred, thick, dashed,
               fill=alertred!6, align=center, inner sep=4pt,
               font=\svfont},
  shiftbox/.style={rectangle, rounded corners=3pt, draw=navyblue, thick, dashed,
                   fill=lightgray!35, align=center, inner sep=4pt,
                   font=\svfont},
  panel/.style={font=\svfont\bfseries, anchor=south west},
]
\node[sig]  (xT) at (0,0)    {$\neural$\\[1pt]\svsmall strong signal};
\node[lat]  (zT) at (3.5,0)  {$\decrep$\\[1pt]\svsmall high-fidelity};
\node[genG] (GT) at (7.0,0)  {$\gen$\\[1pt]\svsmall generator};
\node[img, draw=accentblue, fill=accentblue!8] (IT) at (10.5,0)
     {$\est{I}\!\approx\!\imgcontent$\\[1pt]\svsmall faithful};
\draw[okflow] (xT) -- node[above]{\svsmall high SNR} node[below]{$\decmap$} (zT);
\draw[okflow] (zT) -- node[above]{$\gen$} (GT);
\draw[okflow] (GT) -- (IT);

\node[sig, draw=alertred] (xB) at (0,-3.9) {$\neural$\\[1pt]\svsmall weak / uninformative};
\node[lat, draw=alertred, fill=alertred!8] (zB) at (3.5,-3.9)
     {$\decrep_{\text{class-prior}}$\\[1pt]\svsmall collapsed};
\node[genG, draw=alertred, fill=alertred!10] (GB) at (7.0,-3.9)
     {\svsmall frozen $\gen$\\\svsmall high-capacity};
\node[img, draw=alertred, fill=alertred!8] (IB) at (10.5,-3.9)
     {$\est{I}_{\text{spurious}}$\\[1pt]\svsmall photorealistic};
\draw[badflow] (xB) -- node[above]{\svsmall $\Lam\!\approx\!0$} (zB);
\draw[badflow] (zB) -- node[above]{$\gen$} (GB);
\draw[badflow] (GB) -- (IB);

\draw[alertred, thick, dotted] (IB.south) .. controls +(0,-1.15) and +(0,-1.15) .. (xB.south)
      node[pos=0.5, below=1pt, font=\svsmall, alertred]
      {reconstruction decoupled from $\neural$};

\node[shiftbox] (shift) at ($(zT)!0.5!(zB)$)
     {\textbf{Distribution shift}\\ $p(\decrep\!\mid\!\neural)$: perc.\ $\neq$ imagery};
\node[warn] (diag) at ($(GT)!0.5!(GB)+(1.75,0)$)
     {\textbf{Diagnostic}\\ $\mathcal{E}$ high, $\Lam\!\approx\!0$\\
      (Eq.~\eqref{eq:lambda})};

\draw[navyblue!60, thick, dotted] (shift.south) -- (zB.north);
\draw[alertred!70, thick, dotted] (diag.south) -- (IB.north west);

\begin{scope}[on background layer]
  \node[fit=(xT)(IT), rounded corners, draw=accentblue!30,
        fill=accentblue!4, inner sep=8pt] (pT) {};
  \node[fit=(xB)(IB), rounded corners, draw=alertred!30,
        fill=alertred!4, inner sep=8pt] (pB) {};
\end{scope}
\node[panel, accentblue] at (pT.north west) {Genuine read-out};
\node[panel, alertred]   at (pB.north west) {Prior-driven hallucination};
\end{tikzpicture}
\caption{Genuine read-out (top) versus prior-driven hallucination (bottom),
where an uninformative $\neural$ collapses onto
$\decrep_{\text{class-prior}}$ and the frozen generator still emits a
photorealistic $\est{I}_{\text{spurious}}$ (Eq.~\eqref{eq:spurious}).
Between the panels are the failure driver, a conditional shift in
$p(\decrep\mid\neural)$, and the diagnostic signature of high proxy score
$\mathcal{E}$ with $\Lam\approx0$
(Eq.~\eqref{eq:lambda})~\cite{Shirakawa2025spurious}. Convincing
reconstructions have been obtained from randomised brain data by repeated
generation and post hoc selection~\cite{Shirakawa2025spurious}.}
\label{fig:spurious}
\end{figure*}
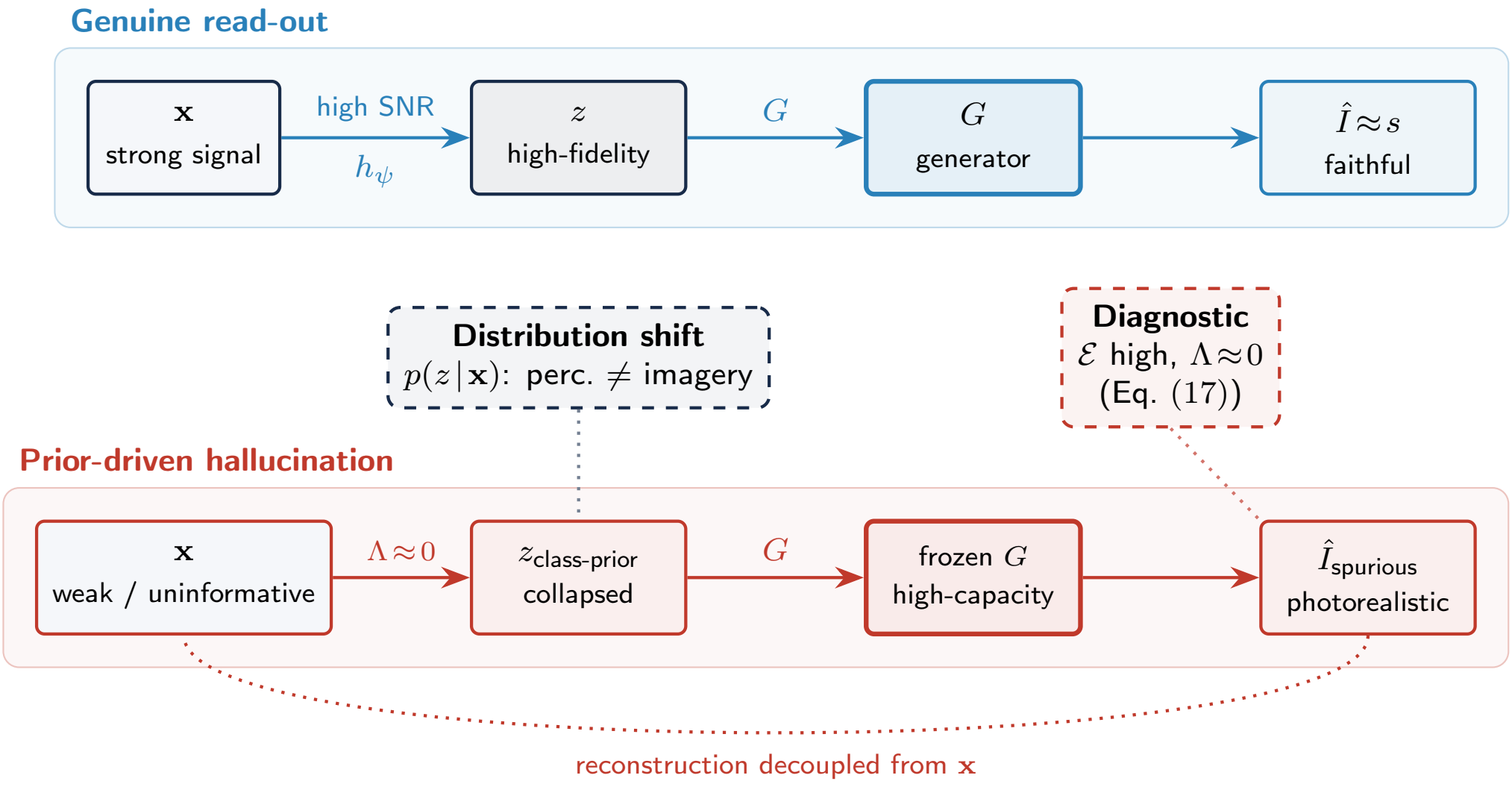

Two issues run through every family: \emph{perception-to-imagery transfer}
and \emph{target-free evaluation} (Sections~\ref{subsec:famIII}
and~\ref{subsec:targetfree}). They meet in the problem of \emph{spurious
reconstruction} (Fig.~\ref{fig:spurious}): a frozen, high-capacity generator
can synthesise a photorealistic image weakly coupled to $\neural$ while
feature, semantic and realism scores are satisfied largely by the
prior~\cite{Shirakawa2025spurious}, a concern a reanalysis has raised for a
published imagery result~\cite{Shirakawa2025advancing}\pre{}.
Section~\ref{subsec:spurious} treats the mechanism, and
Eq.~\eqref{eq:lambda} gives a statistic that separates read-out from prior.

\subsection{Contributions}
\label{subsec:contributions}

\begin{itemize}[leftmargin=1.3em,itemsep=1pt]
\item \textbf{Imagery reconstruction as the object.} Earlier reviews cover
the neuroscience of
imagery~\cite{Kosslyn2001neural,Pearson2015mental,Pearson2019human,DijkstraBosch2019shared}
or seen-image reconstruction~\cite{Guo2025fmri2image,Tanaka2025visual}; here
internally generated vision is treated as a decoding problem in its own
right.
\item \textbf{A taxonomy and an admissibility criterion.} Four families
with bounded composability, and a criterion crossing working-memory
separability with memory source (\lone/\ltwo/\lthree).
\item \textbf{Evaluation without ground truth.} A synthesis of published
imagery performance (Table~\ref{tab:performance}) and, as an author
proposal, a normalised neural dependence that consolidates existing
null-signal controls~\cite{Shirakawa2025spurious,Kneeland2026mirage}
(Eq.~\eqref{eq:lambda}), a diagnostic quantity not yet validated on
independent data.
\item \textbf{A candidate reporting checklist.} Ten proposed reporting items
with illustrative pass criteria (Table~\ref{tab:standards}), an author
proposal rather than an established standard, applied in an exploratory
manner to ten studies to check whether the items are reported in a
checkable form, not to rank study quality (Table~\ref{tab:compliance}),
followed by open problems.
\end{itemize}

\noindent The first three contributions and the performance comparison
synthesise the literature; the neural dependence and the reporting
checklist are proposals of the authors.

\noindent Per-cell evidence for the tables, the electrophysiological
results, the derivation of Eq.~\eqref{eq:pairwise-inflation} and further
figures are given in the appendices following the references.

\FloatBarrier

\section{Scope, Terminology, and Problem Formulation}
\label{sec:scope}

\subsection{Scope}
\label{subsec:scope}

This review covers methods that decode or reconstruct \emph{internally
generated} visual content, as distinct from reconstruction of
\emph{perceived} images~\cite{Kay2008identifying,Miyawaki2008visual}. Most
decoders therefore face a \emph{perception-to-imagery transfer} problem, and
because the imagined content is latent, an
\emph{evaluation-without-ground-truth} problem. Seen-image reconstruction,
such as MindEye~\cite{Scotti2023reconstructing},
MindEye2~\cite{Scotti2024mindeye2}, latent-diffusion
decoders~\cite{Takagi2023highres,Ozcelik2023natural},
MinD-Vis~\cite{Chen2023seeing} and mesh-guided
diffusion~\cite{Zhang2026meshd2r}\pre{}, is treated as a perceptual
baseline whose numbers enter only as the paired vision condition of an
imagery benchmark.

\subsection{Terminology: Defining the In-Scope State}
\label{subsec:terminology}

\paragraph{Voluntary visual mental imagery}
The core object is the \emph{deliberate, top-down generation} of a
quasi-perceptual visual experience \emph{without the corresponding external
stimulus}, under intentional control, typically initiated by a label, a
word or a previously seen exemplar, with retained awareness that the
experience is imagined~\cite{Pearson2019human,Kosslyn2001neural}.
Neighbouring states are separated in Table~\ref{tab:states}, whose ``Scope''
column is implemented at screening by criteria~E2 and~E3
(Section~\ref{subsec:methodology}).

\begin{table}[!htbp]
\centering
\caption{Internally generated visual states and their scope decision. Only
voluntary imagery is core, working memory and dreaming are boundary cases,
and hallucination and attempted generation are out of scope. Perception is
listed as the contrast class.}
\label{tab:states}
\footnotesize
\setlength{\tabcolsep}{3pt}
\renewcommand{\arraystretch}{1.0}
\begin{tabular}{@{}L{2.6cm}cccL{1.4cm}L{2.15cm}@{}}
\toprule
\textbf{State} & \textbf{Volition} & \textbf{Ext.\ stim.} &
\textbf{Pseudo-GT} & \textbf{Aware $\neq$ perc.} & \textbf{Scope} \\
\midrule
Voluntary imagery      & Vol.\   & None          & Yes (cue)  & Yes      & \textbf{Core} \\
\addlinespace[1pt]
Working memory         & Vol.\   & None (removed)& Yes (seen) & Yes      & Boundary \\
\addlinespace[1pt]
Dream imagery          & Invol.\ & None          & No         & Usu.\ no & Boundary \\
\addlinespace[1pt]
Hallucination (clin.)  & Invol.\ & None          & No         & Often no & \textbf{Out} \\
\addlinespace[1pt]
Attempted/covert       & Vol.\   & None          & Variable   & Yes      & Out (mention) \\
\addlinespace[1pt]
Perception (contrast)  & ---     & Present       & Yes (GT)   & ---      & Out (contrast) \\
\bottomrule
\end{tabular}
\end{table}

\paragraph{Imagery, working memory and hallucination}
Imagery and working memory are both voluntary and stimulus-independent at
measurement and share representational format in visual
cortex~\cite{Albers2013shared}, but working memory \emph{maintains} an item
just perceived, which supplies a \emph{de facto} ground truth, whereas
imagery can \emph{generate} content specified only by a symbolic cue; delay-period
decoding~\cite{HarrisonTong2009decoding} is therefore a boundary case.
Hallucination is involuntary, usually lacks experimenter-specified content
and is often believed real, so it is excluded.

\paragraph{A two-dimensional admissibility criterion}
Whether an exemplar is shown inside the imagery trial is a necessary but
not sufficient test: a paradigm may show no exemplar in the trial yet ask
the participant to \emph{retrieve} a picture seen an hour earlier, which
differs from generating content never presented as an image. We therefore
cross it with the \emph{memory source} of the imagined content:

\begin{description}[leftmargin=1.4em,style=nextline,itemsep=1pt]
\item[\lone{} --- maintenance.] An exemplar presented \emph{within the
trial} is held across a delay~\cite{HarrisonTong2009decoding}.
\item[\ltwo{} --- recall of a learned exemplar.] A specific image memorised
\emph{before} the session is recalled on a symbolic cue, as in
NSD-Imagery~\cite{Kneeland2025nsdimagery}, Deeprecon~\cite{Shen2019deep}
and the memorised artworks of~\cite{Naselaris2015voxelwise}; the seen
picture is the image-level scoring target.
\item[\lthree{} --- generation.] Content is specified by a category, concept
or description~\cite{Horikawa2017generic,StecherKaiser2024alpha,Kneeland2025nsdimagery},
so no image-level target is presented.
\end{description}

\noindent Within \lthree{} we distinguish \emph{purely conceptual
generation}, for which no image-level target exists even in principle, from
\emph{compositionally constructed targets}, for which a ground-truth image
can be generated by rule from the cue; the compositional cueing of
Sec.~\ref{subsec:gap1} belongs to the latter.

\noindent The two dimensions are written together
(\sepyes\,\textperiodcentered\,\ltwo{}, \sepno\,\textperiodcentered\,\lone{},
\dots) in Tables~\ref{tab:families} and~\ref{tab:datasets}. The criterion,
for example, classifies the largest public EEG imagery dataset as
entangled, since the target picture is shown inside every imagery
trial~\cite{Gao2026eegvi}. Because purely conceptual \lthree{} content has
no image-level target, image-based scoring is possible only under \lone{},
\ltwo{} or compositionally constructed \lthree{} targets, and the last are
not used in the reviewed corpus. Where an
exemplar is shown inside the trial, the memory source is \lone{} unless the
instruction directs imagery away from it (Fig.~\ref{fig:admissibility});
which cells are empty by definition and which are merely unoccupied is
discussed in \ref{app:emptycells}.

\subsection{Problem Formulation}
\label{subsec:formulation}

Let $\neural$ be the response on a single imagery trial, a voxel pattern for
fMRI or a channel-by-time array for EEG/MEG, often reduced to band-limited
power because the content-bearing imagery signal is oscillatory rather than
time-locked~\cite{Xie2020alpha,StecherKaiser2024alpha}, and let
$\imgcontent$ be the imagined content. No external image $I=\imgcontent$ is
presented: the experimenter observes only a cue $\cue$, which may be a
label, a word or a previously seen exemplar $\exemplar$. The cue is
therefore a \emph{proxy} target and $\imgcontent$ is never observed; an
image-level proxy exists only under \lone{} and \ltwo{}.

A decoder estimates a class, a ranking of candidates in an embedding space,
or a reconstruction in pixel space or a semantic latent such as
CLIP~\cite{Naselaris2011encoding}. Supervision is \textbf{(a)}
imagery-trained on scarce, low-SNR imagery pairs; \textbf{(b)}
perception-transfer, fitting a decoder or a forward encoding model
$g_{\phi}:I\mapsto\neural$ on perception and applying it to imagery, which
relies on representational overlap~\cite{DijkstraBosch2019shared}; or
\textbf{(c)} hybrid. Every method with imagery-reconstruction numbers in
Table~\ref{tab:performance} falls in~(b). Encoding-model
inversion learns $g_{\phi}$ on perception and recovers content under a
natural-image prior~\cite{Naselaris2009bayesian}:
\begin{equation}
\label{eq:inversion}
\est{\imgcontent}
  = \operatorname*{arg\,max}_{\imgcontent}\;
    p\!\left(\neural^{\mathrm{img}} \mid g_{\phi}(\imgcontent)\right)\,
    p(\imgcontent).
\end{equation}

\subsection{A Unified Taxonomy}
\label{subsec:taxonomy}

The primary axis is the \emph{method family}
(Section~\ref{subsec:families}). Six cross-cutting dimensions provide a
shared vocabulary for characterising each family (Table~\ref{tab:families}):
\emph{modality} (fMRI, EEG, MEG, iEEG); \emph{source state} (voluntary
imagery, working-memory maintenance, dreaming); \emph{memory source}
(\lone, \ltwo, \lthree); \emph{supervision source} (imagery-trained,
perception-transfer, hybrid); \emph{output target} (classification,
retrieval, pixel-level or semantic reconstruction); and \emph{stimulus type}
(simple shapes and letters, natural scenes, concept words).

\begin{sidewaystable}
\centering
\setlength{\tabcolsep}{3.5pt}
\renewcommand{\arraystretch}{1.08}
\begin{threeparttable}
\caption{Method families along the dimensions of Sec.~\ref{subsec:taxonomy},
with anchor studies and the admissibility of their imagery conditions.
Modality entries list the modalities supplying anchors as of \searchdate,
and \dag{} marks a preprint. Family~IV is \emph{meta-level} and defines no
decoding objective of its own.}
\label{tab:families}
\scriptsize
\begin{tabularx}{\linewidth}{@{}L{2.75cm}L{2.15cm}L{2.85cm}L{2.55cm}YL{3.55cm}@{}}
\toprule
\textbf{Typical modality} & \textbf{Applicable source states} &
\textbf{Supervision source} & \textbf{Primary output target} &
\textbf{Anchor studies} & \textbf{Admissibility}\textsuperscript{a} \\
\midrule
\famrow{6}{svFamI}{\textbf{Ia.\ Encoding-model inversion into pixel space}}
fMRI only; no EEG/\allowbreak MEG/\allowbreak iEEG anchor as of \searchdateshort &
Vol.\ imagery (scenes, shapes) &
Percep.-trained, inverted (cross-state) &
ID or pixel recon.\ via a retinotopic forward model &
Thirion~\cite{Thirion2006inverse}; Naselaris~\cite{Naselaris2015voxelwise};
Senden~\cite{Senden2019imagery} &
\adm{\sepyes}{\ltwo}~\cite{Naselaris2015voxelwise,Senden2019imagery} \\
\addlinespace[3pt]
\famrow{6}{svFamIb}{\textbf{Ib.\ Encoding-model inversion through a feature space}}
fMRI only &
Vol.\ imagery; retrieved faces &
Percep.-trained, inverted with a deep-feature or generative prior &
Pixel recon.\ from decoded features &
Shen~\cite{Shen2019deep}; Koide-Majima~\cite{KoideMajima2024mental};
Lee \& Kuhl~\cite{LeeKuhl2016reconstructing} &
\adm{\sepyes}{\ltwo}~\cite{Shen2019deep,KoideMajima2024mental,LeeKuhl2016reconstructing} \\
\addlinespace[3pt]
\famrow{6}{svFamII}{\textbf{II.\ Discriminative decoding and representational geometry}}
fMRI \vsep EEG \vsep MEG \vsep iEEG (only family anchored in every modality) &
Vol.\ imagery; WM &
Imagery-trained or cross-state (labels) &
ID\,/ classif.\,/ retrieval; RSA &
fMRI: \cite{Reddy2010reading,Horikawa2017generic,Stokes2009topdown},
\cite{StecherKaiser2026neuro}\pre{}.
EEG: \cite{Xie2020alpha,StecherKaiser2024alpha,Kwon2025functional};
null: \cite{Shatek2019decoding}.
MEG: \cite{Dijkstra2018differential}. iEEG: \cite{Kreiman2000imagery}.
WM boundary: \cite{HarrisonTong2009decoding} &
\adm{\sepyes}{\lthree}~\cite{Horikawa2017generic,StecherKaiser2024alpha},~\cite{StecherKaiser2026neuro}\pre{}\newline
\adm{\sepyes}{\ltwo}~\cite{Reddy2010reading,Stokes2009topdown,Xie2020alpha}\newline
\adm{\sepyes}{\ltwo\textsuperscript{c}}~\cite{Kreiman2000imagery}\newline
\adm{\sepno}{\lone}~\cite{Shatek2019decoding,Kwon2025functional,HarrisonTong2009decoding,Dijkstra2018differential} \\
\addlinespace[3pt]
\famrow{6}{svFamIII}{\textbf{III.\ Generative cross-decoding}}
fMRI; one exploratory EEG attempt, not image-scored
(Sec.~\ref{subsec:eegmeg}) &
Vol.\ imagery (transfer from perception) &
Percep.-transfer\,/ hybrid &
Pixel + semantic recon.\ (diffusion\,/ AR) &
NSD-Imagery~\cite{Kneeland2025nsdimagery}; MIRAGE~\cite{Kneeland2026mirage};
Spera~\cite{Spera2026seeing}\pre{}; EEG, exploratory:
\cite{Shimizu2022improving} &
\adm{\sepyes}{\ltwo}~\cite{Kneeland2025nsdimagery,Kneeland2026mirage,Spera2026seeing}\newline
\adm{\sepyes}{\lthree\textsuperscript{d}}~\cite{Kneeland2025nsdimagery}\newline
\adm{\sepno}{\lone}~\cite{Shimizu2022improving} \\
\addlinespace[3pt]
\famrow{6}{svFamIV}{\textbf{IV.\ Alignment and adaptation} \emph{(meta-level)}}
fMRI (whole-brain), cross-subj.; EEG only via foundation models (outlook) &
Vol.\ imagery; low-SNR regimes &
Imagery$\leftrightarrow$\allowbreak perception alignment; cross-subj.\
adaptation; foundation-model transfer (outlook) &
\emph{Meta}: improves the mapping feeding the targets above &
Hyperalignment~\cite{Haxby2011common,Haxby2020hyperalignment};
MindAligner~\cite{Dai2025mindaligner}; NeuroSwift~\cite{Zhang2025neuroswift};
code conversion~\cite{Wang2025interindividual};
StableMind~\cite{StableMind2026}\pre{}; Spera~\cite{Spera2026seeing}\pre{};
outlook: \cite{Jiang2024labram}, \cite{Yang2024brainmae}\pre{},
\cite{Lane2025scaling}\pre{} &
\sepna\textsuperscript{b}\newline
\adm{\sepyes}{\ltwo}~\cite{Spera2026seeing} \\
\bottomrule
\end{tabularx}
\begin{tablenotes}[flushleft]
\scriptsize
\item[a] Criterion of Sec.~\ref{subsec:terminology}, shared with
Table~\ref{tab:datasets}. Trial level: \sepyes~separable, \sepno~entangled,
\sepunc~unclear, \sepna~no imagery condition. Memory source: \lone~in-trial
maintenance, \ltwo~recall of a pre-session exemplar, \lthree~generation from
a concept or description. Studies with several levels list each.
\item[b] Developed on perceptual, cross-subject or general neural data, so
it inherits the verdict of the imagery dataset it is applied to.
\item[c] Trial-level separable, but the pictures were viewed earlier in the
same session, so the recall interval is far shorter than for the other
\ltwo{} entries.
\item[d] Single-word conceptual stimuli without a ground-truth image, so
reconstructions are human-rated against the word and excluded from
Table~\ref{tab:performance}, Panel~B.
\item[\dag] Preprint at \searchdate. No conclusion or standard rests on a
preprint alone (rule~P3, Sec.~\ref{subsec:methodology}).
\end{tablenotes}
\end{threeparttable}
\end{sidewaystable}

\subsection{Review Methodology}
\label{subsec:methodology}

\paragraph{Type of review and search}
This is a criterion-guided methodological review, that is, a narrative
review with explicit criteria, rather than a PRISMA systematic review: no protocol was pre-registered and no two independent
raters screened. Instead, we state explicit criteria, give a uniform
admissibility verdict for every tabulated study and tie each tabulated cell
to a source sentence (\ref{app:corpus}). We searched Web of Science,
Scopus, PubMed, IEEE~Xplore, DBLP, arXiv and bioRxiv, with Google Scholar
for forward citations, using four query groups: imagery decoding (A),
generative reconstruction (B), evaluation methodology (C) and ethics (D).
A--C supply the tabulated studies, and Section~\ref{subsec:eegmeg} re-runs~A
restricted to EEG, MEG, iEEG, ECoG and sEEG. Searches closed on
\searchdate; the corpus effectively starts with inverse retinotopy in
2006~\cite{Thirion2006inverse}, earlier work entering as
background~\cite{Kosslyn2001neural,Kreiman2000imagery,Marks1973visual}. Only
English full texts were considered.

\paragraph{Inclusion and exclusion criteria}
A record is included if it \textbf{I1} decodes or reconstructs internally
generated visual content from human brain activity measured without the
corresponding stimulus; \textbf{I2} reports a quantitative read-out (an
accuracy, a reconstruction scored by an operator of Eq.~\eqref{eq:eval}, a
representational-geometry statistic, or, for datasets, a technical
validation); \textbf{I3} records humans with fMRI, EEG, MEG, iEEG or single
units; \textbf{I4} is peer-reviewed or a preprint meeting P1--P3; and
\textbf{I5} has an English full text. It is excluded if it \textbf{E1}
concerns perceived stimuli only; \textbf{E2} concerns non-visual content
such as imagined speech or motor imagery (visualised actions instructed as
visual are admitted as boundary cases); \textbf{E3} concerns involuntary or
pathological states; \textbf{E4} is non-human or simulation-only;
\textbf{E5} appears in a non-archival venue; or \textbf{E6} is a methods
paper without an imagery application.

\paragraph{Preprint policy}
A preprint is cited only if no peer-reviewed version existed at
\searchdate{} (P1) and anchors a claim only if no peer-reviewed source
supports it (P2); every preprint anchor is flagged \dag{}, and no standard
or conclusion rests on a preprint alone (P3).

\paragraph{Limitations and disclosure}
Selection was by the first author, with no independent second screener, so
the corpus is a traceable selection whose error has not been estimated, and
its criteria were fixed during the search. Two cited perceptual
methods~\cite{Zhang2025neuroswift,Zhang2026meshd2r}\pre{} share two authors
with this review; they enter under the same criteria as other work, as a
Family~IV anchor and a perceptual baseline respectively.

\FloatBarrier

\section{Background and Cognitive Foundations}
\label{sec:background}

\subsection{Neural Substrates: Visual Mental Imagery versus Perception}
\label{subsec:substrates}

\paragraph{A shared but asymmetric substrate}
Imagery re-enacts perceptual representations across the ventral visual
hierarchy~\cite{Ganis2004brain}. Classifiers trained on seen stimuli
generalise to imagined ones most robustly for high-level
features~\cite{Reddy2010reading,Horikawa2017generic}, shape-specific codes
reinstated during imagery localise to anterior lateral occipital
complex~\cite{Stokes2009topdown}, and imagery and perception share
representations of content and location~\cite{Cichy2012imagery}. Early
visual cortex nonetheless carries depictive imagery information: encoding
models fit on perception identify which of thousands of candidate images a
participant recalls, even from V1 or V2 voxels~\cite{Naselaris2015voxelwise}.
In EEG, shared object representations are carried by parieto-occipital
alpha~\cite{Xie2020alpha}, and medial-temporal neurons responsive to a
picture fire again when it is imagined~\cite{Kreiman2000imagery}.

The overlap is neither complete nor uniform. Imagery is generated from
frontoparietal and higher visual areas projecting back onto sensory
cortex~\cite{DijkstraBosch2019shared,Kosslyn2001neural,Ganis2004brain}, and
generative-feedback accounts explain why seen and mental images can elicit
distinct codes~\cite{Breedlove2020generative}. Early-visual reactivation is
coarser and weaker than in perception, and the share of the perceptual
representation recovered by imagery grows along the
hierarchy~\cite{Horikawa2017generic,DijkstraBosch2019shared}. Imagery
representations also emerge later, are more sustained and are less precise
than perceptual ones~\cite{Dijkstra2018differential,StecherKaiser2024alpha}.

\paragraph{Computational consequence: the transfer problem}
Because semantic representations overlap strongly and retinotopic detail
weakly, decoders transfer most readily at the semantic
level~\cite{Pearson2015mental}, which is why modern pipelines condition on
high-level embeddings. To compare transfer across metrics we use the
\emph{retention ratio}
\begin{equation}
\label{eq:rho}
\rho =
\frac{\mathcal{E}_{\mathrm{img}} - \mathcal{E}_{\mathrm{chance}}}
     {\mathcal{E}_{\mathrm{vis}} - \mathcal{E}_{\mathrm{chance}}},
\end{equation}
the fraction of above-chance vision performance that survives the move to
imagery, with $\mathcal{E}_{\mathrm{chance}}=50\%$ for two-alternative
protocols and $0$ for pixel correlation. On Table~\ref{tab:performance}
(Fig.~\ref{fig:transfer}), MIRAGE retains $\rho=0.83$ of its human-rated
margin but $0.47$ of its pixel correlation, and Brain-Diffuser $0.80$
against $0.60$, whereas iCNN retains $0.48$ of its pixel correlation while
its CLIP score is below chance in both conditions
(Sec.~\ref{subsec:metrics}). $\rho$ is unstable when its denominator is
small, and intervals can be formed only where dispersion is reported
(Panels~C and~D); since human ratings are not purely semantic, the contrast
between streams should not be generalised beyond these methods.

\subsection{Individual Differences and Boundary Conditions}
\label{subsec:individual}

\paragraph{Vividness and aphantasia}
Imagery vividness, usually measured by the VVIQ~\cite{Marks1973visual},
tracks the neural overlap between imagery and perception both trial-wise and
across individuals~\cite{Dijkstra2017vividness}. At the extreme lies
aphantasia, the congenital inability to generate voluntary
imagery~\cite{Zeman2015lives,KeoghPearson2018blind}. Chang
et al.~\cite{Chang2025imageless} decode imagined content from V1 in
aphantasic individuals as well as in controls, yet the patterns do not
cross-decode with perception (26.2\%, $p=0.274$, against 43.9\%,
$p<0.001$, in controls); sound-driven decoding in early visual
cortex~\cite{Vetter2014decoding} is diminished in
aphantasia~\cite{Montabes2024decoding}; decodable V1 representations can
occur without subjective imagery~\cite{Cabbai2024sensory}; and 7\,T fMRI
shows reduced fusiform--frontoparietal connectivity in
aphantasia~\cite{Liu2025visual}. Decodability and phenomenology can
therefore dissociate, so imagery ability should be reported as a covariate
(S4); no study in Table~\ref{tab:performance} relates its scores to it.

\paragraph{Reality monitoring and the working-memory boundary}
Candidate signals that mark an experience as internal are evaluated jointly
by a higher-order circuit~\cite{Dijkstra2022reality}, and the dominant cue
appears to be signal strength, since sufficiently vivid imagery is
misattributed to perception~\cite{DijkstraFleming2023signal}. Content may
thus be strong enough for a classifier yet too weak for experience, and a
decoder is not bound by the reality-monitoring threshold
(Sec.~\ref{subsec:privacy}). Working-memory contents are likewise decodable
from V1--V4~\cite{HarrisonTong2009decoding} in representations partly
shared with imagery~\cite{Albers2013shared}, which is why every tabulated
study carries both components of the admissibility verdict.

\section{A Unified Taxonomy Framework}
\label{sec:taxonomy}

\subsection{The Mind's-Eye Decoding Pipeline}
\label{subsec:pipeline}

\paragraph{A common two-stage map}
Decoding maps $\neural$ to a target through a model-side \emph{decoded
latent} $\decrep$ in a representation space $\mathcal{Z}$, such as a CLIP or
CNN-feature embedding~\cite{Radford2021learning}:
\begin{equation}
\label{eq:pipeline}
\underbrace{\neural}_{\text{neural signal}}
\;\xrightarrow{\;\decmap\;}\;
\underbrace{\est{\decrep}}_{\text{decoded latent }\in\,\mathcal{Z}}
\;\xrightarrow{\;\gen\;}\;
\underbrace{\est{I}}_{\text{reconstruction}},
\qquad
\est{I} = \gen\!\big(\decmap(\neural)\big),
\end{equation}
where $\decmap$ is a neural-to-latent decoder and $\gen$ a generator. With
$\gen=\mathrm{id}$ the latent is used directly for classification or
retrieval, which is where the electrophysiological literature currently sits
(Sec.~\ref{subsec:eegmeg}); in the inverted direction a forward encoding
model $\encmodel:\mathcal{I}\!\to\!\mathcal{X}$ is learned instead.

\paragraph{Two design axes}
Methods differ along two axes that the family labels of
Table~\ref{tab:families} partly conflate (Fig.~\ref{fig:designaxes}). The
\emph{direction of inference} is either \emph{encode-and-invert}, fitting
$\encmodel$ on perception and searching image space to match the response,
or \emph{decode}, mapping the response to a latent with $\decmap$. The
\emph{generator} is none ($\gen=\mathrm{id}$), per-trial \emph{iterative}
optimisation, or a \emph{frozen, amortised} pretrained model applied in one
pass. Family~Ia is encode-and-invert with iterative search; Family~Ib is
decode with iterative generation, since its neural stage is a feature
decoder (Eq.~\eqref{eq:famIb}) and what it inverts is the image-to-feature
network; Family~II is decode without a generator; Family~III is decode with
a frozen amortised generator; and Family~IV, a transform on the input, is
orthogonal to both axes. Ib thus shares its neural stage with Families~II
and~III, which is why its exposure to prior-driven failure resembles III
more than Ia (Sec.~\ref{subsec:spurious}).

\paragraph{Evaluation under a missing target}
Because $\imgcontent$ is unobserved, every method is scored against the
proxy, the cue $\cue$ or the exemplar $\exemplar$, by an operator
\begin{equation}
\label{eq:eval}
\mathcal{E}\big(\est{I}\ \text{or}\ \est{\decrep},\ \exemplar\ \text{or}\ \cue\big)
\in
\left\{\begin{array}{@{}l@{}}
\text{identification},\ \text{feature-metric},\\
\text{human-rater},\ \text{vividness-correspondence}
\end{array}\right\},
\end{equation}
whose \emph{target-free} character, absent in perception where
$\exemplar\!=\!\imgcontent$, is analysed in Section~\ref{sec:evaluation}.

\subsection{The Four Method Families}
\label{subsec:families}

\paragraph{Family~I --- Iterative inversion}
In \textbf{Family~Ia} a forward encoding model $\encmodel$ is fit on
\emph{perception}, mapping image features $\Feat(I)$ to the response, and
imagined content is recovered by inverting it under a prior:
\begin{align}
\label{eq:famI-train}
\est{\phi} &= \argmin_{\phi} \sum_{j} \norm*{\neural_j^{\mathrm{perc}} - \encmodel\big(\Feat(I_j)\big)}_2^2,\\
\label{eq:famI-infer}
\est{I} &= \argmin_{I} \norm*{\neural^{\mathrm{img}} - \est{g}_{\phi}\big(\Feat(I)\big)}_2^2 + \lambda\,\mathcal{R}(I),
\end{align}
where $\mathcal{R}$ is an image prior and \eqref{eq:famI-infer} is the MAP
counterpart of Eq.~\eqref{eq:inversion}. With a fixed, low-level feature
map such as Gabor wavelets or a retinotopic basis, the inversion is a
search over candidates or a direct pixel reconstruction,
\begin{equation}
\label{eq:famIa}
\est{I} = \argmax_{I \in \mathcal{C}}
\operatorname{corr}\big(\neural^{\mathrm{img}},\, \est{g}_{\phi}(\Feat_{\mathrm{low}}(I))\big),
\qquad \Feat_{\mathrm{low}}\ \text{fixed and retinotopic},
\end{equation}
as in~\cite{Thirion2006inverse,Naselaris2015voxelwise,Senden2019imagery}. In
\textbf{Family~Ib} the brain is decoded into the hierarchical features of a
deep network, and pixels or a generator latent are optimised to match them:
\begin{equation}
\label{eq:famIb}
\est{I} = \argmin_{I} \sum_{\ell \in \mathcal{L}}
\norm*{\Feat^{(\ell)}(I) - \est{\Feat}^{(\ell)}}_2^2
+ \lambda\,\mathcal{R}(I),
\qquad
\est{\Feat}^{(\ell)} = \decmap^{(\ell)}(\neural^{\mathrm{img}}),
\end{equation}
as in~\cite{Shen2019deep,KoideMajima2024mental,LeeKuhl2016reconstructing}.
Both variants optimise per trial and are cross-state, since imagery enters
only at inference.

\paragraph{Family~II --- Discriminative decoding and representational geometry}
Here $\gen=\mathrm{id}$ and $\decmap$ predicts a label or regresses onto a
fixed embedding $\decrep_{\mathrm{target}}$ for retrieval:
\begin{equation}
\label{eq:famII}
\est{\psi} = \argmin_{\psi}
\underbrace{\sum \mathrm{CE}\big(\decmap(\neural),\,y\big)}_{\text{classification}}
\quad\text{or}\quad
\argmin_{\psi}
\underbrace{\sum \norm*{\decmap(\neural) - \decrep_{\mathrm{target}}}_2^2}_{\text{retrieval}} .
\end{equation}
Representational similarity analysis (RSA) is a model-free variant that tests
a second-order isomorphism between
neural and feature
geometries~\cite{KriegeskorteMur2008representational,Cichy2012imagery}.

\paragraph{Family~III --- Generative cross-decoding}
A neural-to-latent map is trained on perception or hybrid data to reach the
\emph{conditioning space} of a pretrained generator built on denoising
diffusion~\cite{Ho2020denoising} or latent
diffusion~\cite{Rombach2022highres}, and imagery is cross-decoded through the
frozen generator:
\begin{equation}
\label{eq:famIII}
\begin{aligned}
\est{\psi} &= \argmin_{\psi}\,
\mathbb{E}\Big[\norm{\decmap(\neural^{\mathrm{perc}}) - \decrep(I)}_2^2
+ \beta\,\mathcal{L}_{\mathrm{CLIP}}\big(\decmap(\neural^{\mathrm{perc}}),\,I\big)\Big],\\
\est{I}^{\mathrm{img}} &= \gen\big(\decmap(\neural^{\mathrm{img}})\big),
\end{aligned}
\end{equation}
with $\mathcal{L}_{\mathrm{CLIP}}$ a contrastive term. The imagery step is a
\emph{distribution shift}, so seen-image performance does not guarantee
imagined-image performance~\cite{Kneeland2025nsdimagery,Kneeland2026mirage},
and strong priors make plausible but spurious reconstructions most likely
here~\cite{Shirakawa2025spurious}.

\paragraph{Family~IV --- Alignment and adaptation (meta-level)}
Family~IV learns a transform $\transform$ that aligns representations across
subjects, sessions, sites or states, so that Families~I--III remain usable
with scarce imagery data, either by functional alignment,
\begin{equation}
\label{eq:famIV-align}
\est{\transform} = \argmin_{\transform\in\mathcal{T}} \norm*{\transform(\neural_a) - \neural_b}_2^2
\qquad\text{(e.g.\ hyperalignment~\cite{Haxby2011common,Haxby2020hyperalignment})},
\end{equation}
or by aligning imagery activity into a \emph{frozen} conditioning space,
\begin{equation}
\label{eq:famIV-imagery}
\est{\transform} = \argmin_{\transform} \mathbb{E}\norm*{\decmap\big(\transform(\neural^{\mathrm{img}})\big) - \decrep_{\mathrm{target}}}_2^2
\qquad\text{($\decmap,\gen$ fixed~\cite{Spera2026seeing})},
\end{equation}
together with cross-subject adaptation and, as an outlook, neural foundation
models (Sec.~\ref{subsec:famIV}). It defines no decoding target of its own
and pre-composes with any family.

\FloatBarrier

\section{Method Families in Detail}
\label{sec:methods}

\subsection{Family~I: Iterative Inversion}
\label{subsec:famI}

\paragraph{Estimator}
Reading the data term of Eq.~\eqref{eq:famI-infer} as a Gaussian likelihood
and $\mathcal{R}(I)$ as a negative log-prior exposes Ia as
maximum-a-posteriori inference:
\begin{equation}
\label{eq:map-decomp}
\est{I}
= \argmax_{I}\;
\underbrace{\log p\big(\neural^{\mathrm{img}}\mid \est{g}_{\phi}(\Feat(I))\big)}_{\text{data fidelity (likelihood)}}
+
\underbrace{\log p(I)}_{\mathclap{-\lambda\,\mathcal{R}(I)\ \text{(image prior)}}}.
\end{equation}
No imagery labels enter training; the burden rests on the fidelity of the
forward model and the tractability of the inverse problem.

\paragraph{Representative work}
Inverse retinotopy set the template by inverting a perception-fit forward
model~\cite{Thirion2006inverse}. Voxel-wise encoding with Gabor and semantic
tuning then showed that low-level features of \emph{remembered} scenes are
present in imagery: the encoding model identified which of five memorised
artworks a participant recalled among random photographs well above
chance~\cite{Naselaris2015voxelwise,Naselaris2011encoding}. The same logic
recovers imagined letters~\cite{Senden2019imagery}. In Family~Ib, iCNN
optimises pixels so that their features in a deep backbone such as
VGG~\cite{Simonyan2015very} match brain-decoded features, optionally under a
deep generative prior~\cite{Shen2019deep,Goodfellow2014generative,Kingma2014auto};
imagined artificial shapes gave rudimentary but recognisable reconstructions,
whereas imagined natural images were reported as not well
reconstructed~\cite{Shen2019deep}. A Bayesian reformulation with a VQGAN
prior and decoded CLIP and VGG features~\cite{KoideMajima2024mental} raises
pairwise identification of imagined natural images from 50.4\% to 75.6\%,
with two qualifications (Table~\ref{tab:performance}, notes~d and~e): conv1
is excluded, and each pipeline is scored with its own weighted similarity, a
circularity raised by a reanalysis~\cite{Shirakawa2025advancing}\pre{}.

\paragraph{Trade-offs}
Ia recovers only what a retinotopic model encodes, so a high score requires
genuine low-level information, but imagery's early-visual signal is coarse
and weak (Sec.~\ref{subsec:substrates}), and the need for a spatially
resolved forward model confines the family to fMRI. Ib gains expressivity
from a learned feature decoder and a prior and is correspondingly exposed to
the prior--evidence tension of Sec.~\ref{subsec:spurious}. Both are
non-convex, so results depend on initialisation, prior weight and restarts,
and the choice of which run to display is addressed by standard~S7.

\subsection{Family~II: Discriminative Decoding and Representational Geometry}
\label{subsec:famII}

\paragraph{Estimators}
The direct mode optimises Eq.~\eqref{eq:famII}; the model-free mode
correlates representational dissimilarity matrices (RDMs),
\begin{equation}
\label{eq:rsa}
\rho_{\mathrm{RSA}}
= \operatorname{corr}\!\Big(\mathrm{RDM}\big(\neural\big),\;
\mathrm{RDM}\big(\Feat(I)\big)\Big).
\end{equation}

\paragraph{Representative work}
Reddy et al.~\cite{Reddy2010reading} trained classifiers on four object
categories and tested every combination of perception and imagery:
within-perception 67\%, within-imagery 50\% and cross-decoding 47\% and
52\%, against chance at 25\%, with only perception-to-perception above
chance in retinotopic cortex. Horikawa and Kamitani~\cite{Horikawa2017generic}
removed the closed-set restriction by decoding hierarchical DNN features,
identifying categories not used in training. Their imagery condition is
\lthree{} by instruction, but participants were first shown exemplar images
of each category, and the imagined categories are those of the perceptual
test session, so recall of seen exemplars cannot be excluded. The same
approach separates imagery from perception of real
objects~\cite{Lee2012disentangling}, recovers sleep-onset and dream
content~\cite{Horikawa2013neural,HorikawaKamitani2017dream}, and relates
decodable patterns to reported image
qualities~\cite{StecherKaiser2026neuro}\pre{}. In the reviewed corpus this is
the only family anchored in every modality reviewed, with
EEG~\cite{Xie2020alpha,StecherKaiser2024alpha},
MEG~\cite{Dijkstra2018differential} and single-unit
anchors~\cite{Kreiman2000imagery,Cerf2010online} and a published EEG
null~\cite{Shatek2019decoding} (Sec.~\ref{subsec:eegmeg}).

Kwon et al.~\cite{Kwon2025functional} decode four imagined actions from
EEG (72.3\% against 25\% chance); because participants were instructed to
imagine without muscle engagement, we admit the study under E2 as a
boundary case (Sec.~\ref{subsec:eegmeg}).

\paragraph{Trade-offs}
Because semantic representations overlap strongly across states and
classification needs only category-level separability, Family~II is
data-efficient and comparatively robust to low SNR. Its ceiling has the same
origin: the output is a label or an embedding, retrieval is bounded by the
candidate set, RSA yields no read-out, and identification depends on the
distractor set (Sec.~\ref{subsec:metrics}). This is the output type that EEG
and MEG currently support.

\subsection{Family~III: Generative Cross-Decoding}
\label{subsec:famIII}

\paragraph{Mechanics and anchors}
Family~III maps the response to the conditioning code of a frozen generator,
typically a CLIP image or text
embedding~\cite{Rombach2022highres,Radford2021learning,Ramesh2022hierarchical,Xu2023versatile,Pernias2024wurstchen},
so imagery decoders inherit a high-capacity prior and reconstruct in one
pass. NSD-Imagery evaluates NSD-trained
decoders~\cite{Ozcelik2023natural,Scotti2023reconstructing,Scotti2024mindeye2,Shen2019deep,Takagi2023highres}
on paired imagery and finds seen-image performance largely
\emph{decoupled} from imagined-image performance, with simple linear or
multimodal decoders generalising better than complex
ones~\cite{Kneeland2025nsdimagery}. MIRAGE ridge-regresses into multimodal
text-plus-image embeddings that condition a frozen cascaded diffusion model
and improves imagery reconstruction under human forced
choice~\cite{Kneeland2026mirage}. Outside fMRI there is one exploratory EEG
attempt~\cite{Shimizu2022improving}.

\paragraph{The nature of the shift}
The map $\decmap$ is fit on perception and deployed on imagery, and whether
that change is attenuation or a change of code determines which remedies
apply. Write the response in state $r\in\{\mathrm{perc},\mathrm{img}\}$ as
$\neural^{(r)} = A^{(r)} f^{(r)}(\imgcontent) + \boldsymbol{\varepsilon}^{(r)}$,
with $A^{(r)}$ a region-specific gain. Under the shared-code hypothesis
$\mathrm{H}_0$, $f^{\mathrm{img}}=f^{\mathrm{perc}}$ and imagery differs only
in gain and noise; under $\mathrm{H}_1$ the code itself changes. Even under
$\mathrm{H}_0$ with $A^{\mathrm{img}}<A^{\mathrm{perc}}$ the Bayes-optimal
$\mathbb{E}[\decrep\mid\neural]$ differs through shrinkage, so a change in
$p(\decrep\mid\neural)$ alone does not separate the two cases; and because
the latent available on imagery trials is $\decrep(\exemplar)$ rather than
$\decrep(\imgcontent)$, the apparent imagery conditional also moves with the
proxy bias of Sec.~\ref{subsec:targetfree}.

The neural evidence is consistent with $\mathrm{H}_1$ in early visual cortex
and compatible with $\mathrm{H}_0$ plus gain elsewhere: the overlap is
attenuated in early visual cortex~\cite{Horikawa2017generic,DijkstraBosch2019shared};
perception-trained classifiers transfer in object-responsive but not in
retinotopic cortex~\cite{Reddy2010reading}; imagery aligns with late rather
than full perceptual
processing~\cite{Dijkstra2018differential,Xie2020alpha,StecherKaiser2024alpha};
and aphantasic imagery is decodable in V1 yet does not cross-decode with
perception~\cite{Chang2025imageless}, which gain alone explains less
naturally than an altered code.

Published benchmark values add a diagnostic signal (Fig.~\ref{fig:transfer}).
They are compiled across two
reports~\cite{Kneeland2025nsdimagery,Kneeland2026mirage} and compared under a
common human forced-choice framing; this is a cross-study published-value
comparison, not a unified benchmark experiment. In these values, MindEye2
falls from 75.54\% to 50.21\%, near the 50\% chance level, on simple shapes,
and from the highest complex-scene score in the compilation, 90.56\%, to
64.83\%~\cite{Kneeland2025nsdimagery}. Across the five methods that differ
from chance the drop spans 5.9 to 26.1 points and the ordering of methods
changes, while MIRAGE scores within 0.1 points of the best vision value and
loses 5.9 points on imagery~\cite{Kneeland2026mirage}. The change in
ordering is consistent with $\mathrm{H}_1$ but also with attenuation acting
on heterogeneous estimators, so it does not discriminate between them; that
vision rankings do not carry over to imagery is a testable hypothesis
suggested by these values rather than an established result. A direct test
is possible with existing data (Sec.~\ref{subsec:gap2}).

\paragraph{Spurious reconstruction}
The general mechanism is analysed in Sec.~\ref{subsec:spurious}. The risk is
highest here because a frozen, high-capacity generator can satisfy feature,
semantic and realism scores from little neural
information~\cite{Shirakawa2025spurious}. MIRAGE addressed this by drawing
human-experiment distractors from decoded random brain patterns, so that
target and distractor carry the prior equally~\cite{Kneeland2026mirage};
this approximates the between-trial null of Eq.~\eqref{eq:lambda}, which is
why Table~\ref{tab:compliance} credits the study partially on S1. Since the
risk grows as the signal weakens, null-signal controls would be a
precondition for extending the family to EEG or MEG.

\subsection{Family~IV: Alignment and Adaptation}
\label{subsec:famIV}

\paragraph{Composability and its limits}
Let $M$ be the base mapping of any family (the inversion operator of
Family~I, $\decmap$ of Family~II or $\gen\circ\decmap$ of Family~III).
Alignment leaves the family's loss untouched and substitutes
$\neural\mapsto\transform(\neural)$, giving $M\circ\transform$:
\begin{equation}
\label{eq:compose}
\begin{alignedat}{2}
\est{I}
&= \big(\gen\circ\decmap\circ\transform\big)\big(\neural^{\mathrm{img}}\big)
&\qquad&\text{(with Family~III)},\\[2pt]
\est{I}
&= \argmin_{I}\norm*{\transform(\neural^{\mathrm{img}}) - \encmodel(\Feat(I))}_2^2 + \lambda\,\mathcal{R}(I)
&\qquad&\text{(with Family~I)}.
\end{alignedat}
\end{equation}
The composition is always well formed, because $\transform$ is fitted
separately. Whether it preserves what $M$ does depends on $\transform$
mapping the target domain into the region on which $M$ was fitted, which
presupposes a shared latent geometry. Across subjects within one state this
is well supported~\cite{Haxby2011common,Haxby2020hyperalignment}; across
states it is what $\mathrm{H}_1$ (Sec.~\ref{subsec:famIII}) denies, and
alignment into a perception-derived space can import a perceptual bias. If
$\transform$ projects onto the subspace that $\decmap$ reads most
confidently, composition can raise proxy scores while lowering the neural
dependence of Eq.~\eqref{eq:lambda}, so imagery-specific alignment should be
reported with the null-signal control applied \emph{after} composition.

\paragraph{Representative work}
Hyperalignment learns a common space by Procrustes-style
transforms~\cite{Haxby2011common,Haxby2020hyperalignment}; learned
alignment~\cite{Dai2025mindaligner}, lightweight subject
adaptation~\cite{Zhang2025neuroswift}, code conversion without shared
stimuli~\cite{Wang2025interindividual} and source-free or unseen-subject
decoding~\cite{Kong2025generalizing,StableMind2026} extend it to limited
data. The imagery-specific instance of Eq.~\eqref{eq:famIV-imagery} aligns
imagery activity into a frozen diffusion space~\cite{Spera2026seeing}\pre{},
while neural foundation models~\cite{Jiang2024labram,Yang2024brainmae,Lane2025scaling}\pre{}
have so far been evaluated on general tasks rather than imagery. No
published imagery score isolates an alignment layer against the same
pipeline without it, so the family has no entry in
Table~\ref{tab:performance}.

\FloatBarrier

\section{Datasets, Paradigms, and Benchmarks}
\label{sec:datasets}

\subsection{Experimental Paradigms}
\label{subsec:paradigms}

\emph{Cued imagery} dominates: a label $\cue$ or a previously seen exemplar
$\exemplar$ specifies the content on each
trial~\cite{Horikawa2017generic,Shen2019deep}. \emph{Free imagery},
accessible only through retrospective report, is closer to real-world use
but hard to decode or score, and \emph{working-memory delay}
paradigms~\cite{HarrisonTong2009decoding} are the \lone{} paradigms of
Sec.~\ref{subsec:terminology}. When the cue is an image just viewed, a
perceptual trace of $\exemplar$, a maintenance signal and the generated
imagery are entangled in $\neural$, and a decoder scored against
$\exemplar$ cannot separate read-out from carry-over. This is most
pronounced outside fMRI: the largest public EEG imagery dataset presents the
target picture for 4\,s before each 4\,s imagery period~\cite{Gao2026eegvi},
and the action-imagery study a 5\,s video before
imagery~\cite{Kwon2025functional}; both designs aim to remove the
afterimage, not the maintenance signal.

\subsection{Modality-Specific Characteristics}
\label{subsec:modalities}

\textbf{fMRI} offers the retinotopic access that underwrites pixel- and
feature-level reconstruction, but limited per-trial SNR, which is why every
benchmark in Table~\ref{tab:performance} averages repetitions; \textbf{7\,T
fMRI} improves SNR enough to reveal phenotype
differences~\cite{Liu2025visual}. \textbf{EEG and MEG} provide millisecond
resolution suited to the late, sustained dynamics of
imagery~\cite{Dijkstra2018differential}, but volume conduction and artefacts
make image-level reconstruction much harder. \textbf{Intracranial
recording} combines millisecond and cellular resolution but samples medial
temporal cortex in clinical populations and has yielded no public imagery
corpus~\cite{Kreiman2000imagery,Cerf2010online}. Design matters as much as
modality: in a block design slow drifts are confounded with class labels,
and a reanalysis~\cite{Li2020perils} of an early 40-class EEG
dataset~\cite{Spampinato2017deep} showed accuracy falling to chance under a
rapid-event design.

\subsection{Principal Datasets and Reported Performance}
\label{subsec:datasetstable}

\begin{sidewaystable}
\centering
\setlength{\tabcolsep}{3pt}
\renewcommand{\arraystretch}{1.0}
\begin{threeparttable}
\caption{Principal imagery datasets and the perceptual corpora used for
cross-decoding, with the admissibility criterion of
Sec.~\ref{subsec:terminology}. Datasets are open unless stated; coverage
closed on \searchdate.}
\label{tab:datasets}
\scriptsize
\begin{tabularx}{\linewidth}{@{}L{2.55cm}L{1.85cm}L{1.5cm}YL{2.7cm}L{2.5cm}@{}}
\toprule
\textbf{Dataset} & \textbf{Modality} & \textbf{Subjects} &
\textbf{Imagery task and cue} & \textbf{Target status} &
\textbf{Admissibility}\textsuperscript{a} \\
\midrule
\famrow{6}{svGrey}{\textit{Imagery-focused, fMRI}}
NSD-Imagery~\cite{Kneeland2025nsdimagery} & fMRI 7T & 8 &
6 simple shapes, 6 complex scenes (5 from shared1000), 6 concept words;
memorised letter cues; 432 trials per subject\textsuperscript{b} &
Proxy $\cue$; image target except concepts &
\adm{\sepyes}{\ltwo} simple, complex\newline
\adm{\sepyes}{\lthree} conceptual \\
\addlinespace[1.5pt]
GOD, Horikawa \& Kamitani~\cite{Horikawa2017generic} & fMRI 3T & 5 &
Category-name cue, eyes closed; arbitrary exemplars of the category &
Proxy $\cue$; seen-image GT for training &
\adm{\sepyes}{\lthree} \\
\addlinespace[1.5pt]
Deeprecon~\cite{Shen2019deep} & fMRI 3T & 3\textsuperscript{c} &
Word cues for 25 images memorised beforehand (10 natural, 15 artificial) &
Proxy $\cue$; seen-image GT for training &
\adm{\sepyes}{\ltwo} \\
\addlinespace[2pt]
\famrow{6}{svGrey}{\textit{Imagery-focused, EEG / MEG / iEEG}}
VI-BCI EEG~\cite{Gao2026eegvi} & EEG, 32 ch & 22 (2\,sess.) &
10 pictures; 17\,s trial: picture 4\,s, mask, imagery 4\,s; VVIQ &
Proxy $\exemplar$; class label only &
\adm{\sepno}{\lone} \\
\addlinespace[1.5pt]
Action-imagery EEG~\cite{Kwon2025functional} & EEG, 64 ch & 15 &
4 upper-limb actions; 17\,s trial: video 5\,s, rest 5\,s, imagery 5\,s;
E2 boundary case; not openly released &
Proxy $\exemplar$; class label only &
\adm{\sepno}{\lone} \\
\addlinespace[1.5pt]
Semantic-concept EEG~\cite{Wilson2023semantic} & EEG, 124 ch & 12 &
3 concepts (pictorial, orthographic, auditory), each perceived, then
imagined in the trial; VVIQ, BAIS-V &
Proxy $\exemplar$; class label only &
\adm{\sepno}{\lone} \\
\addlinespace[1.5pt]
EEG\,+\,fNIRS imagery~\cite{Rybar2025simultaneous} & EEG, 64 ch; fNIRS &
12 (+7\,EEG) &
36 animals or tools; 0.6\,s picture and mask, then imagery of one's own
exemplar; animal\,/\,tool labels &
Proxy $\exemplar$; class label only &
\adm{\sepno}{\lthree}\textsuperscript{d} \\
\addlinespace[1.5pt]
MEG / iEEG imagery & MEG; iEEG & \sepna &
None openly released as of \searchdateshort; none outside fMRI pairs
imagery with image-level targets &
\sepna & \sepna \\
\addlinespace[2pt]
\famrow{6}{svGrey}{\textit{Perceptual corpora used as cross-decoding substrate}}
NSD~\cite{Allen2022massive} & fMRI 7T & 8 &
Natural scenes, up to 30{,}000 trials per subject &
Perceptual GT & \sepna \\
BOLD5000~\cite{Chang2019bold5000} & fMRI 3T & 4 &
${\sim}5{,}000$ natural images & Perceptual GT & \sepna \\
THINGS~\cite{Hebart2023things,Grootswagers2022human,Gifford2022large} &
fMRI; MEG; EEG & 3; 4; 50, 10 &
THINGS object images; EEG by rapid serial visual presentation &
Perceptual GT & \sepna \\
EEG-ImageNet\textsuperscript{$\ast$}~\cite{Spampinato2017deep} & EEG & 6 &
Block-design viewing of 40 ImageNet classes & Perceptual GT & \sepna \\
\bottomrule
\end{tabularx}
\begin{tablenotes}[para,flushleft]
\scriptsize
\item[a] Trial level (\sepyes~separable, \sepno~entangled, \sepunc~unclear,
\sepna~none) crossed with memory source (\lone/\ltwo/\lthree); evidence in
Table~\ref{tab:corpus}.
\item[b] Vision\,+\,imagery; 144 further attention-run trials are released
but excluded by the authors.
\item[c] Five in later public sessions of the same
laboratory~\cite{Wang2025interindividual}.
\item[d] A picture is shown in the trial, but participants visualise their
own exemplar of the concept (Sec.~\ref{subsec:terminology}).
\item[$\ast$] Block design with class-contiguous trials, hence a leakage
risk~\cite{Li2020perils}; historical benchmark only.
\end{tablenotes}
\end{threeparttable}
\end{sidewaystable}

Purpose-built imagery datasets are small in trials and stimuli rather than
in participants: NSD-Imagery has 432 vision and imagery trials over 18
stimuli per subject~\cite{Kneeland2025nsdimagery}, against up to 30{,}000
perception trials in NSD~\cite{Allen2022massive}. The three fMRI imagery
datasets cue learned material symbolically and are separable at the trial
level, whereas all four EEG datasets of Table~\ref{tab:datasets} are
entangled and supply only class labels, so no electrophysiological analogue
of NSD-Imagery was identified.

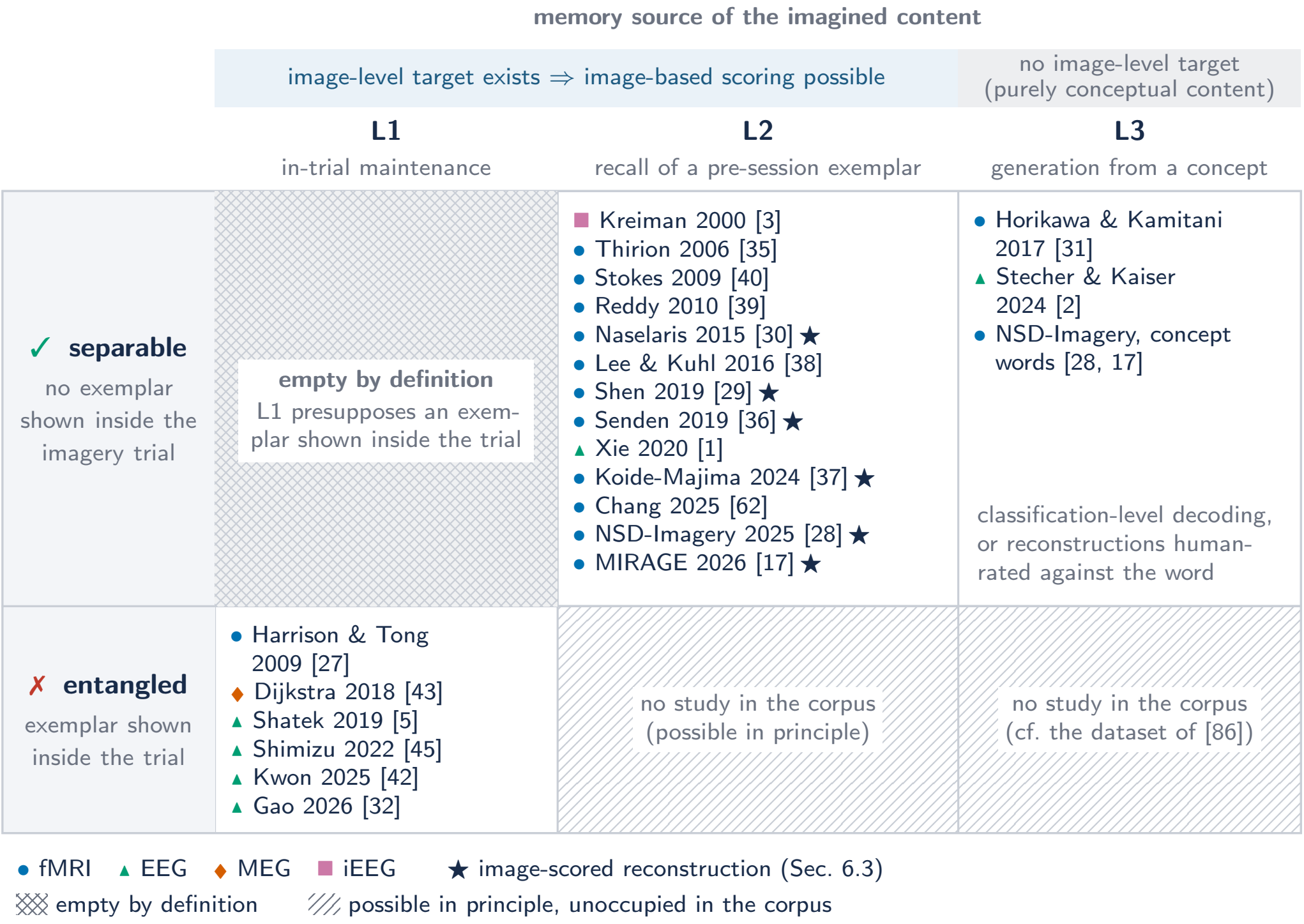
\begin{figure*}[!htbp]
\centering
\begingroup
\newcommand{\amF}{\textcolor{svfMRI}{$\bullet$}}
\newcommand{\amE}{\textcolor{svEEG}{\raisebox{.1ex}{$\scriptstyle\blacktriangle$}}}
\newcommand{\amM}{\textcolor{svMEG}{$\scriptstyle\blacklozenge$}}
\newcommand{\amI}{\textcolor{sviEEG}{\raisebox{.1ex}{$\scriptscriptstyle\blacksquare$}}}
\newcommand{\amSt}{\,{\color{svInk}$\bigstar$}}
\begin{tikzpicture}[sv, x=1cm, y=1cm]
  \def\hw{2.2}  
  \def\cw{3.55}  
  \def\rA{4.05}  
  \def\rB{2.35}  
  \pgfmathsetmacro{\xa}{\hw}
  \def\cwB{4.15} 
  \pgfmathsetmacro{\xb}{\hw+\cw}
  \pgfmathsetmacro{\xc}{\hw+\cw+\cwB}
  \pgfmathsetmacro{\xd}{\hw+2*\cw+\cwB}

  \fill[svBlue!10] (\xa,\rA+1.12) rectangle (\xc,\rA+1.72);
  \node[font=\svsmall,text=svBlue!70!black] at ($(\xa,\rA+1.42)!.5!(\xc,\rA+1.42)$)
       {image-level target exists $\Rightarrow$ image-based scoring possible};
  \fill[svGrey!12] (\xc,\rA+1.12) rectangle (\xd,\rA+1.72);
  \node[font=\svsmall,text=svGrey,align=center] at ($(\xc,\rA+1.42)!.5!(\xd,\rA+1.42)$)
       {no image-level target\\[-1pt] (purely conceptual content)};

  \foreach \x/\w/\t/\s in {\xa/\cw/L1/{in-trial maintenance},
                        \xb/\cwB/L2/{recall of a pre-session exemplar},
                        \xc/\cw/L3/{generation from a concept}}{
    \node[anchor=base, font=\svfont\bfseries] at (\x+\w/2,\rA+0.78) {\t};
    \node[anchor=base, svnote] at (\x+\w/2,\rA+0.42) {\s};
  }
  \node[anchor=south, font=\svsmall\bfseries, text=svGrey] at ($(\xa,\rA+1.78)!.5!(\xd,\rA+1.78)$)
       {memory source of the imagined content};

  \draw[svRule] (0,0) rectangle (\xd,\rA+0.25);
  \draw[svRule] (0,-\rB) rectangle (\xd,0);
  \foreach \x in {\xa,\xb,\xc} \draw[svRule] (\x,-\rB) -- (\x,\rA+0.25);
  \fill[svPanel] (0,-\rB) rectangle (\hw,\rA+0.25);
  \draw[svRule] (0,-\rB) rectangle (\xd,\rA+0.25);
  \draw[svRule] (0,0) -- (\xd,0);

  \node[text width=\hw cm-4pt, align=center] at (\hw/2,\rA/2+0.1)
     {\textbf{\mbox{\sepyes\; separable}}\\[2pt]{\svsmall\color{svGrey}no exemplar shown inside the imagery trial}};
  \node[text width=\hw cm-4pt, align=center] at (\hw/2,-\rB/2)
     {\textbf{\mbox{\sepno\; entangled}}\\[2pt]{\svsmall\color{svGrey}exemplar shown inside the trial}};

  \fill[svGrey!10] (\xa,0) rectangle ++(\cw,\rA+0.25);
  \fill[pattern=crosshatch, pattern color=svRule] (\xa,0) rectangle ++(\cw,\rA+0.25);
  \node[svnote,fill=white,inner sep=3pt,text width=\cw cm-0.7cm] at (\xa+\cw/2,\rA/2)
     {\textbf{empty by definition}\\[1pt]\lone{} presupposes an exemplar shown inside the trial};

  \foreach \xx/\ww in {\xb/\cwB, \xc/\cw}{
    \fill[pattern=north east lines, pattern color=svRule] (\xx,-\rB) rectangle ++(\ww,\rB);
  }
  \node[svnote,fill=white,inner sep=2pt] at (\xb+\cwB/2,-\rB/2) {no study in the corpus\\(possible in principle)};
  \node[svnote,fill=white,inner sep=2pt] at (\xc+\cw/2,-\rB/2) {no study in the corpus\\(cf.\ the dataset of \cite{Rybar2025simultaneous})};

  \node[anchor=north west, font=\svsmall, align=left, inner sep=4pt] at (\xb,\rA+0.2) {%
    \amI\ Kreiman 2000 \cite{Kreiman2000imagery}\\
    \amF\ Thirion 2006 \cite{Thirion2006inverse}\\
    \amF\ Stokes 2009 \cite{Stokes2009topdown}\\
    \amF\ Reddy 2010 \cite{Reddy2010reading}\\
    \amF\ Naselaris 2015 \cite{Naselaris2015voxelwise}\amSt\\
    \amF\ Lee \& Kuhl 2016 \cite{LeeKuhl2016reconstructing}\\
    \amF\ Shen 2019 \cite{Shen2019deep}\amSt\\
    \amF\ Senden 2019 \cite{Senden2019imagery}\amSt\\
    \amE\ Xie 2020 \cite{Xie2020alpha}\\
    \amF\ Koide-Majima 2024 \cite{KoideMajima2024mental}\amSt\\
    \amF\ Chang 2025 \cite{Chang2025imageless}\\
    \amF\ NSD-Imagery 2025 \cite{Kneeland2025nsdimagery}\amSt\\
    \amF\ MIRAGE 2026 \cite{Kneeland2026mirage}\amSt};

  \node[anchor=north west, font=\svsmall, align=left, inner sep=4pt] at (\xc,\rA+0.2) {%
    \amF\ Horikawa \& Kamitani\\\phantom{\amF}\ 2017 \cite{Horikawa2017generic}\\
    \amE\ Stecher \& Kaiser\\\phantom{\amE}\ 2024 \cite{StecherKaiser2024alpha}\\
    \amF\ NSD-Imagery, concept\\\phantom{\amF}\ words \cite{Kneeland2025nsdimagery,Kneeland2026mirage}};
  \node[anchor=south west, svnote, align=left, text width=\cw cm-10pt] at (\xc+0.05,0.08)
    {classification-level decoding, or reconstructions human-rated against the word};

  \node[anchor=north west, font=\svsmall, align=left, inner sep=4pt] at (\xa,-0.05) {%
    \amF\ Harrison \& Tong\\\phantom{\amF}\ 2009 \cite{HarrisonTong2009decoding}\\
    \amM\ Dijkstra 2018 \cite{Dijkstra2018differential}\\
    \amE\ Shatek 2019 \cite{Shatek2019decoding}\\
    \amE\ Shimizu 2022 \cite{Shimizu2022improving}\\
    \amE\ Kwon 2025 \cite{Kwon2025functional}\\
    \amE\ Gao 2026 \cite{Gao2026eegvi}};

  \node[anchor=north west, font=\svsmall, align=left] at (0,-\rB-0.12)
    {\amF\ fMRI\quad \amE\ EEG\quad \amM\ MEG\quad \amI\ iEEG\qquad
     $\bigstar$\ image-scored reconstruction (Sec.~\ref{subsec:datasetstable})\\[2pt]
     \tikz[baseline=-.6ex]\fill[pattern=crosshatch, pattern color=svGrey] (0,-.1) rectangle (.32,.12);\ empty by definition\qquad
     \tikz[baseline=-.6ex]\fill[pattern=north east lines, pattern color=svGrey] (0,-.1) rectangle (.32,.12);\ possible in principle, unoccupied in the corpus};
\end{tikzpicture}
\endgroup
\caption{The imagery corpus (Table~\ref{tab:corpus}) placed in the two-dimensional
admissibility criterion of Sec.~\ref{subsec:terminology}: trial-level
separability (rows) crossed with the memory source of the imagined content
(columns). Marker shape gives the modality and $\bigstar$ an image-scored
reconstruction. The cross-hatched cell is empty by definition, because
\lone{} presupposes an exemplar shown inside the trial; the hatched cells are
possible in principle but unoccupied in the corpus. In the reviewed corpus,
every image-scored reconstruction is of the \ltwo{} kind and no \lone{}
paradigm has been carried to image-scored reconstruction; purely conceptual
\lthree{} content has no image-level target.}
\label{fig:admissibility}
\end{figure*}

\paragraph{What image-based scoring can and cannot test}
Apart from compositionally constructed \lthree{} targets, which no study in
the reviewed corpus uses, image-based scoring is available only under
\lone{} or \ltwo{}, and the reviewed corpus occupies only \ltwo{}
(Fig.~\ref{fig:admissibility}): wherever an
imagery reconstruction is scored against an image, as for Deeprecon's
memorised pictures~\cite{Shen2019deep} and their
reanalysis~\cite{KoideMajima2024mental}, NSD-Imagery's simple and complex
stimuli~\cite{Kneeland2025nsdimagery,Kneeland2026mirage,Spera2026seeing},
the memorised artworks of~\cite{Naselaris2015voxelwise} and the letters
of~\cite{Senden2019imagery}, the participant recalls an exemplar shown
before the session. In the reviewed corpus, no \lone{} paradigm has been carried to
image-scored reconstruction, and no study has built a scorable image target
for \lthree{} content, although \lthree{} is decoded at the classification
level~\cite{Horikawa2017generic,StecherKaiser2024alpha} and NSD-Imagery's
conceptual reconstructions are human-rated against
words~\cite{Kneeland2025nsdimagery,Kneeland2026mirage}. The gap is thus that
generation has not been put to an image-level test, rather than that it has
failed one. This matters because retrieving a stored exemplar can, in the
limit, be satisfied by an expressive prior from a category label
(Sec.~\ref{subsec:spurious}), whereas generated content has no stored target
to collapse onto (Sec.~\ref{subsec:gap1}).

\paragraph{Internal validity of the imagery benchmarks}
Five of NSD-Imagery's six complex stimuli come from the NSD shared1000 set,
which participants viewed repeatedly~\cite{Kneeland2025nsdimagery}; this is
not leakage, since the set is held out from training, but it places the
condition at the recall end of \ltwo{} in a semantically clustered
corpus~\cite{Shirakawa2025spurious}. Deeprecon's artificial shapes lie
outside the natural-image training distribution~\cite{Shen2019deep} and are
therefore the more diagnostic half of Panel~D. Neither benchmark decodes
single trials, so every figure in Table~\ref{tab:performance} is a
multi-trial quantity (note~g).

\begin{sidewaystable}
\centering
\setlength{\tabcolsep}{3pt}
\renewcommand{\arraystretch}{0.95}
\begin{threeparttable}
\caption{Reported performance on paired vision/imagery benchmarks, all from
fMRI perception-transfer decoders. Rows compare within a panel only
(note~e); values are verbatim except V$-$I and $\rho$ (notes~a,~f).}
\label{tab:performance}
\scriptsize
\newcommand{\pnl}[1]{\multicolumn{11}{@{}p{\linewidth}@{}}{\RaggedRight\textit{#1}}\\}
\newcommand{\sub}[1]{\multicolumn{1}{c}{\textcolor{svGrey}{#1}}}
\newcommand{\subm}[2]{\multicolumn{#1}{c}{\textcolor{svGrey}{#2}}}
\newcommand{\ns}[1]{\textcolor{svGrey}{#1}}
\begin{tabular*}{\linewidth}{@{\extracolsep{\fill}}L{2.75cm}ccc@{\hspace{3pt}}cccc@{\hspace{3pt}}ccL{2.25cm}@{}}
\toprule
 & \multicolumn{3}{c}{\textbf{Vision}} & \multicolumn{4}{c}{\textbf{Imagery}}
 & \textbf{V$-$I}\textsuperscript{a} & $\boldsymbol{\rho}$\textsuperscript{f} & \textbf{Source} \\
\cmidrule(lr){2-4}\cmidrule(lr){5-8}
\pnl{Panel A --- NSD-Imagery; human 2AFC, $n=500$ raters; trained on subj.\ 1, 2, 5, 7; chance $50\%$; \% overall (simple / complex / conceptual)\textsuperscript{b}}
\textbf{Method} & \sub{overall} & \sub{simple} & \sub{complex} & \sub{overall} & \sub{simple} & \sub{complex} & \sub{conceptual} & & & \\
MIRAGE          & 84.22 & 80.77 & 87.66 & 78.30 & 73.93 & 83.19 & 77.68 & $+5.9$  & 0.83 & \cite{Kneeland2026mirage} Tab.~2 \\
MindEye1        & 84.29 & 79.32 & 89.32 & 73.00 & 71.01 & 82.28 & 65.68 & $+11.3$ & 0.67 & \cite{Kneeland2025nsdimagery} Tab.~2 \\
MindEye2        & 83.05 & 75.54 & \textbf{90.56} & 56.96 & 50.21 & 64.83 & 55.74 & $+26.1$ & 0.21 & \cite{Kneeland2025nsdimagery} Tab.~2 \\
Brain-Diffuser  & 80.13 & 71.79 & 88.46 & 73.95 & 68.20 & 82.70 & 71.01 & $+6.2$  & 0.80 & \cite{Kneeland2025nsdimagery} Tab.~2 \\
iCNN            & 74.52 & 75.85 & 73.19 & 66.15 & 66.81 & 70.04 & 61.60 & $+8.4$  & 0.66 & \cite{Kneeland2025nsdimagery} Tab.~2 \\
\ns{Takagi \& Nishimoto} & \ns{49.58}\textsuperscript{c} & \ns{48.95} & \ns{50.20}
                & \ns{52.89}\textsuperscript{c} & \ns{57.26} & \ns{48.31} & \ns{53.14}
                & \ns{$-3.3$} & \ns{n.d.} & \cite{Kneeland2025nsdimagery} Tab.~2 \\
\midrule
\pnl{Panel B --- NSD-Imagery; automated metrics on simple + complex stimuli; PixCorr (higher better) and CLIP two-way \% (chance $50\%$)\textsuperscript{b}}
\textbf{Method} & \sub{PixCorr} & \sub{CLIP \%} & & \sub{PixCorr} & \sub{CLIP \%} & & & & \sub{PixCorr\,/\,CLIP} & \\
MIRAGE          & 0.221 & 66.69 & & 0.104 & 57.46 & & & $+9.2$  & 0.47\,/\,0.45 & \cite{Kneeland2026mirage} Tab.~1 \\
MindEye1        & 0.218 & 65.34 & & 0.086 & 54.72 & & & $+10.6$ & 0.39\,/\,0.31 & \cite{Kneeland2026mirage} Tab.~1 \\
MindEye2        & 0.161 & 65.93 & & 0.036 & 50.78 & & & $+15.2$ & 0.22\,/\,0.05 & \cite{Kneeland2026mirage} Tab.~1 \\
Brain-Diffuser  & 0.107 & 58.31 & & 0.064 & 54.07 & & & $+4.2$  & 0.60\,/\,0.49 & \cite{Kneeland2026mirage} Tab.~1 \\
iCNN            & 0.224 & 49.03 & & 0.108 & \textbf{41.72} & & & $+7.3$ & 0.48\,/\,n.d. & \cite{Kneeland2026mirage} Tab.~1 \\
Takagi \& Nishimoto & $-0.013$ & 43.01 & & $-0.006$ & 40.08 & & & $+2.9$ & n.d. & \cite{Kneeland2025nsdimagery} Tab.~1 \\
\midrule
\pnl{Panel C --- Deeprecon; pairwise identification, each pipeline scored by its \emph{own} similarity\textsuperscript{e}; conv2--fc6\textsuperscript{d}; 3 subjects; chance $50\%$; \% $\pm$ SD}
\textbf{Method} & \subm{2}{natural} & \sub{artificial} & \subm{2}{natural} & \subm{2}{artificial} & & & \\
Koide-Majima & \multicolumn{2}{c}{$90.7 \pm 13.5$} & $89.6 \pm 8.9$ &
  \multicolumn{2}{c}{$75.6 \pm 22.3$} & \multicolumn{2}{c}{$72.9 \pm 23.2$} &
  $+15.1$ & 0.63 & \cite{KoideMajima2024mental} \\
iCNN, re-run & \multicolumn{2}{c}{$64.3 \pm 30.6$} & $66.7 \pm 32.9$ &
  \multicolumn{2}{c}{$50.4 \pm 32.2$} & \multicolumn{2}{c}{$51.4 \pm 31.8$} &
  $+13.9$ & 0.03 & \cite{KoideMajima2024mental} \\
\midrule
\pnl{Panel D --- Deeprecon, iCNN as reported in~\cite{Shen2019deep}; 3 subjects; chance $50\%$; identification \% by pixel correlation\,/\,human judgement ($n=13$ raters)}
\textbf{Method} & \sub{pixel-wise} & \sub{human} & & \sub{pixel-wise} & \sub{human} & & & & & \\
Natural, with DGN & 76.1 & 97.0 & &
  \multicolumn{4}{c}{\multirow{2}{*}{qualitative only\textsuperscript{i}}} &
  \multirow{2}{*}{\sepna} & \multirow{2}{*}{n.d.} & \cite{Shen2019deep} Figs.~3B, 8C \\
\hspace{1em}without DGN & 79.7 & 96.0 & & \multicolumn{4}{c}{} & & & \\
Artificial shapes & 70.5 & 91.0 & & \textbf{51.9} & 83.2 & & &
  $+18.6$\,/\,$+7.8$ & 0.09\,/\,0.81 & \cite{Shen2019deep} Figs.~6C, 8D \\
\bottomrule
\end{tabular*}
\begin{tablenotes}[para,flushleft]
\scriptsize
\item[a] Vision minus imagery, in the units of the two preceding columns
(CLIP in B, natural images in C, pixel-wise\,/\,human in D).
\item[b] Conceptual stimuli lack a vision condition; Panel~B covers simple
and complex stimuli only.
\item[c] Not different from chance ($p=0.891$ vision, $p=0.133$ imagery);
all other Panel~A entries $p<0.001$.
\item[d] From \cite{KoideMajima2024mental}, Figs.~2e, 3e, S7, S9--10;
conv1 excluded, a diagnostic contrast by the authors, not a like-for-like
replication (Sec.~\ref{subsec:famI}).
\item[e] \textbf{No cross-panel comparison}: panels differ in raters,
distractors, scoring space, stimuli and subjects, and Panel~C scores each
pipeline with its own objective (Sec.~\ref{subsec:spurious}); the
unreproduced brain-correlation columns of \cite{Kneeland2025nsdimagery}
and~\cite{Kneeland2026mirage} differ slightly. Panel~A compiles values
published in these two reports, a published-value comparison rather than a
single benchmark run.
\item[f] Eq.~\eqref{eq:rho}; PixCorr\,/\,CLIP (B), pixel-wise\,/\,human
(D); n.d.: vision not above chance.
\item[g] All values are trial-averaged (Sec.~\ref{subsec:datasetstable}).
\item[h] Not a leaderboard: \cite{Kneeland2026mirage} and~\cite{Shen2019deep}
report further methods or conditions not shown.
\item[i] Imagined natural images were not well reconstructed and are shown
only qualitatively; the absence is the finding.
\end{tablenotes}
\end{threeparttable}
\end{sidewaystable}

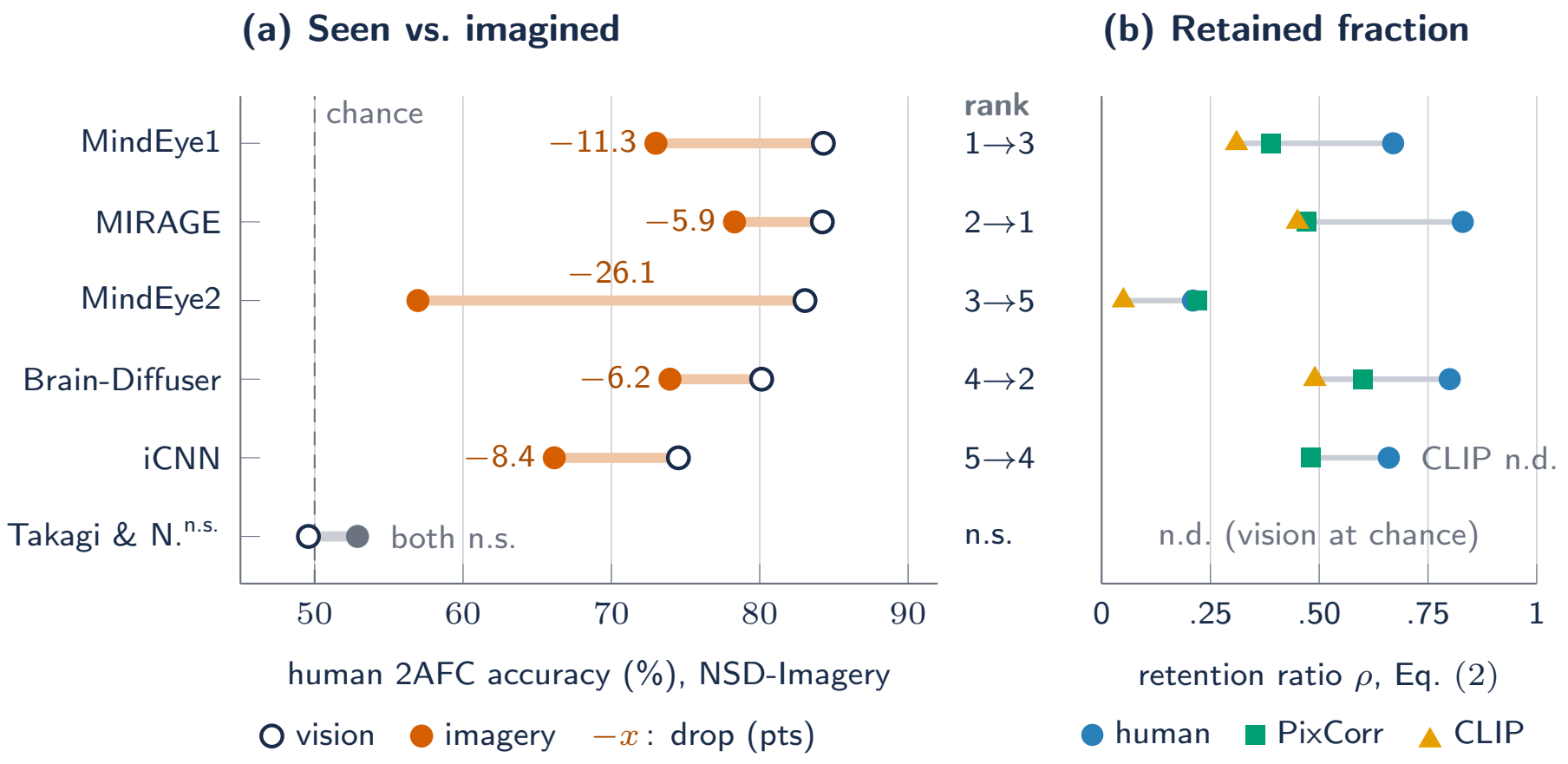
\begin{figure*}[!htbp]
\centering
\begin{tikzpicture}[sv]
\pgfplotsset{
  svaxis/.style={
    width=6.9cm, height=5.3cm, scale only axis=false,
    ymin=0.4, ymax=6.6, ytick={1,...,6},
    axis line style={svGrey}, tick style={svGrey},
    tick label style={font=\svsmall, text=svInk},
    label style={font=\svsmall, text=svInk},
    xmajorgrids, grid style={svRule, line width=.3pt},
    axis x line*=bottom, axis y line*=left, clip=false,
    title style={font=\svfont\bfseries, text=svInk, anchor=south west, at={(0,1)}, xshift=-4pt},
  }}
\begin{axis}[svaxis, name=A,
    xmin=45, xmax=92, xtick={50,60,70,80,90},
    xlabel={human 2AFC accuracy (\%), NSD-Imagery},
    yticklabels={Takagi \& N.\textsuperscript{n.s.},iCNN,Brain-Diffuser,MindEye2,MIRAGE,MindEye1},
    title={(a) Seen vs.\ imagined},
  ]
  \draw[svGrey, svdash] (axis cs:50,0.4) -- (axis cs:50,6.6);
  \node[svnote, anchor=north west, inner sep=1pt] at (axis cs:50.4,6.62) {chance};
  \draw[line width=2.2pt, svFamIII!35] (axis cs:84.29,6) -- (axis cs:73.0,6);
  \fill[white, draw=svInk, line width=.8pt] (axis cs:84.29,6) circle (2.3pt);
  \fill[svFamIII] (axis cs:73.0,6) circle (2.5pt);
  \node[font=\svsmall, text=svFamIII!80!black, anchor=east] at (axis cs:73.0,6) {$-$11.3\hspace{3pt}};
  \node[font=\svsmall, text=svInk, anchor=west] at (axis cs:92.5,6) {1$\to$3};
  \draw[line width=2.2pt, svFamIII!35] (axis cs:84.22,5) -- (axis cs:78.3,5);
  \fill[white, draw=svInk, line width=.8pt] (axis cs:84.22,5) circle (2.3pt);
  \fill[svFamIII] (axis cs:78.3,5) circle (2.5pt);
  \node[font=\svsmall, text=svFamIII!80!black, anchor=east] at (axis cs:78.3,5) {$-$5.9\hspace{3pt}};
  \node[font=\svsmall, text=svInk, anchor=west] at (axis cs:92.5,5) {2$\to$1};
  \draw[line width=2.2pt, svFamIII!35] (axis cs:83.05,4) -- (axis cs:56.96,4);
  \fill[white, draw=svInk, line width=.8pt] (axis cs:83.05,4) circle (2.3pt);
  \fill[svFamIII] (axis cs:56.96,4) circle (2.5pt);
  \node[font=\svsmall, text=svFamIII!80!black, anchor=south, inner sep=1.5pt] at (axis cs:70.0,4.08) {$-$26.1};
  \node[font=\svsmall, text=svInk, anchor=west] at (axis cs:92.5,4) {3$\to$5};
  \draw[line width=2.2pt, svFamIII!35] (axis cs:80.13,3) -- (axis cs:73.95,3);
  \fill[white, draw=svInk, line width=.8pt] (axis cs:80.13,3) circle (2.3pt);
  \fill[svFamIII] (axis cs:73.95,3) circle (2.5pt);
  \node[font=\svsmall, text=svFamIII!80!black, anchor=east] at (axis cs:73.95,3) {$-$6.2\hspace{3pt}};
  \node[font=\svsmall, text=svInk, anchor=west] at (axis cs:92.5,3) {4$\to$2};
  \draw[line width=2.2pt, svFamIII!35] (axis cs:74.52,2) -- (axis cs:66.15,2);
  \fill[white, draw=svInk, line width=.8pt] (axis cs:74.52,2) circle (2.3pt);
  \fill[svFamIII] (axis cs:66.15,2) circle (2.5pt);
  \node[font=\svsmall, text=svFamIII!80!black, anchor=east] at (axis cs:66.15,2) {$-$8.4\hspace{3pt}};
  \node[font=\svsmall, text=svInk, anchor=west] at (axis cs:92.5,2) {5$\to$4};
  \draw[line width=2.2pt, svGrey!35] (axis cs:49.58,1) -- (axis cs:52.89,1);
  \fill[white, draw=svInk, line width=.8pt] (axis cs:49.58,1) circle (2.3pt);
  \fill[svGrey] (axis cs:52.89,1) circle (2.5pt);
  \node[svnote, anchor=west] at (axis cs:52.89,1) {\hspace{3pt}both n.s.};
  \node[font=\svsmall, text=svInk, anchor=west] at (axis cs:92.5,1) {n.s.};
  \node[font=\svsmall\bfseries, text=svGrey, anchor=west] at (axis cs:92.5,6.5) {rank};
\end{axis}
\begin{axis}[svaxis, name=B, at={(A.east)}, anchor=west, xshift=1.25cm,
    width=4.9cm,
    xmin=0, xmax=1, xtick={0,0.25,0.5,0.75,1},
    xticklabels={0,.25,.50,.75,1},
    xlabel={retention ratio $\rho$, Eq.~\eqref{eq:rho}},
    yticklabels={}, ytick style={draw=none},
    title={(b) Retained fraction},
  ]
  \draw[svRule, line width=1.2pt] (axis cs:0.31,6) -- (axis cs:0.67,6);
  \fill[svBlue] (axis cs:0.67,6) circle (2.4pt);
  \fill[svFamII] ($(axis cs:0.39,6)+(-2.1pt,-2.1pt)$) rectangle ++(4.2pt,4.2pt);
  \node[regular polygon, regular polygon sides=3, fill=svGold, inner sep=1.05pt] at (axis cs:0.31,6) {};
  \draw[svRule, line width=1.2pt] (axis cs:0.45,5) -- (axis cs:0.83,5);
  \fill[svBlue] (axis cs:0.83,5) circle (2.4pt);
  \fill[svFamII] ($(axis cs:0.47,5)+(-2.1pt,-2.1pt)$) rectangle ++(4.2pt,4.2pt);
  \node[regular polygon, regular polygon sides=3, fill=svGold, inner sep=1.05pt] at (axis cs:0.45,5) {};
  \draw[svRule, line width=1.2pt] (axis cs:0.05,4) -- (axis cs:0.22,4);
  \fill[svBlue] (axis cs:0.21,4) circle (2.4pt);
  \fill[svFamII] ($(axis cs:0.22,4)+(-2.1pt,-2.1pt)$) rectangle ++(4.2pt,4.2pt);
  \node[regular polygon, regular polygon sides=3, fill=svGold, inner sep=1.05pt] at (axis cs:0.05,4) {};
  \draw[svRule, line width=1.2pt] (axis cs:0.49,3) -- (axis cs:0.8,3);
  \fill[svBlue] (axis cs:0.8,3) circle (2.4pt);
  \fill[svFamII] ($(axis cs:0.6,3)+(-2.1pt,-2.1pt)$) rectangle ++(4.2pt,4.2pt);
  \node[regular polygon, regular polygon sides=3, fill=svGold, inner sep=1.05pt] at (axis cs:0.49,3) {};
  \draw[svRule, line width=1.2pt] (axis cs:0.48,2) -- (axis cs:0.66,2);
  \fill[svBlue] (axis cs:0.66,2) circle (2.4pt);
  \fill[svFamII] ($(axis cs:0.48,2)+(-2.1pt,-2.1pt)$) rectangle ++(4.2pt,4.2pt);
  \node[svnote, anchor=west] at (axis cs:0.6900000000000001,2) {CLIP n.d.};
  \node[svnote] at (axis cs:0.5,1) {n.d.\ (vision at chance)};
  \coordinate (leg) at (axis cs:0,0.4);
\end{axis}
\node[anchor=north west, font=\svsmall, text=svInk]
  at ($(A.south west)+(0.0,-0.9)$)
  {\tikz[baseline=-.6ex]\draw[fill=white,draw=svInk,line width=.8pt] (0,0) circle (2.3pt);\ vision\quad
   \tikz[baseline=-.6ex]\fill[svFamIII] (0,0) circle (2.5pt);\ imagery\quad
   \textcolor{svFamIII!80!black}{$-x$}\,: drop (pts)};
\node[anchor=north west, font=\svsmall, text=svInk]
  at ($(B.south west)+(-0.3,-0.9)$)
  {\tikz[baseline=-.6ex]\fill[svBlue] (0,0) circle (2.4pt);\ human\quad
   \tikz[baseline=-.6ex]\fill[svFamII] (-2.1pt,-2.1pt) rectangle (2.1pt,2.1pt);\ PixCorr\quad
   \tikz[baseline=-.6ex]\node[regular polygon,regular polygon sides=3,fill=svGold,inner sep=1.05pt]{};\ CLIP};
\end{tikzpicture}
\caption{Perception-to-imagery transfer computed from the published values
of Table~\ref{tab:performance}, compiled across two
reports~\cite{Kneeland2025nsdimagery,Kneeland2026mirage} under a common human
forced-choice framing; this is a published-value comparison, not a unified
benchmark experiment. \textbf{(a)} Human forced-choice accuracy on seen
(open) against imagined (filled) images, where the length of each bar is the
seen-to-imagined drop and the rank column shows the change in method
ordering within this compilation, a diagnostic signal discussed in
Sec.~\ref{subsec:famIII}; Takagi \& Nishimoto is plotted in grey because
neither value differs from chance. \textbf{(b)} The retention ratio $\rho$
of Eq.~\eqref{eq:rho} for three protocols, with the iCNN CLIP value absent
rather than imputed because its vision score does not exceed chance.}
\label{fig:transfer}
\end{figure*}

\subsection{EEG/MEG/iEEG Imagery: What Exists and What Does Not}
\label{subsec:eegmeg}

In the reviewed corpus, image-level reconstruction scored against
image-level targets has been reported only from fMRI, while electrophysiology has focused on
classification, representational analysis and BCI control (the corpus by
year and modality is shown in Fig.~\ref{fig:timeline}). Part of this division
is principled, since Family~Ia needs a spatially resolved forward model that
sensor-level EEG and MEG do not provide. For semantic-level generative
decoding, however, the main constraints appear to be missing image-level
imagery targets and single-trial SNR: semantic reconstruction of \emph{perceived}
images already succeeds from MEG~\cite{Benchetrit2024brain} and
EEG~\cite{Li2024visual}, and the semantic stream is the one that transfers
to imagery (Sec.~\ref{subsec:substrates}).\footnote{This subsection rests on
Group~A of the search strategy of Sec.~\ref{subsec:methodology}, re-run with
the modality clause restricted to EEG, MEG, iEEG, ECoG and sEEG and closed on
the same date. Class counts, chance levels and validation units of all
electrophysiological studies are listed in Table~\ref{tab:ephys}.}

\paragraph{What has been established}
Imagined content is decodable with an oscillatory carrier: EEG
cross-classifies imagined and perceived objects in the parieto-occipital
alpha band~\cite{Xie2020alpha}, and extends to whole scenes and graded scene
properties~\cite{StecherKaiser2024alpha}. Imagery has a distinct temporal
profile, its MEG representations emerging later, more sustained and less
precise than perceptual ones~\cite{Dijkstra2018differential}. Reinstatement
is visible in single neurons, which patients can drive volitionally to
control a display~\cite{Kreiman2000imagery,Cerf2010online}, the only
closed-loop imagery demonstration we found. Visual imagery has also been
evaluated as an EEG-BCI control strategy with mixed
outcomes~\cite{Kosmyna2018attending,Lee2020neural,Kilmarx2024evaluating},
and a time-resolved EEG study found no above-chance decoding during imagery
at any time point~\cite{Shatek2019decoding}, consistent with the recoverable
signal lying in band-limited power over a long window.

\paragraph{Why reconstruction is harder here}
For Family~I the limit is physical: a sensor-level topography is a
volume-conducted mixture whose cortical source is ill-posed, and the
decodable imagery signal is band-limited
power~\cite{Xie2020alpha,StecherKaiser2024alpha}, which supports closed-set
classification but cannot constrain pixels through
Eq.~\eqref{eq:map-decomp}. For Family~III the limits are data and SNR:
imagery offers no event to lock to beyond the cue and its onset jitters
across trials, so averaging blurs the dynamics of interest, and jitter alone
can mimic a sustained representation~\cite{DijkstraBosch2019shared}.

\paragraph{Ocular state and leakage}
Imagery is often performed with eyes closed and perception with eyes
open~\cite{Reddy2010reading,Horikawa2017generic}, and eye closure raises
posterior alpha, the band that carries decodable imagery content, so eye
state should be reported (S10). Long trials, class-blocked runs and a fixed
cue--imagery--rest template also expose imagery paradigms to the temporally
autocorrelated leakage shown for block-designed EEG~\cite{Li2020perils}; for
example, one study augments trials before a 60/20/20
split~\cite{Kwon2025functional}, and another cuts each imagery period into
segments before splitting~\cite{Shimizu2022improving}. Splitting and
permuting by imagery episode or run addresses this (S8).

\paragraph{What is missing}
No openly released MEG or intracranial visual-imagery dataset was identified
as of \searchdate, and the public EEG
datasets~\cite{Gao2026eegvi,Wilson2023semantic,Rybar2025simultaneous} carry
class labels rather than image-level targets. The one EEG synthesis
attempt~\cite{Shimizu2022improving} scores its images only through a
classifier, slightly above the 2.5\% chance of 40 classes, and shows the
target before each imagery period (\lone), so it remains exploratory. An
electrophysiological dataset would enable Family~III if it (a)~cued
material learned before the session symbolically; (b)~paired each trial with
its image-level target and released the stimuli; (c)~recorded a
within-subject perception condition; (d)~logged eye state and vividness per
trial; (e)~randomised class order and released episode labels; and
(f)~added an \lthree{} condition.

\FloatBarrier

\section{Evaluation Methodology}
\label{sec:evaluation}

\subsection{The Target-Free Evaluation Setting}
\label{subsec:targetfree}

\begin{figure*}[!htbp]
\centering
\begin{tikzpicture}[sv, x=1cm, y=1cm]
\begin{scope}
  \node[font=\svfont\bfseries, anchor=west] at (-0.2,4.55) {(a) The proxy gap};
  \node[svbox, minimum width=1.7cm, minimum height=0.85cm] (Ic) at (0.85,0.6)
        {$I_c$ or $c$\\[-1pt]{\svsmall\color{svGrey}proxy}};
  \node[svbox, minimum width=1.7cm, minimum height=0.85cm, draw=svBlue, thick] (Ih) at (5.15,0.6)
        {$\hat I$\\[-1pt]{\svsmall\color{svGrey}reconstruction}};
  \node[svbox, svdash, minimum width=1.9cm, minimum height=0.85cm, fill=svPanel] (s) at (3.0,3.55)
        {$s$\\[-1pt]{\svsmall\color{svGrey}imagined content}\\[-3pt]{\svsmall\color{svGrey}never observed}};
  \draw[line width=1.1pt, svBlue] (Ic) -- (Ih);
  \node[font=\svsmall, text=svBlue, fill=white, inner sep=1.5pt] at (3.0,0.6) {$d(\hat I, I_c)$};
  \node[font=\svsmall, text=svBlue] at (3.0,0.25) {measured};
  \draw[line width=1.1pt, svGrey, svdash] (Ih) -- node[right=3pt, font=\svsmall, text=svGrey, align=left]
        {$d(\hat I, s)$\\desired,\\unobservable} (s);
  \draw[line width=1.1pt, svRed] (Ic) -- node[left=3pt, font=\svsmall, text=svRed, align=right]
        {$\Delta(I_c,s)$\\proxy gap} (s);
  \node[font=\svsmall, text=svInk] at (3.0,-0.4)
        {$\bigl|\,d(\hat I,I_c)-d(\hat I,s)\,\bigr|\;\le\;\Delta(I_c,s)$\quad Eq.~\eqref{eq:eval-bias}};
  \begin{scope}[shift={(0.75,-1.3)}]
    \node[font=\svsmall, anchor=east] at (0,0) {$\mathbb{E}[\Delta]$};
    \fill[svRed!25] (0.15,-0.1) rectangle (0.95,0.1);
    \fill[svRed!50] (1.70,-0.1) rectangle (3.0,0.1);
    \fill[svRed!80] (3.75,-0.1) rectangle (5.25,0.1);
    \node[font=\svsmall, anchor=north] at (0.55,-0.12) {\textsf{L1}};
    \node[font=\svsmall, anchor=north] at (2.35,-0.12) {\textsf{L2}};
    \node[font=\svsmall, anchor=north] at (4.50,-0.12) {\textsf{L3}};
    \node[font=\svsmall] at (1.32,0) {$<$};
    \node[font=\svsmall] at (3.37,0) {$<$};
  \end{scope}
  \node[svnote, anchor=north west, align=left, text width=6.0cm] at (-0.2,-1.95)
        {Ordering of $\mathbb{E}[\Delta]$ by memory source is coarse; the bias is systematic, not zero-mean.};
\end{scope}
\begin{scope}[shift={(7.2,0)}]
  \node[font=\svfont\bfseries, anchor=west] at (-0.75,4.55) {(b) Neural dependence};
  \def\base{-0.3}
  \def\ch{0.55}
  \def\ya{3.45} \def\yb{2.65} \def\yc{1.65}
  \draw[svGrey] (0,\base) -- (0,4.0);
  \draw[svGrey] (0,\base) -- (3.55,\base);
  \node[svnote, rotate=90, anchor=south] at (-0.55,1.9) {proxy score $\mathcal{E}(\,\cdot\,, I_c)$};
  \draw[svGrey, svdash] (0,\ch) -- (3.55,\ch) node[svnote, anchor=west] {$\mathcal{E}_{\mathrm{chance}}$};
  \fill[svBlue]    (0.25,\base) rectangle (0.95,\ya);
  \fill[svBlue!55] (1.40,\base) rectangle (2.10,\yb);
  \fill[svBlue!25] (2.55,\base) rectangle (3.25,\yc);
  \draw[svGrey, svdash] (0,\ch) -- (3.55,\ch);
  \node[font=\svsmall, anchor=north] at (0.60,\base) {$\mathbf{x}$};
  \node[font=\svsmall, anchor=north] at (1.75,\base) {$\mathbf{x}^{\mathrm{within}}_{\mathrm{null}}$};
  \node[font=\svsmall, anchor=north] at (2.90,\base) {$\mathbf{x}^{\mathrm{between}}_{\mathrm{null}}$};
  \draw[svRule] (0.95,\ya) -- (3.6,\ya);
  \draw[svRule] (2.10,\yb) -- (3.6,\yb);
  \draw[svRule] (3.25,\yc) -- (3.6,\yc);
  \draw[decorate, decoration={brace, amplitude=3pt}, svInk]
        (3.6,\ya) -- (3.6,\yb) node[midway, right=4pt, font=\svsmall] {$\Lambda_{\mathrm{within}}$};
  \draw[decorate, decoration={brace, amplitude=3pt}, svInk]
        (3.6,\yb) -- (3.6,\yc) node[midway, right=4pt, font=\svsmall, align=left]
        {$\Lambda_{\mathrm{between}}-\Lambda_{\mathrm{within}}$\\[-1pt]{\color{svGrey}category alone}};
  \draw[svRed, line width=.8pt, {Bar[width=4pt]}-{Bar[width=4pt]}] (0.12,\ch) -- (0.12,\ya);
  \node[font=\svsmall, text=svRed, rotate=90, anchor=south] at (0.12,2.0) {denominator of $\tilde\Lambda$};
  \node[svnote, anchor=north west, align=flush left, text width=5.6cm] at (-0.75,-0.95)
        {$\mathbf{x}$: real trial; within-null: a different stimulus of the same category or set; between-null: a trial of any other
         stimulus. $\tilde\Lambda=\Lambda/(\mathcal{E}(\hat I(\mathbf{x}),I_c)-\mathcal{E}_{\mathrm{chance}})$,
         Eq.~\eqref{eq:lambda}. Only $\Lambda_{\mathrm{within}}>0$ certifies within-category dependence. Heights are schematic.};
\end{scope}
\end{tikzpicture}
\caption{Target-free evaluation. \textbf{(a)}~The proxy gap of
Eq.~\eqref{eq:eval-bias}: the evaluator measures $d(\est{I},\exemplar)$, the
desired $d(\est{I},\imgcontent)$ is unobservable, and their difference is
bounded by $\Delta(\exemplar,\imgcontent)$, whose expectation the
memory-source levels order coarsely. \textbf{(b)}~Reading the neural
dependence of Eq.~\eqref{eq:lambda} from null re-runs of the same pipeline:
$\Lam_{\mathrm{within}}$ against a same-category null, the share
$\Lam_{\mathrm{between}}-\Lam_{\mathrm{within}}$ carried by category alone,
and the denominator of $\tilde{\Lam}$. Bar heights are schematic.}
\label{fig:proxygap}
\end{figure*}
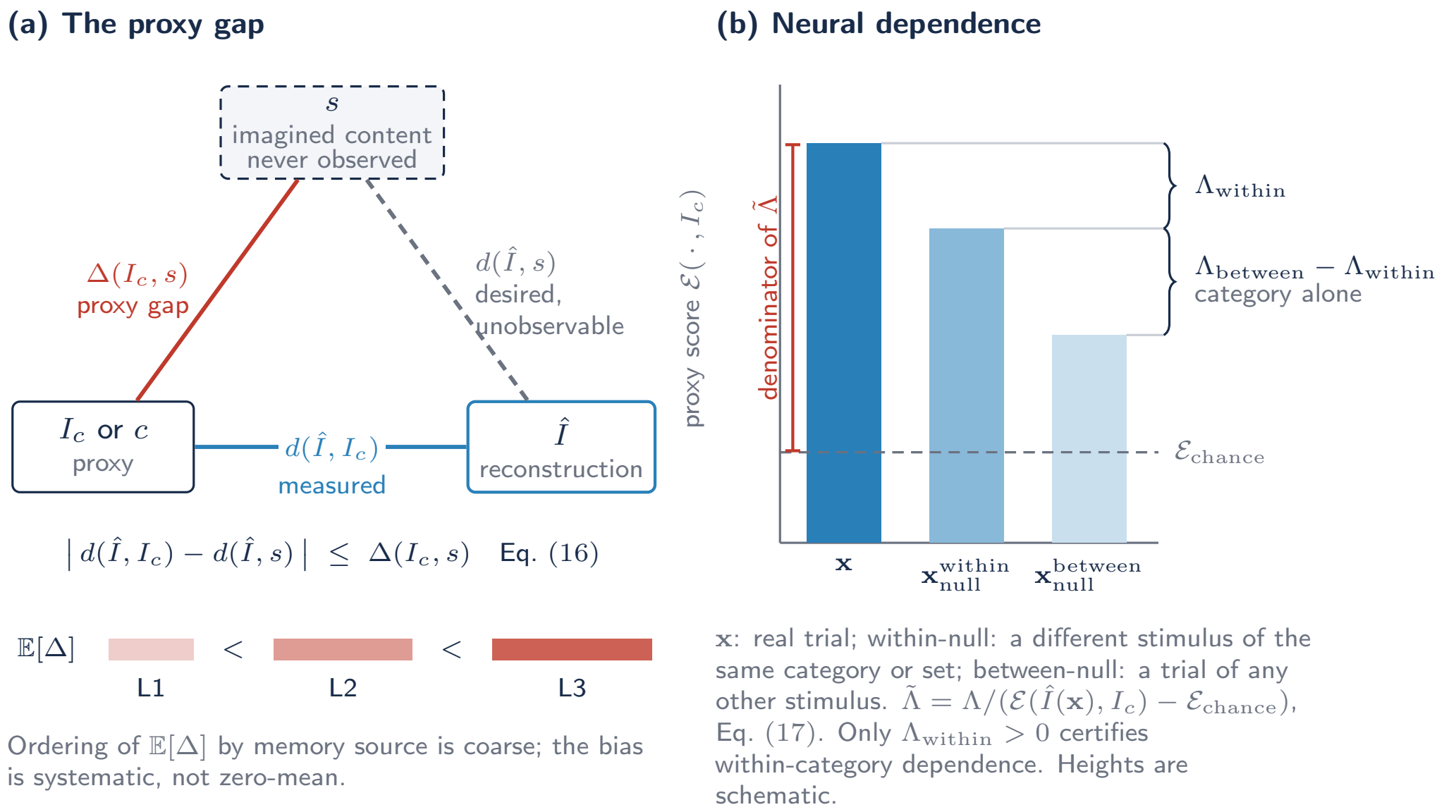

\paragraph{The proxy gap}
In perception $\exemplar=\imgcontent$, so a distance between $\est{I}$ and
$\exemplar$ is a distance to the ground truth. In imagery the evaluator can
compare $\est{I}$ only with the proxy through the operator of
Eq.~\eqref{eq:eval}, and for a distance-type operator $d$ the triangle
inequality bounds the discrepancy by the distance between cue and content
(Fig.~\ref{fig:proxygap}a):
\begin{equation}
\label{eq:eval-bias}
\bigl|\, d\big(\est{I},\exemplar\big) - d\big(\est{I},\imgcontent\big) \,\bigr|
\;\le\; d\big(\exemplar,\imgcontent\big) \;=:\; \Delta\big(\exemplar,\imgcontent\big).
\end{equation}
For similarity-type operators, such as identification, CLIP scores and human
ratings, no such bound holds in general, and the relevant quantity is the
\emph{proxy bias}
$b(\est{I};\exemplar,\imgcontent)=\mathcal{E}(\est{I},\exemplar)-\mathcal{E}(\est{I},\imgcontent)$.
The bias depends on the reconstruction itself and is systematic rather than
zero-mean, varying with imagery ability, cueing and memory source
(Secs.~\ref{subsec:paradigms} and~\ref{subsec:individual}), so it cannot be
averaged away and confounds comparison across paradigms. The memory-source
levels order $\mathbb{E}[\Delta]$ coarsely: smallest for \lone{}, where the
proxy was on screen moments earlier, larger for \ltwo{}, and largest for
purely conceptual \lthree{}, where the proxy is a word and the exemplar is the participant's
own.

\paragraph{Neural dependence}
Eq.~\eqref{eq:eval-bias} prices the proxy but does not indicate whether a
reconstruction used the brain at all. Consolidating the randomised-data
control of~\cite{Shirakawa2025spurious} and the randomly decoded distractors
of~\cite{Kneeland2026mirage}, we propose the \emph{neural dependence} of a
reconstruction procedure, a diagnostic quantity rather than a validated
metric, defined as the drop in its score when the pairing between
neural data and target is destroyed with everything else fixed, together
with its normalised form (Fig.~\ref{fig:proxygap}b):
\begin{subequations}
\label{eq:lambda}
\begin{align}
\Lam &=
\mathbb{E}\big[\mathcal{E}\big(\est{I}(\neural),\,\exemplar\big)\big]
-
\mathbb{E}\big[\mathcal{E}\big(\est{I}(\neural_{\mathrm{null}}),\,\exemplar\big)\big],
\label{eq:lambda-abs}\\
\tilde{\Lam} &=
\frac{\Lam}{\mathbb{E}\big[\mathcal{E}\big(\est{I}(\neural),\,\exemplar\big)\big]-\mathcal{E}_{\mathrm{chance}}},
\label{eq:lambda-norm}
\end{align}
\end{subequations}
with expectations over trials and the generator's stochasticity and the
same seeds for real and null inputs. $\tilde{\Lam}$, the fraction of the
above-chance score attributable to trial-specific signal, is intended to be
comparable across operators and studies and is reported unclipped. As a
quantity proposed here, $\Lam$ has not been validated on independent data:
its sensitivity, its specificity and its robustness to cue leakage and to
eye state remain to be established. The null is
\emph{test-time mismatch}: the trained pipeline is driven with another
trial's activity (training-label permutation answers a different question
and is reported separately). With
$\neural^{\mathrm{between}}_{\mathrm{null}}$ a trial of any other stimulus
and $\neural^{\mathrm{within}}_{\mathrm{null}}$ a trial of a different
stimulus from the same category or set, a pipeline that decodes category
alone achieves $\Lam_{\mathrm{between}}>0$, whereas only
$\Lam_{\mathrm{within}}>0$ indicates within-category dependence. $\Lam$ can
be estimated from existing data by re-running the pipeline on null inputs;
label outputs, cue leakage and cost are treated in \ref{app:nullimpl}.
Reporting
$(\mathcal{E},\Lam_{\mathrm{between}},\Lam_{\mathrm{within}},\tilde{\Lam})$
with an episode-level permutation test is proposed item~S1; it is designed
to separate the two panels of Fig.~\ref{fig:spurious}, where a spurious reconstruction holds
$\mathcal{E}$ high while $\Lam\approx0$. Even $\Lam_{\mathrm{within}}>0$ is
necessary rather than sufficient, since it does not certify the visual
detail displayed. In the exploratory checklist application below, an S1
entry records only whether a corresponding null control is reported, not
whether that control is adequate.

\paragraph{Output dimension collapse}
If the neural-to-latent map is a ridge regression from training activity
$X_{\mathrm{tr}}$ onto training latents $Y_{\mathrm{tr}}$, every prediction
is a combination of the training latents:
\begin{equation}
\label{eq:collapse}
\est{\decrep}_{\mathrm{te}}
= Y_{\mathrm{tr}}^{\top}\,
  \underbrace{X_{\mathrm{tr}}\big(X_{\mathrm{tr}}^{\top}X_{\mathrm{tr}}+\lambda I\big)^{-1}\neural_{\mathrm{te}}}_{\mathclap{=\,\mathbf{m}\,\in\,\mathbb{R}^{N}}}
= \sum_{i=1}^{N} m_i\, \decrep^{(i)}_{\mathrm{tr}}.
\end{equation}
The linear span is restrictive only when $N$ is smaller than the latent
dimension; the relevant constraint is effective: shrinkage concentrates
$\mathbf{m}$ on the dominant directions of the training targets, so that
when those targets are clustered the predictions fall near the subspace
spanned by the cluster centroids. Shirakawa et al.\ call this \emph{output
dimension collapse} and find that the CLIP text features of NSD images form
roughly 40 clusters with substantial train--test
overlap~\cite{Shirakawa2025spurious}. Two quantities measure it:
\begin{equation}
\label{eq:effdim}
\kappa_{r} = 1-\frac{\sum_{j}\norm*{(\mathrm{Id}-P_{r})\big(\decrep^{(j)}_{\mathrm{te}}-\bar{\decrep}_{\mathrm{tr}}\big)}_2^2}
                    {\sum_{j}\norm*{\decrep^{(j)}_{\mathrm{te}}-\bar{\decrep}_{\mathrm{tr}}}_2^2},
\qquad
r_{\mathrm{eff}}(Z) = \frac{\big(\operatorname{tr} C_{Z}\big)^{2}}{\operatorname{tr}\big(C_{Z}^{2}\big)},
\end{equation}
where $P_{r}$ projects onto the $r$ leading principal directions of the
centred training latents, with $r=r_{95}$ explaining 95\% of their variance,
and $C_{Z}$ is the covariance of a set of latents $Z$. A small $r_{95}$ with
$\kappa_{r_{95}}$ near $0.95$ marks a test set inside a low-dimensional,
clustered training distribution; a ratio
$r_{\mathrm{eff}}(\est{Z}_{\mathrm{te}})/r_{\mathrm{eff}}(Z_{\mathrm{te}})$
well below 1 marks predictions less diverse than their targets (S6).
This argument applies directly to linear/ridge latent mappings; for
nonlinear decoders it is a heuristic diagnostic. Where it applies, collapse
compounds the proxy gap: a translator confined to a few clusters and scored
against a cluster-level proxy can appear successful while recovering little
that is trial-specific.

\subsection{A Taxonomy of Evaluation Metrics}
\label{subsec:metrics}

\paragraph{(1) Identification and retrieval}
Identification asks whether $\est{I}$ or $\decrep$ is closer to the correct
proxy than to distractors, with chance $1/n$ for $n$ alternatives, and its
main pitfall is the distractor set. If the test set has $k$ equally sized
categories and the decoder recovers category but nothing within it, so that
a same-category distractor is identified at chance and a different-category
one with probability $q\in[0.5,1]$, the large-$n$ pairwise accuracy is
\begin{equation}
\label{eq:pairwise-inflation}
\mathrm{Acc} = \frac{1}{k}\cdot 0.5 + \Big(1-\frac{1}{k}\Big)\, q ,
\end{equation}
derived with its finite-sample form in \ref{app:pairwise} and plotted
in Fig.~\ref{fig:inflation}~\cite{Shirakawa2025spurious}. With $q=1$ it is
75\% for $k=2$ and 95\% for $k=10$, so a pairwise accuracy alone is not
evidence about the particular image. Protocol choice also matters: under
human forced choice and an automated CLIP two-way comparison, both at a
nominal 50\% chance, iCNN reaches 66.15\% and 41.72\% on the same imagery
reconstructions~\cite{Kneeland2025nsdimagery,Kneeland2026mirage}. An
accuracy reported without its $n$, distractor rule and scoring space
therefore cannot be ranked against another.

\paragraph{(2) Low-level and feature-metric alignment}
Feature metrics score $\mathcal{E}$ as a distance in a fixed representation:
pixel correlation and SSIM~\cite{Wang2004image}; features from
AlexNet~\cite{Krizhevsky2012imagenet}, residual networks~\cite{He2016deep},
DINO~\cite{Caron2021emerging} or EVA-02~\cite{EvaProject2023eva02}; learned
perceptual distances~\cite{Zhang2018unreasonable}; and CLIP
scores~\cite{Radford2021learning}. FID~\cite{Heusel2017gans} and the
Inception Score~\cite{Salimans2016improved} gauge realism, not fidelity to
$\imgcontent$. For imagery, low-level fidelity and semantics are only weakly
coupled (Sec.~\ref{subsec:substrates}),
\begin{equation}
\label{eq:twostream}
\mathcal{E}_{\text{low}}\big(\est{I},\exemplar\big)
\not\Rightarrow
\mathcal{E}_{\text{high}}\big(\est{I},\exemplar\big),
\end{equation}
so the two streams should be reported separately. Panel~B of
Table~\ref{tab:performance} shows the dissociation: iCNN has the highest
pixel correlation in both conditions (0.224, 0.108) and the lowest CLIP
score of the five significant methods (49.03\%, 41.72\%).

\paragraph{(3) Human rating and (4) phenomenological correspondence}
Raters reward naturalness, so a fluent, prior-driven synthesis can score well
while weakly tied to $\neural$. Drawing the distractor from a reconstruction
of a \emph{different} brain pattern by the \emph{same}
pipeline~\cite{Kneeland2025nsdimagery,Kneeland2026mirage} equalises the
prior and estimates $\Lam_{\mathrm{between}}$; a same-category distractor
would estimate $\Lam_{\mathrm{within}}$. Decodability can also be scored
against experience, trial-wise or via the VVIQ~\cite{Marks1973visual}; a
vividness correlation supports genuine read-out~\cite{Dijkstra2017vividness},
but its absence does not
disqualify~\cite{Chang2025imageless,Cabbai2024sensory}. NSD-Imagery records
per-trial vividness~\cite{Kneeland2025nsdimagery}, yet no study in
Table~\ref{tab:performance} reports this analysis.

\subsection{Spurious Reconstruction}
\label{subsec:spurious}

\paragraph{The mechanism}
A frozen, high-capacity text-to-image prior (Eq.~\eqref{eq:famIII}) can
synthesise a photorealistic, category-consistent image from little
information in $\neural$, so the output becomes a blend of
\emph{classification} into a trained category and \emph{prior-driven
generation}~\cite{Shirakawa2025spurious}. As the neural dependence falls, it
collapses onto the prior's typical sample for the inferred class
(Fig.~\ref{fig:spurious}):
\begin{equation}
\label{eq:spurious}
\est{I} = \gen\!\big(\decmap(\neural)\big)
\;\xrightarrow[\;\Lam\,\to\,0\;]{}\;
\gen\big(\decrep_{\text{class-prior}}\big),
\end{equation}
which remains fluent while decoupled from $\neural$, and which is testable
because its antecedent is estimated by the null-signal control.

\paragraph{Why standard metrics may not detect it}
Feature and CLIP scores can be met by the class-typical sample, realism
ratings reward the prior's fluency, and cross-category identification can
be solved by category alone (Eq.~\eqref{eq:pairwise-inflation}). Convincing
reconstructions have been obtained even from \emph{randomised} brain data by
sampling repeatedly and keeping the best output~\cite{Shirakawa2025spurious}.
A reanalysis of a published imagery result raises concerns about selective
multi-run reporting and circular evaluation~\cite{Shirakawa2025advancing}\pre{};
under P3 it carries no conclusion alone, but Panel~C of
Table~\ref{tab:performance} does score each pipeline with its own weighted
similarity~\cite{KoideMajima2024mental} and is therefore a within-study
contrast rather than an accuracy comparable with Panels~A, B or~D. As modern
priors make realistic outputs common regardless of
fidelity~\cite{Shirakawa2025spurious,Vigotsky2024mental}, realism is weak
evidence, which is why $\Lam$, insensitive to realism, is reported instead.

\paragraph{The recovery check}
Feeding the \emph{true} latent features of a target image into the
generator tests whether the image comes back; if it does not, the generator
bounds the pipeline whatever the decoder does. Text-guided pipelines return
semantically similar but perceptually different images in this check,
whereas feature-matching inversion passes it~\cite{Shirakawa2025spurious}.
For imagery only the generator-side half transfers: the check is run on the
\emph{perceptual} stimuli of the same corpus, with the paired vision
condition as an upper bound under the same generator and subjects (S5).

\paragraph{The chain of failure}
A clustered corpus induces collapse (Eq.~\eqref{eq:collapse}); collapsed
latents are rendered as fluent class-typical images
(Eq.~\eqref{eq:spurious}); categorical test structure and
fluency-rewarding raters inflate their scores
(Eq.~\eqref{eq:pairwise-inflation}); and the score is then read as fidelity.
Each link has a standard in Table~\ref{tab:standards} (Fig.~\ref{fig:causalchain}): S6 for the corpus, S2 and S5 for translator and generator, S3 and
S7 for the metric, and S1 for the inference. Together they operationalise
for imagery the correctives proposed by Shirakawa et al., namely diverse
training data, compositional representations and evaluation designed to
expose prior fill-in~\cite{Shirakawa2025spurious}.

\subsection{A Candidate Reporting Checklist}
\label{subsec:guidelines}

We distil the analysis into a candidate reporting checklist of ten proposed
items, each stated in Table~\ref{tab:standards} as what to report, how to
compute it and an illustrative pass criterion, and apply it in an
exploratory manner to ten studies in Table~\ref{tab:compliance}. The items
are proposals of the authors, not established standards.

\begin{table*}[!htbp]
\centering
\setlength{\tabcolsep}{3.5pt}
\renewcommand{\arraystretch}{1.0}
\begin{threeparttable}
\caption{A candidate reporting checklist for imagery decoding and
reconstruction: ten proposed reporting items with illustrative pass
criteria. The items are proposals of the authors, not established
standards. S1--S8 apply to any imagery decoding study, S9 to any study
claiming imagery rather than maintenance, and S10 to EEG and MEG.}
\label{tab:standards}
\scriptsize
\begin{tabularx}{\linewidth}{@{}L{1.9cm}YYL{2.3cm}@{}}
\toprule
\textbf{Proposed item} & \textbf{Compute} & \textbf{Illustrative pass criterion} &
\textbf{Failure it closes} \\
\midrule
\textbf{S1} Null-signal baseline &
$\Lam$, $\tilde{\Lam}$ (Eq.~\eqref{eq:lambda}) under test-time mismatch:
re-run the deployed pipeline on another trial's $\neural$ with generator,
samples and seeds fixed; between-trial and within-category nulls for
generative and retrieval pipelines; label permutation reported separately &
$\Lam_{\mathrm{within}}>0$ and $\Lam_{\mathrm{between}}>0$, each with an
episode-level permutation test, $p<0.05$ over $B$ stated permutations
($B\ge1000$ where feasible); for label outputs, above-chance held-out
accuracy under the same test; never $\mathcal{E}$ alone &
Prior-driven synthesis scored as read-out (Eq.~\eqref{eq:spurious}) \\
\addlinespace[1pt]
\textbf{S2} Zero-shot categories &
Evaluate on categories, and where possible exemplars, held out from
decoder training &
Above chance on held-out categories, reported separately &
Classification mistaken for generalisation
\cite{Shirakawa2025spurious,Horikawa2017generic} \\
\addlinespace[1pt]
\textbf{S3} Dual-stream metrics &
Low-level (PixCorr, SSIM~\cite{Wang2004image}) and semantic (CLIP) scores
computed separately &
Both reported, not merged into a composite nor selected post hoc &
Dissociation hidden by one scalar (Eq.~\eqref{eq:twostream}) \\
\addlinespace[1pt]
\textbf{S4} Vividness covariate &
Trait (VVIQ) and, where available, per-trial vividness; score regressed on it &
Distribution and score--vividness relation reported, even when null &
Proxy gap confounded with method quality (Eq.~\eqref{eq:eval-bias}) \\
\addlinespace[1pt]
\textbf{S5} Generator recovery &
Feed true latents of the \emph{perceptual} stimuli into the frozen
generator and score recovery with the reconstruction metrics &
Recovery exceeds the reported reconstruction score, or the generator bound
is stated &
Generator loss attributed to the decoder~\cite{Shirakawa2025spurious} \\
\addlinespace[1pt]
\textbf{S6} Train--test structure &
Semantic clusters of the training targets and train--test overlap;
$r_{95}$, $\kappa_{r_{95}}$ and the effective-rank ratio
$r_{\mathrm{eff}}(\est{Z}_{\mathrm{te}})/r_{\mathrm{eff}}(Z_{\mathrm{te}})$
(Eq.~\eqref{eq:effdim}) &
Reported; if overlap is substantial or the ratio well below 1, S2 also on
a cluster-disjoint split &
Output dimension collapse (Eq.~\eqref{eq:collapse}) \\
\addlinespace[1pt]
\textbf{S7} No best-of-$N$ selection &
Fix samples per trial and the selection rule before evaluation &
Rule stated; qualitative examples drawn at random; seed variability
reported &
Selective reporting~\cite{Shirakawa2025advancing} \\
\addlinespace[1pt]
\textbf{S8} Episode-level validation &
Split and permute by imagery episode, run or session; augment after
splitting &
No episode on both sides of a split; cross-block or cross-session result
for block designs &
Autocorrelation leakage~\cite{Li2020perils,Kwon2025functional,Shimizu2022improving} \\
\addlinespace[1pt]
\textbf{S9} Admissibility statement &
Trial-level separability crossed with memory source (\lone/\ltwo/\lthree),
Sec.~\ref{subsec:terminology} &
Both stated, with the trial timeline that justifies them &
Maintenance or recall reported as generation \\
\addlinespace[1pt]
\textbf{S10} Eye state &
Eye state in imagery and in perception; eye tracking or EOG regression
where feasible &
Stated for both; any difference addressed for the alpha band &
Alpha carrying ocular state, not content \\
\bottomrule
\end{tabularx}
\end{threeparttable}
\end{table*}

\begin{table*}[!htbp]
\centering
\setlength{\tabcolsep}{0pt}
\renewcommand{\arraystretch}{1.35}
\begin{threeparttable}
\caption{Exploratory application of the proposed reporting checklist of
Table~\ref{tab:standards} to ten studies, scored from the published paper and
supplementary material only, with per-cell evidence in
\ref{app:compliance}. \cyes{} the paper reports the item in a checkable
form, \cpart{} partially, \cno{} not reported, \cna{} not applicable. This is
a single exploratory scoring pass, not a quality ranking: it does not
evaluate the studies themselves, only whether the proposed items are
reported in a checkable form. Inter-rater reliability was not estimated,
and several studies predate the concerns motivating S1, S5 and S6.}
\label{tab:compliance}
\scriptsize
\begin{tabular*}{\linewidth}{@{\extracolsep{\fill}}>{\RaggedRight\arraybackslash}m{3.9cm}>{\centering\arraybackslash}m{1.08cm}*{10}{M{0.72cm}}@{}}
\toprule
\textbf{Study} & \textbf{Family} &
\textbf{S1} & \textbf{S2} & \textbf{S3} & \textbf{S4} & \textbf{S5} &
\textbf{S6} & \textbf{S7} & \textbf{S8} & \textbf{S9} & \textbf{S10} \\
\midrule
Naselaris et al.\ 2015~\cite{Naselaris2015voxelwise} & \textcolor{svFamI}{\textbf{Ia}} &
\Cp & \Cy & \Cp & \Cn & \Ca & \Cn & \Cy & \Cy & \Cy & \Ca \\
Senden et al.\ 2019~\cite{Senden2019imagery} & \textcolor{svFamI}{\textbf{Ia}} &
\Cn & \Cp & \Cp & \Cn & \Ca & \Cn & \Cp & \Cy & \Cy & \Ca \\
Shen et al.\ 2019~\cite{Shen2019deep} & \textcolor{svFamIb}{\textbf{Ib}} &
\Cn & \Cy & \Cp & \Cp & \Cn & \Cn & \Cp & \Cy & \Cy & \Ca \\
Koide-Majima et al.\ 2024~\cite{KoideMajima2024mental} & \textcolor{svFamIb}{\textbf{Ib}} &
\Cn & \Cy & \Cp & \Cn & \Cn & \Cn & \Cn & \Cy & \Cy & \Ca \\
\addlinespace[2pt]
Reddy et al.\ 2010~\cite{Reddy2010reading} & \textcolor{svFamII}{\textbf{II}} &
\Cp & \Cn & \Ca & \Cn & \Ca & \Cn & \Ca & \Cy & \Cy & \Cy \\
Horikawa \& Kamitani 2017~\cite{Horikawa2017generic} & \textcolor{svFamII}{\textbf{II}} &
\Cp & \Cy & \Ca & \Cp & \Ca & \Cn & \Ca & \Cy & \Cy & \Cy \\
Xie et al.\ 2020~\cite{Xie2020alpha} & \textcolor{svFamII}{\textbf{II}} &
\Cy & \Cn & \Ca & \Cn & \Ca & \Cn & \Ca & \Cy & \Cy & \Cp \\
Kwon et al.\ 2025~\cite{Kwon2025functional} & \textcolor{svFamII}{\textbf{II}} &
\Cn & \Cn & \Ca & \Cn & \Ca & \Cn & \Ca & \Cn & \Cn & \Cn \\
\addlinespace[2pt]
Kneeland et al.\ 2025~\cite{Kneeland2025nsdimagery} & \textcolor{svFamIII}{\textbf{III}} &
\Cp & \Cp & \Cp & \Cn & \Cn & \Cn & \Cy & \Cy & \Cy & \Ca \\
Kneeland et al.\ 2026~\cite{Kneeland2026mirage} & \textcolor{svFamIII}{\textbf{III}} &
\Cp & \Cp & \Cp & \Cn & \Cn & \Cn & \Cy & \Cy & \Cy & \Ca \\
\bottomrule
\end{tabular*}
\begin{tablenotes}[flushleft]
\scriptsize
\item \textbf{Reading the columns.} S1 is \cyes{} where the null described for
proposed item S1 in Table~\ref{tab:standards} is reported, which records that
a corresponding control is reported, not that it is adequate: for label outputs, whose mismatch null
scores chance by construction, a permutation test of held-out accuracy at
the trial level of a randomised design, as in the sign-permutation cluster
statistics of \cite{Xie2020alpha}. \cpart{} marks a between-trial null
without the within-category one, as in the randomly decoded distractors
of~\cite{Kneeland2026mirage}, or permutation tests against chance whose
relation to the test-time mismatch of Eq.~\eqref{eq:lambda} the report does
not make checkable. S2 is \cyes{} where held-out
categories are evaluated by
design~\cite{Horikawa2017generic,Shen2019deep,KoideMajima2024mental,Naselaris2015voxelwise}
and \cpart{} where test stimuli are held out yet belong to the training
cluster
structure~\cite{Kneeland2025nsdimagery,Kneeland2026mirage}. S3 is \cna{}
where the output is a label; elsewhere it is scored from the published
descriptions of metrics and distractor protocols rather than a complete
re-check against the distractor-rule requirement of
Sec.~\ref{subsec:guidelines}, so studies reporting both metric streams are
entered as \cpart{}. S8 and S9 are \cno{} for
\cite{Kwon2025functional} (Sec.~\ref{subsec:famII}), and S10 is \cna{} for
fMRI, \cyes{} where eye state is specified and
monitored~\cite{Reddy2010reading,Horikawa2017generic}, \cpart{} where its
alpha-band implication goes unanalysed~\cite{Xie2020alpha}.
\item \textbf{Summary.} Of the two proposed items that no study reports even
partially in this application, S6 requires
only an analysis of the stimulus corpus and of predictions already computed,
and S5 only access to the generator, so both could be supplied
retrospectively from existing data.
\end{tablenotes}
\end{threeparttable}
\end{table*}

\paragraph{Provenance, scoring and clarifications}
S1--S4 were fixed before the exploratory application and S5--S10 added
during it, so the application is not fully independent of the corpus it
scores. The distractor-rule
requirement of S3 below was specified after the initial scoring; S3 cells
are therefore scored from published distractor descriptions rather than a
complete re-check, and are entered conservatively (Table~\ref{tab:compliance}). A study reports all
proposed items in a fully checkable form when every applicable item is
\cyes{}. Three readings apply to the illustrative pass criteria: S3 also requires the number of alternatives $n$, the
distractor-sampling rule and the proportion of same-category distractors
(Eq.~\eqref{eq:pairwise-inflation}); the S5 recovery score should be
reported as the pipeline's ceiling, with the ratio of reconstruction to
recovery score; and S1 should state the cost of the null re-runs and, for
imagery-trained decoders with a cue on screen, include the cue-only control
(\ref{app:nullimpl}).

\paragraph{The exploratory application}
In this exploratory application, checkable reporting is more frequent where
a proposed item matches existing practice, as with S2 in work building on
zero-shot generic decoding~\cite{Horikawa2017generic}, and less frequent
where no convention requires the quantity: no study reports S5 or S6, one
reports S1 in a fully checkable form and five partially, and S3 is reported
at most partially under the conservative scoring described above. These
observations describe reporting in ten studies, not the quality of those
studies or of the field.

\FloatBarrier

\section{Open Problems and Future Directions}
\label{sec:challenges}

\subsection{Data Scarcity and the Generative Gap}
\label{subsec:gap1}

The reconstruction values in Table~\ref{tab:performance} rest on two
benchmarks and eleven participants~\cite{Kneeland2025nsdimagery,Shen2019deep},
only four of whom supply the decoders evaluated on NSD-Imagery, and outside
fMRI no imagery dataset with image-level targets was identified
(Sec.~\ref{subsec:eegmeg}); the missing
ingredient there is the target rather than the sample size.

\paragraph{Generation}
The \lthree{} case, content composed rather than retrieved, has no
image-based evaluation (Sec.~\ref{subsec:datasetstable}). Three routes could
make it scorable. \emph{(i) Compositional cueing}: cue novel combinations of
familiar attributes, such as a red circle above a blue square, so that a
ground-truth image can be built from the cue without being shown.
\emph{(ii) Post hoc externalisation}: have participants draw, select or
describe their exemplar after the trial; decoding recalled video content
into descriptive text~\cite{Horikawa2025mind} suggests that a verbal target
is feasible. \emph{(iii) Within-participant consistency}: score whether a
reconstruction matches the same participant's other instances of a repeated
\lthree{} cue better than other participants' instances, an identification
problem with a well-defined chance level and no image target. Route~(i)
yields compositionally constructed \lthree{} targets in the sense of
Sec.~\ref{subsec:terminology}, whereas routes~(ii) and~(iii) address purely
conceptual content.

More perceptual data inherits the cross-state shift and does not raise the
effective dimension of predictions fitted on a clustered corpus
(Eq.~\eqref{eq:effdim}); cross-state alignment into a space where a trained
decoder and a frozen generator remain valid~\cite{Spera2026seeing}\pre{} is
a promising alternative, provided the composed system is reported with S1.

\subsection{The Nature of the Cross-State Shift}
\label{subsec:gap2}

Section~\ref{subsec:famIII} posed the question as a shared code with
state-specific gain ($\mathrm{H}_0$) against an altered code
($\mathrm{H}_1$)~\cite{DijkstraBosch2019shared,Pearson2019human}.
NSD-Imagery and Deeprecon both contain matched perception and imagery trials
within subjects, so the question can be addressed without new data
(Fig.~\ref{fig:h0h1}). One would fit the perception mapping
$W^{\mathrm{perc}}$ and compare, on held-out imagery trials, a
\emph{gain-only} model $\operatorname{diag}(\mathbf{a})\,W^{\mathrm{perc}}$
with region- or voxel-wise gains against a freely refitted
$W^{\mathrm{img}}$ under matched regularisation, by cross-validated
prediction of the proxy latent, separately for early visual and high-level
ventral cortex. A gain-only win supports $\mathrm{H}_0$, and a refit win
beyond what its extra parameters buy supports $\mathrm{H}_1$. Reddy
et al.\ ran a classification-era version of this
comparison~\cite{Reddy2010reading}; we found no study repeating it for
generative decoders. A theory of achievable fidelity under such a shift would need to
reproduce the retention ratios of Fig.~\ref{fig:transfer}(b), including a
seen-to-imagined drop that, in the published values, differs by a factor of
4.4 across methods trained on the same corpus, and the two-stream dissociation
(Eq.~\eqref{eq:twostream}) suggests stream-specific bounds, tighter for
semantics than for pixels.

\subsection{Individual Phenotypes and Fine-Grained Dynamics}
\label{subsec:gap3}

Because the proxy bias depends on imagery ability, which spans a continuum
from aphantasia to hyper-vivid
imagery~\cite{Zeman2015lives,Marks1973visual,Dijkstra2017vividness},
modelling ability is a prerequisite for unbiased evaluation, yet no study in
Table~\ref{tab:performance} reports scores by ability, even where per-trial
labels exist~\cite{Kneeland2025nsdimagery}. 7\,T fMRI~\cite{Liu2025visual},
intracranial recordings~\cite{Dijkstra2018differential,Kreiman2000imagery}
and intracranial foundation models~\cite{Zhang2023brant} offer finer access
to phenotype differences and imagery's sustained dynamics.

\subsection{Real-Time Closed-Loop Use}
\label{subsec:gap4}

Nearly all work is offline, whereas communication and assistive use require
a closed loop. The amortised pass of Eq.~\eqref{eq:famIII} suits real time
better than per-trial optimisation, though diffusion
generation~\cite{Rombach2022highres} still incurs latency. Every number in
Table~\ref{tab:performance} is a multi-trial average, whereas a closed loop
has none to average, which should lower $\Lam$ and raise the prior's share
of the output; reporting single-trial performance would be a natural first
step. The one closed-loop demonstration~\cite{Cerf2010online} used a
four-way choice, and EEG-BCI work with visual imagery reports similarly few
degrees of freedom~\cite{Kosmyna2018attending,Lee2020neural,Kilmarx2024evaluating}.
Because a hallucinated reconstruction fed back to a user could mislead them,
the null-signal control is here also a safety consideration.

\section{Ethical Considerations}
\label{sec:ethics}

The literature in this section is the normative Group~D of
Sec.~\ref{subsec:methodology}, to which the study-level criteria do not
apply.

\subsection{Mental Privacy}
\label{subsec:privacy}

A perceived image existed in front of the person; an imagined one need never
have entered the world. Imagery decoding therefore concerns content that is
private in a strong sense~\cite{Schacter2007cognitive}, and without an
external record the person cannot contest what a decoder claims to have
read~\cite{Ienca2017towards}. Since a decoder is not bound by the
reality-monitoring threshold (Sec.~\ref{subsec:individual}), a result may
also concern content the person would sincerely deny having
imagined~\cite{Chang2025imageless,Cabbai2024sensory}.

\subsection{Consent and Cross-Subject Decoding}
\label{subsec:capability}

A practical safeguard against unconsented decoding has been that a decoder
needs hours of the person's own labelled data; semantic language decoding
from fMRI failed when trained on other people or when the participant
resisted~\cite{Tang2023semantic}. Family~IV reduces that
requirement~\cite{Haxby2011common,Wang2025interindividual,Kong2025generalizing,Zhang2025neuroswift}.
These results are perceptual and consented, so the concern is prospective,
but a method that reduces the data needed from a person also reduces what
that person must agree to. Similarly, every reconstruction in
Table~\ref{tab:performance} is trained on perceptual data, such as
NSD~\cite{Allen2022massive}, consented to as a perception experiment;
consent documents could state whether derived models may be applied to
internally generated content.

\subsection{Neurorights and Regulation}
\label{subsec:neurorights}

The neurorights agenda holds that existing rights may be insufficient for
technologies that access neural activity
directly~\cite{Yuste2017four,Ienca2017towards}. Chile added to its
constitution in 2021 a requirement that technological development safeguard
brain activity and the information derived from
it~\cite{Chile2021ley21383}, applied by its Supreme Court in 2023 to a
consumer EEG device~\cite{CornejoPlaza2024chilean}; UNESCO adopted a
recommendation on the ethics of neurotechnology in
2025~\cite{UNESCO2025recommendation}; and Colorado and California treat
neural data as sensitive in their privacy
statutes~\cite{Colorado2024hb1058,California2024sb1223}. Because mental
states can also be inferred from non-neural signals, whether ``neural data''
is the right regulatory target is debated~\cite{Magee2024beyond}; imagery
decoding, whose target is a mental state, sits on that boundary, which
suggests framing safeguards around consent and
purpose~\cite{Goering2021recommendations}.

\subsection{Misuse of Hallucinatory Reconstructions}
\label{subsec:forensic}

A reconstruction collapsed onto a class prior could be misread as a picture
of what a person remembered or intended. The best results in
Table~\ref{tab:performance}, 78.30\% for MIRAGE~\cite{Kneeland2026mirage}
and 83.2\% for imagined shapes~\cite{Shen2019deep}, are two-alternative,
multi-trial figures from cooperative participants imagining pre-specified
content; they do not support inference about uncooperative subjects, single
trials or unspecified content, and on this record generative imagery
reconstructions are not suitable as evidence of memory or
intent~\cite{Yuste2017four,Goering2021recommendations}. In the assessment of
unresponsive patients, cited as motivation in the benchmark
literature~\cite{Kneeland2025nsdimagery}, errors can run both ways: a fluent
image generated largely by the prior could be taken as evidence of
awareness, and a failed reconstruction, which can have many causes unrelated
to the person, as evidence of its absence.

\section{Conclusion}
\label{sec:conclusion}

Reconstructing what a person imagines differs from reconstructing what a
person sees: imagery is generated top-down, is weak in early visual cortex
and is never observed, so every decoder is trained and judged through a
proxy. Three observations follow from the published record reviewed here.
In a published-value comparison, method ordering changes between seen and
imagined images, which suggests that vision performance is not a reliable
proxy for imagery performance; whether the neural code itself changes can be
tested with existing data. Image-based scoring is confined to maintained or
recalled exemplars and compositionally constructed targets, the reviewed
corpus occupies only recall, and content generated from a concept has not
yet been tested at the image level. Within the reviewed corpus, image-scored
reconstruction has been reported only from fMRI, a pattern consistent with
the lack of image-level targets in electrophysiological imagery data.

Because a high-capacity generator can produce convincing images from little
neural information, proxy fidelity alone cannot separate read-out from prior
fill-in. Reporting the normalised neural dependence, a diagnostic quantity
proposed here, with a within-category null, alongside the other proposed
items of Table~\ref{tab:standards}, is one way to address this. In this
exploratory application, none of the ten studies reports all proposed items
in a fully checkable form; this is not a quality ranking, and the two items
that no study reports (S5 and S6) could be computed retrospectively. Because the
object is private experience, progress should be accompanied by protections
for mental privacy and by caution against treating generative
reconstructions as evidence.

\bibliographystyle{elsarticle-num}
\bibliography{cas-refs}

\clearpage
\appendix

\section{The Corpus and Its Per-Cell Evidence}
\label{app:corpus}

Table~\ref{tab:corpus} itemises the imagery studies and datasets of
Tables~\ref{tab:families}, \ref{tab:datasets} and~\ref{tab:compliance} of the main
text with the evidence behind each admissibility verdict;
\ref{app:reviews} and \ref{app:compliance} do the same for
Tables~\ref{tab:reviews} and~\ref{tab:compliance}, and Section~\ref{app:ephys}
collects the electrophysiological results. For GOD~\cite{Horikawa2017generic},
the \lthree{} verdict reflects the instruction to imagine arbitrary
exemplars; participants had first been shown exemplars of each category, and
the fifty categories are those of the perceptual test session
(Sec.~\ref{subsec:famII}).

\begin{sidewaystable}
\centering
\setlength{\tabcolsep}{3.5pt}
\renewcommand{\arraystretch}{1.0}
\begin{threeparttable}
\caption{The imagery corpus satisfying I1--I5 of
Sec.~\ref{subsec:methodology}, with the evidence for each admissibility
verdict. ``Trial'' gives the trial-level verdict and ``Src'' the memory
source, and the evidence column paraphrases the sentence or figure from
which the verdict is derived. Supporting methodological and normative
references are cited in the text but are not part of this corpus.}
\label{tab:corpus}
\scriptsize
\begin{tabularx}{\linewidth}{@{}L{3.6cm}L{1.15cm}cccY@{}}
\toprule
\textbf{Study / dataset} & \textbf{Modality} & \textbf{Family} &
\textbf{Trial} & \textbf{Src} & \textbf{Evidence for the verdict} \\
\midrule
Thirion et al.\ 2006~\cite{Thirion2006inverse} & fMRI & Ia & \sepyes & \ltwo &
Imagery of previously presented Gabor-like patterns, cued without
re-presentation \\
Stokes et al.\ 2009~\cite{Stokes2009topdown} & fMRI & II & \sepyes & \ltwo &
Letter cue (X or O) initiates imagery of a shape learned before the run; no
exemplar in the trial \\
Reddy et al.\ 2010~\cite{Reddy2010reading} & fMRI & II & \sepyes & \ltwo &
Auditory name cues; participants close their eyes before imagery blocks and
open them before perception blocks, verified by eye tracking \\
Naselaris et al.\ 2015~\cite{Naselaris2015voxelwise} & fMRI 7T & Ia & \sepyes & \ltwo &
Imagery runs cue mental images of five artworks memorised in advance while
fixating a blank grey screen \\
Lee \& Kuhl 2016~\cite{LeeKuhl2016reconstructing} & fMRI & Ib & \sepyes & \ltwo &
Faces retrieved from paired-associate learning completed before scanning \\
Horikawa \& Kamitani 2017~\cite{Horikawa2017generic} & fMRI & II & \sepyes & \lthree &
Cue is an array of 50 object names; participants imagine arbitrary
exemplars of the red-marked category with eyes closed for 15\,s \\
Shen et al.\ 2019~\cite{Shen2019deep} & fMRI & Ib & \sepyes & \ltwo &
Word cues for 25 images memorised before the experiment; word--image pairs
viewable only during inter-run rest \\
Senden et al.\ 2019~\cite{Senden2019imagery} & fMRI & Ia & \sepyes & \ltwo &
Imagery of letter shapes learned before the imagery runs \\
Shatek et al.\ 2019~\cite{Shatek2019decoding} & EEG & II & \sepno & \lone &
Four images shown in sequence within each trial, then a numeric retro-cue
names which to imagine \\
\bottomrule
\end{tabularx}
\end{threeparttable}
\end{sidewaystable}

\begin{sidewaystable}
\centering
\setlength{\tabcolsep}{3.5pt}
\renewcommand{\arraystretch}{1.0}
\begin{threeparttable}
\raggedright\small\textbf{Table~\ref{tab:corpus} (continued).}\par\medskip
\scriptsize
\begin{tabularx}{\linewidth}{@{}L{3.6cm}L{1.15cm}cccY@{}}
\toprule
\textbf{Study / dataset} & \textbf{Modality} & \textbf{Family} &
\textbf{Trial} & \textbf{Src} & \textbf{Evidence for the verdict} \\
\midrule
Xie et al.\ 2020~\cite{Xie2020alpha} & EEG & II & \sepyes & \ltwo &
Imagery cued symbolically for objects learned before the imagery blocks \\
Chang et al.\ 2025~\cite{Chang2025imageless} & fMRI & II & \sepyes & \ltwo &
Lateralised imagery of learned patterns on cue; aphantasia and control
groups \\
Stecher \& Kaiser 2024~\cite{StecherKaiser2024alpha} & EEG & II & \sepyes & \lthree &
Three-sentence written scene descriptions; the matching photographs are
shown only afterwards, in a separate perception task \\
Gao et al.\ 2026~\cite{Gao2026eegvi} & EEG & --- & \sepno & \lone &
Each 17\,s trial: 3\,s fixation, 4\,s picture, mask, 4\,s imagery, 4\,s rest \\
Kwon et al.\ 2025~\cite{Kwon2025functional} & EEG & II & \sepno & \lone &
Each 17\,s trial: 2\,s rest, 5\,s action video, 5\,s rest (stated purpose:
afterimage removal), 5\,s imagery \\
Shimizu \& Srinivasan 2022~\cite{Shimizu2022improving} & EEG & III & \sepno & \lone &
500\,ms image, 500\,ms noise mask, then 10\,s imagery, within each trial \\
Kneeland et al.\ 2025~\cite{Kneeland2025nsdimagery} & fMRI 7T & III & \sepyes & \mbox{\ltwo/\lthree} &
Only a single-letter cue appears during imagery; all 18 cue--stimulus pairs
memorised before scanning. Conceptual stimuli are words with no image \\
Kneeland et al.\ 2026~\cite{Kneeland2026mirage} & fMRI 7T & III & \sepyes & \mbox{\ltwo/\lthree} &
Same benchmark and protocol as above \\
Koide-Majima et al.\ 2024~\cite{KoideMajima2024mental} & fMRI & Ib & \sepyes & \ltwo &
Reanalysis of the Deeprecon imagery data; cueing as
in~\cite{Shen2019deep} \\
Dijkstra et al.\ 2018~\cite{Dijkstra2018differential} & MEG & II & \sepno & \lone &
Retro-cue task: a face and a house are shown for 0.8\,s each, then a noise
mask and a 0.5\,s cue naming which of the two to imagine for 3.5\,s \\
Kreiman et al.\ 2000~\cite{Kreiman2000imagery} & iEEG & II & \sepyes & \ltwo &
Patients imagine pictures viewed earlier in the same session; no exemplar
during the imagery period \\
Harrison \& Tong 2009~\cite{HarrisonTong2009decoding} & fMRI & II & \sepno & \lone &
Delay-period decoding of a just-seen grating (boundary case by design) \\
\bottomrule
\end{tabularx}
\begin{tablenotes}[flushleft]
\scriptsize
\item Three verdicts of Tables~\ref{tab:families} and~\ref{tab:datasets}
are justified in this note rather than in rows. The two EEG datasets
of~\cite{Wilson2023semantic,Rybar2025simultaneous}, examined trial by trial
from their data descriptors:
\cite{Wilson2023semantic} is \adm{\sepno}{\lone}, since each concept is
perceived and then imagined within the trial, the auditory condition asking
for imagery in the voice of the speaker presented;
\cite{Rybar2025simultaneous} is \adm{\sepno}{\lthree}, since each trial
shows a 0.6\,s picture and a mask before the imagery period but participants
are told to visualise their own representation of the concept rather than
the picture shown (Sec.~\ref{subsec:terminology}). The
preprint~\cite{StecherKaiser2026neuro}\pre{} is \adm{\sepyes}{\lthree},
since imagery is generated from short text prompts and no image is shown.
\end{tablenotes}
\end{threeparttable}
\end{sidewaystable}

\section{Evidence for Table~\ref{tab:reviews}}
\label{app:reviews}

Each verdict in Table~\ref{tab:reviews} rests on the presence or absence of a
dedicated section in the review concerned.
\cite{Kosslyn2001neural} is organised by neural substrate and contains no
section on decoding method, evaluation or governance.
\cite{Pearson2015mental} is organised by mechanism and clinical
application, with decoding appearing as evidence within the mechanism
sections.
\cite{Pearson2019human} contains a substantial treatment of objective
measurement, which we score as partial on the target-free column because it
concerns measuring imagery rather than evaluating reconstructions, and no
section on method taxonomy, standards or ethics.
\cite{DijkstraBosch2019shared} is organised by the dimensions of
perception--imagery overlap and closes with outstanding questions rather
than standards.
\cite{Vigotsky2024mental} is a validity critique with a sustained treatment
of what decoders do and do not establish, which we score \hasY{} on
target-free evaluation, while whether its cautions amount to an auditable
protocol is a judgement we record as partial.
\cite{Guo2025fmri2image} organises fMRI-to-image work as a three-stage
pipeline, a unified taxonomy, and treats perception throughout.
\cite{Tanaka2025visual} is organised by latent representation,
compositionality and modularity and raises zero-shot and spurious
prediction for seen images, both scored partial for the reasons in
note~(d) of Table~\ref{tab:reviews}.

\section{Evidence for Table~\ref{tab:compliance}}
\label{app:compliance}

\emph{S1.} \cite{Xie2020alpha} tests held-out pairwise accuracy against
chance with sign-permutation cluster statistics; for label outputs this is
the mismatch null of Eq.~\eqref{eq:lambda} (Sec.~\ref{app:nullimpl}), hence
\cyes{}. \cite{Kneeland2026mirage} constructs human-experiment distractors
by decoding randomly selected brain activity patterns, so that target and
distractor are equally shaped by the prior, and separately reports
reconstruction quality across guidance strengths; this estimates
$\Lam_{\mathrm{between}}$ but not $\Lam_{\mathrm{within}}$, hence \cpart{}.
\cite{Naselaris2015voxelwise}, \cite{Reddy2010reading} and
\cite{Horikawa2017generic} report permutation tests against chance but do
not make their relation to a test-time mismatch through the deployed
pipeline checkable, hence \cpart{}.
\emph{S2.} \cite{Horikawa2017generic} evaluates categories absent from
decoder training by construction; \cite{Shen2019deep} and
\cite{KoideMajima2024mental} evaluate artificial shapes with decoders
trained only on natural images; \cite{Naselaris2015voxelwise} identifies
against large sets of previously unseen photographs. The NSD-Imagery
entries are \cpart{}, since their test stimuli are held out from training
but five of six complex stimuli come from NSD shared1000 and the corpus is
cluster-structured (Sec.~\ref{subsec:datasetstable}).
\emph{S3.} Studies that tabulate both a low-level and a semantic
metric~\cite{Shen2019deep,Kneeland2025nsdimagery,Kneeland2026mirage} are
scored \cpart{}: S3 also requires the number of alternatives, the
distractor-sampling rule and the proportion of same-category distractors
(Sec.~\ref{subsec:guidelines}), and these cells are scored from the
published descriptions of the distractor protocol rather than a complete
re-check against that requirement. \cpart{} is also given where a single
identification metric accompanies a qualitative assessment, and \cna{} where
the output is a label.
\emph{S4.} \cite{Shen2019deep} collects a five-point vividness rating on
every imagery trial and reports its distribution but does not relate it to
decoding score, and \cite{Horikawa2017generic} likewise collects a vividness
judgement, so both are \cpart{}. \cite{Kneeland2025nsdimagery} collects a
binary vividness response per trial, but because no analysis against it is
reported we score \cno{}.
\emph{S5, S6.} Neither is reported in the main paper or supplement of any of
the ten; both require only the generator, the stimulus corpus and
predictions already computed, so they could be supplied retrospectively.
\emph{S7.} \cyes{} where the number of samples per trial and the selection
rule are stated and best/median/worst cases are
shown~\cite{Kneeland2025nsdimagery,Kneeland2026mirage}, or where the
reconstruction is deterministic given the data~\cite{Naselaris2015voxelwise};
\cpart{} where multiple runs are described but the figure-selection rule is
not; \cno{} where neither is stated.
\emph{S8.} \cno{} for \cite{Kwon2025functional}, where fivefold
Gaussian-noise augmentation precedes a 60/20/20 split. Elsewhere \cyes{}:
the fMRI studies split by run or session, \cite{Shatek2019decoding}
cross-validates at the trial level and \cite{Xie2020alpha} on pseudo-trials
averaged from disjoint sets of trials, both within a randomised design.
\emph{S9.} \cyes{} wherever the trial timeline is reported in enough detail
to derive the verdict in Table~\ref{tab:corpus}; \cno{} for
\cite{Kwon2025functional}, which reports the timeline but does not address
what the in-trial cue video implies for maintenance.
\emph{S10.} \cyes{} where eye state is specified and
monitored~\cite{Reddy2010reading,Horikawa2017generic}; \cpart{} for
\cite{Xie2020alpha}, where eye state is specified but the alpha-band
implication is not analysed; \cno{} for \cite{Kwon2025functional}, which
reports eyes open during imagery without a corresponding statement for the
cue period or an artefact analysis; \cna{} for fMRI studies.

\section{Electrophysiological Imagery Results}
\label{app:ephys}

Table~\ref{tab:ephys} gathers the electrophysiological imagery studies of
Sections~\ref{subsec:famII} and~\ref{subsec:eegmeg}, with the
class count, chance level and validation unit that
Sec.~\ref{subsec:metrics} requires before an accuracy can be interpreted.
Qualitative entries restate findings given in the main text, and
admissibility follows Table~\ref{tab:corpus}. Cells are taken from the
original reports and data descriptors; where an item could not be located
there, the cell says so.

\begin{sidewaystable}
\centering
\setlength{\tabcolsep}{3.5pt}
\renewcommand{\arraystretch}{1.0}
\begin{threeparttable}
\caption{Electrophysiological imagery decoding at a glance. ``Validation
unit'' is the level at which data were split or permuted (standard~S8), and
``Adm.'' the admissibility verdict of Sec.~\ref{subsec:terminology}. No row
reports image-level reconstruction scored against an image-level target.}
\label{tab:ephys}
\scriptsize
\begin{tabularx}{\linewidth}{@{}L{3.0cm}L{1.35cm}L{3.0cm}L{1.1cm}YL{2.6cm}L{2.3cm}L{1.15cm}@{}}
\toprule
\textbf{Study} & \textbf{Mod.} & \textbf{Imagined content (classes $k$)} &
\textbf{Chance} & \textbf{Reported result} & \textbf{Validation unit} &
\textbf{Eye state} & \textbf{Adm.} \\
\midrule
\famrow{8}{svGrey}{\textit{EEG content decoding}}
Xie et al.\ 2020~\cite{Xie2020alpha} & EEG, 64 ch &
Twelve everyday objects, cued by spoken words after a training phase (12,
pairwise) & 50\% &
Above chance in the alpha band; cross-classifies with late stages of
perception & Pseudo-trials averaged from disjoint trials,
leave-one-pseudo-trial-out; sign-permutation cluster test &
Fixation cross in both tasks; alpha implication not analysed &
\adm{\sepyes}{\ltwo} \\
\addlinespace[1pt]
Stecher \& Kaiser 2024~\cite{StecherKaiser2024alpha} & EEG, 64 ch &
Sixteen scenes from three-sentence descriptions (16, mean pairwise) & 50\% &
Individual scenes and graded scene properties decoded from alpha power &
Leave-one-trial-per-scene-out; TFCE with 10{,}000 sign permutations &
Open, fixating a central dot & \adm{\sepyes}{\lthree} \\
\addlinespace[1pt]
Shatek et al.\ 2019~\cite{Shatek2019decoding} & EEG, 64 ch &
One of four in-trial images (two categories, two exemplars each),
retro-cued by serial position (2) & 50\% &
No above-chance decoding during imagery at any time point; perception
decoded robustly & Leave-two-trials-out, one per category; per-participant
label permutation (1{,}000) &
Open; fixation cross, imagery projected into an on-screen frame &
\adm{\sepno}{\lone} \\
\addlinespace[1pt]
Kwon et al.\ 2025~\cite{Kwon2025functional} & EEG, 64 ch &
Upper-limb actions after an in-trial cue video (4) & 25\% &
72.3\% mean accuracy; leave-one-subject-out also reported &
Augmented trial (augmentation before a 60/20/20 split) & Eyes open in
imagery; cue period not stated & \adm{\sepno}{\lone} \\
\addlinespace[1pt]
Shimizu \& Srinivasan 2022~\cite{Shimizu2022improving} & EEG &
Images shown for 500\,ms before imagery (40) & 2.5\% &
GAN images from imagery-period EEG classified slightly above chance by a
ResNet & 500\,ms segment within an episode & Open (instructed) &
\adm{\sepno}{\lone} \\
\addlinespace[2pt]
\bottomrule
\end{tabularx}
\end{threeparttable}
\end{sidewaystable}

\begin{sidewaystable}
\centering
\setlength{\tabcolsep}{3.5pt}
\renewcommand{\arraystretch}{1.0}
\begin{threeparttable}
\raggedright\small\textbf{Table~\ref{tab:ephys} (continued).}\par\medskip
\scriptsize
\begin{tabularx}{\linewidth}{@{}L{3.0cm}L{1.35cm}L{3.0cm}L{1.1cm}YL{2.6cm}L{2.3cm}L{1.15cm}@{}}
\toprule
\textbf{Study} & \textbf{Mod.} & \textbf{Imagined content (classes $k$)} &
\textbf{Chance} & \textbf{Reported result} & \textbf{Validation unit} &
\textbf{Eye state} & \textbf{Adm.} \\
\midrule
\famrow{8}{svGrey}{\textit{EEG datasets and BCI control}}
Gao et al.\ 2026~\cite{Gao2026eegvi} & EEG, 32 ch &
Ten pictures in three categories, each shown for 4\,s in-trial; classified
within category (3, 3, 4) & 33\%, 33\%, 25\% &
Dataset, 22 participants in two sessions; VVIQ administered; EEGNet 62.0\%
(objects) to about 75\% (animals, figures), CSP\,+\,KNN 20--26 points lower &
Random within-subject 80/20 split of trials, per session &
Not stated in the sources examined & \adm{\sepno}{\lone} \\
\addlinespace[1pt]
Wilson et al.\ 2023~\cite{Wilson2023semantic} & EEG, 124 ch &
Three concepts (penguin, guitar, flower) in pictorial, orthographic and
auditory form, each imagined after being perceived in the trial (3) &
50\% (task) &
Dataset, 12 participants; technical validation decodes imagination versus
perception (75\% for the visual modalities), not content &
Stratified 5-fold cross-validation over trials, 50 repetitions &
Not specified; room darkened & \adm{\sepno}{\lone} \\
\addlinespace[1pt]
Rybar et al.\ 2025~\cite{Rybar2025simultaneous} & EEG, 64 ch; fNIRS &
Animals versus tools (18 concepts each), visualised after a 0.6\,s
in-trial picture and mask, alongside silent naming, auditory and tactile
imagery (2) & 50\% &
Dataset, 12 participants (EEG and fNIRS) plus 7 (EEG, auditory imagery
only); technical validation reports evoked responses, not decoding &
None reported in the descriptor &
Open; eye movements to be minimised & \adm{\sepno}{\lthree} \\
Kosmyna et al.\ 2018~\cite{Kosmyna2018attending} & EEG &
Flower versus hammer, each imagined after observing the same image in the
trial (2) & 50\% &
The two imagined images not separated above chance (about 52\%); imagery
versus observation 71\%, imagery versus rest 77\% &
10-fold chronological cross-validation, 5-trial margin &
Not stated in the sources examined & \adm{\sepno}{\lone} \\
\addlinespace[1pt]
Lee et al.\ 2020~\cite{Lee2020neural} & EEG &
Twelve words or phrases plus rest, imagined as visual scenes after an
in-trial visual cue (13) & 7.7\% &
$40.1\pm4.2\%$ for visual imagery, with every participant above the
confidence level & Not stated in the sources examined &
Not stated in the sources examined & \adm{\sepno}{\lone} \\
\addlinespace[1pt]
\bottomrule
\end{tabularx}
\end{threeparttable}
\end{sidewaystable}

\begin{sidewaystable}
\centering
\setlength{\tabcolsep}{3.5pt}
\renewcommand{\arraystretch}{1.0}
\begin{threeparttable}
\raggedright\small\textbf{Table~\ref{tab:ephys} (continued).}\par\medskip
\scriptsize
\begin{tabularx}{\linewidth}{@{}L{3.0cm}L{1.35cm}L{3.0cm}L{1.1cm}YL{2.6cm}L{2.3cm}L{1.15cm}@{}}
\toprule
\textbf{Study} & \textbf{Mod.} & \textbf{Imagined content (classes $k$)} &
\textbf{Chance} & \textbf{Reported result} & \textbf{Validation unit} &
\textbf{Eye state} & \textbf{Adm.} \\
\midrule
\famrow{8}{svGrey}{\textit{EEG datasets and BCI control (continued)}}
Kilmarx et al.\ 2024~\cite{Kilmarx2024evaluating} & EEG &
Four image categories; short-term imagery after the image, long-term
imagery after an auditory cue (4) & 25\% &
Observation 42.1\%, short-term imagery 30.1\%, long-term imagery 26.7\%,
against a significance threshold of 29.6\% &
Leave-one-block-out (5-fold) & Not stated in the sources examined &
\adm{\sepno}{\lone}\newline\adm{\sepyes}{\ltwo} \\
\addlinespace[2pt]
\famrow{8}{svGrey}{\textit{MEG and intracranial}}
Dijkstra et al.\ 2018~\cite{Dijkstra2018differential} & MEG, 275 ch &
Face versus house (eight exemplars each), retro-cued after both were shown
in the trial (2) & 50\% &
Category decodable from 540\,ms after the cue; imagery representations
later, more sustained and less temporally specific than perceptual ones &
5-fold cross-validation over trials; cluster-based permutation (1{,}000) &
Fixation bull's-eye; eye tracking, eye position not predictive of category
during imagery & \adm{\sepno}{\lone} \\
\addlinespace[1pt]
Kreiman et al.\ 2000~\cite{Kreiman2000imagery} & iEEG, single units &
Pictures viewed earlier in the session & \sepna &
Picture-selective medial-temporal neurons reinstated during imagery &
\sepna & Closed during imagery & \adm{\sepyes}{\ltwo} \\
\addlinespace[1pt]
Cerf et al.\ 2010~\cite{Cerf2010online} & iEEG, single units &
Fading a two-image hybrid toward one of four pre-identified images &
\sepna &
Volitional, closed-loop control of pre-identified neurons; target reached on
596 of 864 trials (69.0\%) & Online &
Open, viewing the hybrid image & \sepno{} (stimulus on screen) \\
\bottomrule
\end{tabularx}
\begin{tablenotes}[flushleft]
\scriptsize
\item \sepna{} = not applicable; for~\cite{Cerf2010online} no nominal chance
level applies, since a trial ends when the target is reached or on time-out.
Entries without a number restate findings reported in the main text, and no
value has been estimated. ``Not stated in
the sources examined'' means the item was not found in the report or data
descriptor consulted. The flower--hammer imagery figure
for~\cite{Kosmyna2018attending} is as summarised
in~\cite{Kilmarx2024evaluating}; for~\cite{Kilmarx2024evaluating} the
short-term condition shows the image inside the trial and the long-term
condition recalls it from memory on an auditory cue.
\end{tablenotes}
\end{threeparttable}
\end{sidewaystable}

\FloatBarrier

\section{Inflation of Pairwise Identification under Categorical Structure}
\label{app:pairwise}

This section derives Eq.~\eqref{eq:pairwise-inflation}
together with its finite-sample form, which quantify how far a pairwise
identification score can exceed chance when the decoder recovers category
but no within-category detail~\cite{Shirakawa2025spurious}.

Let $\mathbf{y}$ be a test sample, $\est{\mathbf{y}}$ the prediction for
it, and $\mathbf{y}^{-}$ a candidate drawn from the remaining test samples.
Define the indicator
\begin{equation}
\label{eq:S-indicator}
S(\est{\mathbf{y}},\mathbf{y},\mathbf{y}^{-}) =
\begin{cases}
1, & \operatorname{sim}(\est{\mathbf{y}},\mathbf{y}) \ge \operatorname{sim}(\est{\mathbf{y}},\mathbf{y}^{-}),\\
0, & \text{otherwise},
\end{cases}
\end{equation}
for an arbitrary similarity $\operatorname{sim}(\cdot,\cdot)$. The pairwise
identification accuracy over $n$ test samples is
\begin{equation}
\label{eq:acc-def}
\mathrm{Acc} = \frac{1}{n(n-1)}\sum_{i=1}^{n}\sum_{j\neq i}
S\big(\est{\mathbf{y}}^{(i)},\mathbf{y}^{(i)},\mathbf{y}^{(j)}\big).
\end{equation}
Now model a decoder that recovers semantic category but not appearance.
Partition the triplets into the set $Z$ in which test and candidate belong
to the same category and its complement $\bar{Z}$, and assume
\begin{equation}
\label{eq:model-assumption}
\mathbb{E}_{Z}\big[S\big] = 0.5,
\qquad
\mathbb{E}_{\bar{Z}}\big[S\big] = q, \quad q\in[0.5,1].
\end{equation}
The first equation says that two samples of the same category are
indistinguishable to this decoder, the second that samples of different
categories are discriminated with probability $q$.

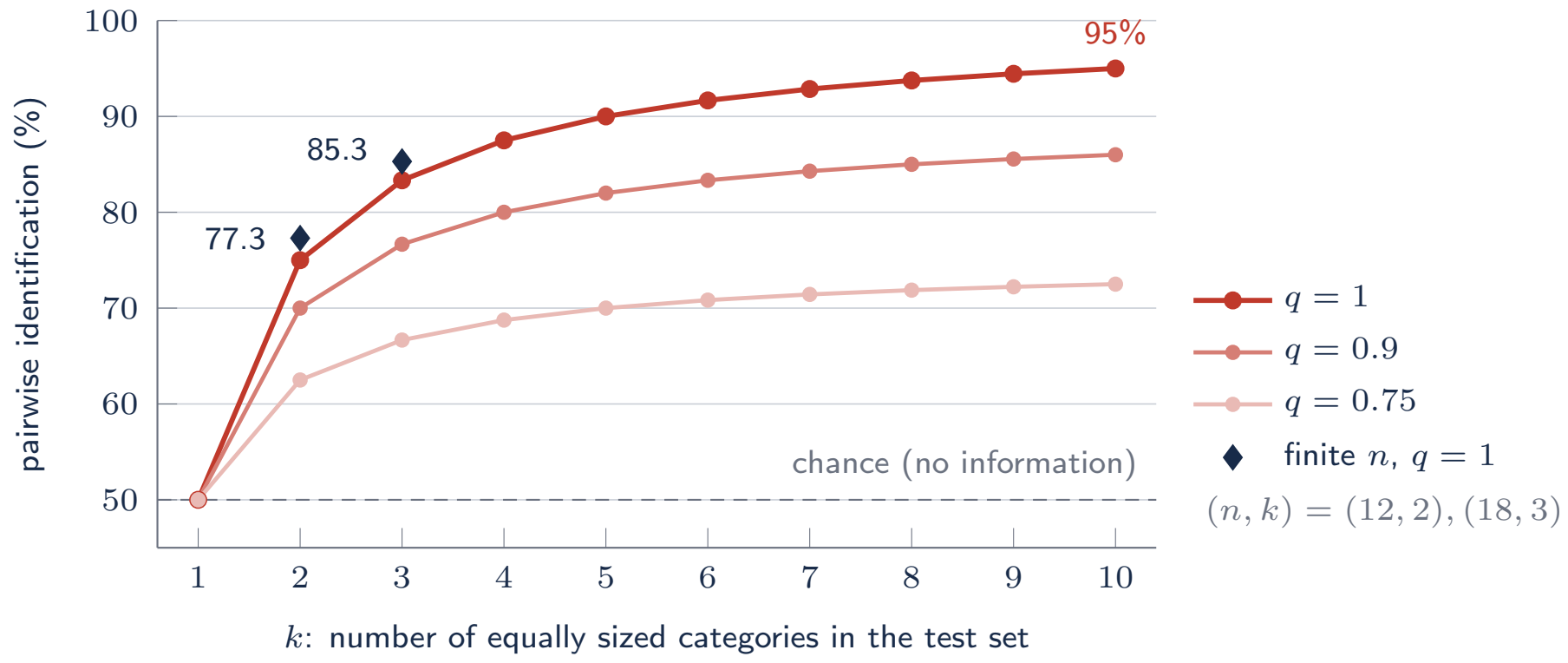
\begin{figure*}[!htbp]
\centering
\begin{tikzpicture}[sv]
\begin{axis}[
   width=9.2cm, height=5.6cm,
   xmin=0.6, xmax=10.4, xtick={1,...,10},
   ymin=45, ymax=100, ytick={50,60,...,100},
   xlabel={$k$: number of equally sized categories in the test set},
   ylabel={pairwise identification (\%)},
   axis line style={svGrey}, tick style={svGrey},
   tick label style={font=\svsmall, text=svInk},
   label style={font=\svsmall, text=svInk},
   ymajorgrids, grid style={svRule, line width=.3pt},
   axis x line*=bottom, axis y line*=left, clip=false,
   legend style={font=\svsmall, draw=none, fill=none, at={(1.02,0.0)}, anchor=south west,
                 legend cell align=left, row sep=1pt},
]
  \addplot[svGrey, svdash, domain=0.6:10.4, forget plot] {50};
  \node[svnote, anchor=south east] at (axis cs:10.4,50) {chance (no information)};
  \addplot[svRed, line width=1.1pt, mark=*, mark size=1.4pt, samples at={1,...,10}]
     {100*(0.5/x + (1-1/x)*1.00)};
  \addlegendentry{$q=1$}
  \addplot[svRed!65, line width=.9pt, mark=*, mark size=1.2pt, samples at={1,...,10}]
     {100*(0.5/x + (1-1/x)*0.90)};
  \addlegendentry{$q=0.9$}
  \addplot[svRed!35, line width=.9pt, mark=*, mark size=1.2pt, samples at={1,...,10}]
     {100*(0.5/x + (1-1/x)*0.75)};
  \addlegendentry{$q=0.75$}
  \addplot[only marks, mark=diamond*, mark size=2.6pt, svInk] coordinates {(2,77.3) (3,85.3)};
  \addlegendentry{finite $n$, $q=1$}
  \addlegendimage{empty legend}\addlegendentry{\hspace{-.6cm}\textcolor{svGrey}{$(n,k)=(12,2),(18,3)$}}
  \node[font=\svsmall, anchor=east, text=svInk] at (axis cs:1.85,77.3) {77.3};
  \node[font=\svsmall, anchor=east, text=svInk] at (axis cs:2.85,86.6) {85.3};
  \node[font=\svsmall, anchor=south, text=svRed] at (axis cs:10,95.5) {95\%};
\end{axis}
\end{tikzpicture}
\caption{Pairwise identification reached by a decoder with no within-category
information, Eq.~\eqref{eq:pairwise-inflation}, as a function of the number
$k$ of equally sized categories, for three values of the between-category
discrimination~$q$. Diamonds mark the finite-sample values derived in
\ref{app:pairwise} for the NSD-Imagery designs ($n=18$, $k=3$ and $n=12$,
$k=2$) at $q=1$.}
\label{fig:inflation}
\end{figure*}

Suppose the test set comprises $k$ equally sized categories of $n/k$
samples. For a uniformly drawn candidate $\mathbf{y}^{-}\neq\mathbf{y}$, the
probability that it falls in the same category as $\mathbf{y}$ is
$\pi_{n,k} = (n/k-1)/(n-1)$, so taking expectations in
Eq.~\eqref{eq:acc-def} and substituting Eq.~\eqref{eq:model-assumption}
gives the exact finite-sample accuracy
\[
\mathrm{Acc}_{n,k} \;=\; 0.5\,\pi_{n,k} \;+\; q\,\bigl(1-\pi_{n,k}\bigr).
\]
As $n$ grows, $\pi_{n,k}\to1/k$ and
\begin{equation}
\label{eq:acc-derived}
\mathrm{Acc}
= \frac{1}{k}\cdot 0.5 + \Big(1-\frac{1}{k}\Big)\, q ,
\end{equation}
which is Eq.~\eqref{eq:pairwise-inflation}
(Fig.~\ref{fig:inflation}).

Three readings follow. Setting $q=1$, that is perfect between-category
discrimination with zero within-category information, gives
$\mathrm{Acc}=1-1/(2k)$ in the limit: 75\% for two categories, 90\% for five
and 95\% for ten. A reported pairwise accuracy therefore carries little
information about within-category fidelity unless $k$ is small and known.
Conversely, an observed accuracy on a test set of $n$ samples in $k$
categories is consistent with $q=(\mathrm{Acc}-0.5\,\pi_{n,k})/(1-\pi_{n,k})$
and no within-category information at all. Finally, the small-sample
correction is not negligible at the sizes imagery benchmarks use. For
NSD-Imagery's eighteen stimuli in three sets of six ($n=18$, $k=3$),
$\pi=5/17\approx0.29$, so $q=1$ yields $\mathrm{Acc}\approx85.3\%$ rather
than the asymptotic 83.3\%; for the twelve simple and complex stimuli scored
in Panel~B of Table~\ref{tab:performance} ($n=12$, $k=2$), $\pi=5/11\approx0.45$
and $\mathrm{Acc}\approx77.3\%$ rather than 75\%. These figures apply when
distractors are drawn across sets, which should be checked against the
distractor rule of each benchmark~\cite{Kneeland2025nsdimagery,Kneeland2026mirage};
when distractors are restricted to a set, the relevant $k$ is the number of
semantic categories within it, which neither benchmark defines. The NSD
corpus from which the complex stimuli are drawn also resolves into roughly
forty semantic clusters~\cite{Shirakawa2025spurious}. This is why the
distractor rule must be reported with S3 (Sec.~\ref{subsec:guidelines}), and
why hard within-category distractors are preferable wherever the stimulus
set permits them.

\FloatBarrier

\section{Notes Supporting the Main Text}
\label{app:notes}

\subsection{Structural and Empirical Empty Cells of the Admissibility Criterion}
\label{app:emptycells}

The two dimensions of the admissibility criterion
(Sec.~\ref{subsec:terminology}) are not fully independent. Because \lone{} is
defined by presentation inside the trial, the cell
\sepyes\,\textperiodcentered\,\lone{} is empty by definition, not for want
of studies. The cells \sepno\,\textperiodcentered\,\ltwo{} and
\sepno\,\textperiodcentered\,\lthree{} are possible in principle, by
re-presenting a pre-learned exemplar inside the trial or by showing an
exemplar and asking the participant to imagine their own, as the
visual-imagery task of the EEG--fNIRS dataset
of~\cite{Rybar2025simultaneous} does; they are merely unoccupied in the
corpus. Where an exemplar is shown inside the trial, the memory source is
therefore \lone{} unless the instruction directs imagery away from the
exemplar shown, and the verdicts of Tables~\ref{tab:families}
and~\ref{tab:datasets} and of Fig.~\ref{fig:admissibility} follow this rule.

\subsection{Implementation of the Null-Signal Control}
\label{app:nullimpl}

The normalised neural dependence $\tilde{\Lam}$ of Eq.~\eqref{eq:lambda} is
defined when its denominator is positive. It is not confined to $[0,1]$: it
exceeds 1 when the null scores below chance, which flags a bias in the
operator itself of the kind displayed by the below-chance CLIP scores of
Table~\ref{tab:performance}, Panel~B, and it is negative when real input
scores below the null; it is therefore reported unclipped. Test-time
mismatch may also use phase-scrambled activity in place of another trial's
activity.

\emph{Label outputs.} For label outputs ($\gen=\mathrm{id}$) a mismatched
trial is scored against the label of the trial it replaces, so its expected
score is $\mathcal{E}_{\mathrm{chance}}$ by construction; $\Lam$ then
equals the above-chance margin, $\tilde{\Lam}=1$, and the control reduces to
a valid permutation test of held-out accuracy.

\emph{Permutation unit and cost.} Permutations are drawn at the level of
the imagery episode (S8), and the cost of the null re-runs is bounded by
re-running only the stages downstream of the substituted input with cached
seeds.

\emph{Cue leakage.} One leak survives every mismatch null: where a
stimulus-specific cue remains on screen during imagery, as the
single-letter cues of NSD-Imagery do, an imagery-trained decoder can obtain
$\Lam_{\mathrm{within}}>0$ from the response to the cue. Such decoders
should add a control decoded from a cue-only window or under a reassigned
cue--stimulus mapping.

\subsection{The Two Design Axes}
\label{app:designaxes}

\begin{figure*}[!htbp]
\centering
\begingroup
\newcommand{\chip}[2]{\tikz[baseline=-.65ex]\node[svtag,fill=#1,font=\sffamily\tiny\bfseries]{#2};}
\begin{tikzpicture}[sv, x=1cm, y=1cm,
   fam/.style={rounded corners=3pt, draw=#1, fill=#1!8, line width=.9pt},
   famlbl/.style={font=\svfont\bfseries, text=#1!85!black, anchor=north west}]
  \def\rh{2.0}     
  \def\cw{4.8}    
  \def\xo{3.75}    
  \node[font=\svsmall\bfseries, text=svGrey] at (\xo+\cw,3*\rh+0.95) {direction of inference};
  \node[font=\svfont\bfseries] at (\xo+0.5*\cw,3*\rh+0.45) {encode-and-invert};
  \node[svnote] at (\xo+0.5*\cw,3*\rh+0.12) {fit $g_\phi$ on perception, search image space};
  \node[font=\svfont\bfseries] at (\xo+1.5*\cw,3*\rh+0.45) {decode};
  \node[svnote] at (\xo+1.5*\cw,3*\rh+0.12) {map the response to a latent with $h_\psi$};
  \node[font=\svsmall\bfseries, text=svGrey] at (2.68,3*\rh+0.45) {kind of generator};
  \foreach \r/\t/\s in {2/{none}/{$G=\mathrm{id}$},
                        1/{per-trial\\iterative}/{optimise image\\[-2pt] or latent per trial},
                        0/{frozen,\\amortised}/{pretrained,\\[-2pt] one pass}}{
    \node[font=\svfont\bfseries, align=center, anchor=south] at (2.68,\r*\rh+0.45*\rh) {\t};
    \node[svnote, anchor=north, align=center] at (2.68,\r*\rh+0.40*\rh) {\s};
  }
  \foreach \r in {0,1,2} \foreach \c in {0,1}
     \draw[svRule] (\xo+\c*\cw,\r*\rh) rectangle ++(\cw,\rh);
  \foreach \r in {0,2}{
     \fill[pattern=north east lines, pattern color=svRule] (\xo+0.02,\r*\rh+0.02) rectangle ++(\cw-0.04,\rh-0.04);
     \node[svnote, fill=white, inner sep=2pt] at (\xo+0.5*\cw,\r*\rh+0.5*\rh) {no family};
  }
  \draw[fam=svFamI] (\xo+0.12,1*\rh+0.12) rectangle ++(\cw-0.24,\rh-0.24);
  \node[famlbl=svFamI] at (\xo+0.2,2*\rh-0.18) {Ia \;encoding-model inversion};
  \node[font=\svsmall, align=left, anchor=north west, text width=\cw cm-0.5cm] at (\xo+0.2,2*\rh-0.62)
     {candidate search or pixel optimisation, Eqs.~\eqref{eq:famI-infer}--\eqref{eq:famIa};
      a high score needs genuine low-level information};
  \node[anchor=south east] at (\xo+\cw-0.18,1*\rh+0.18) {\chip{svfMRI}{fMRI}};
  \draw[fam=svFamIb] (\xo+\cw+0.12,1*\rh+0.12) rectangle ++(\cw-0.24,\rh-0.24);
  \node[famlbl=svFamIb] at (\xo+\cw+0.2,2*\rh-0.18) {Ib \;decoded-feature inversion};
  \node[font=\svsmall, align=left, anchor=north west, text width=\cw cm-0.5cm] at (\xo+\cw+0.2,2*\rh-0.62)
     {feature decoder + image prior, Eq.~\eqref{eq:famIb};
      admits the prior--evidence tension};
  \node[anchor=south east] at (\xo+2*\cw-0.18,1*\rh+0.18) {\chip{svfMRI}{fMRI}};
  \draw[fam=svFamII] (\xo+\cw+0.12,2*\rh+0.12) rectangle ++(\cw-0.24,\rh-0.24);
  \node[famlbl=svFamII] at (\xo+\cw+0.2,3*\rh-0.18) {II \;discriminative / RSA};
  \node[font=\svsmall, align=left, anchor=north west, text width=\cw cm-0.5cm] at (\xo+\cw+0.2,3*\rh-0.62)
     {label or embedding, Eqs.~\eqref{eq:famII}, \eqref{eq:rsa};
      ceiling set by the candidate set};
  \node[anchor=south east] at (\xo+2*\cw-0.18,2*\rh+0.18)
     {\chip{svfMRI}{fMRI}\,\chip{svEEG}{EEG}\,\chip{svMEG}{MEG}\,\chip{sviEEG}{iEEG}};
  \draw[fam=svFamIII] (\xo+\cw+0.12,0.12) rectangle ++(\cw-0.24,\rh-0.24);
  \node[famlbl=svFamIII] at (\xo+\cw+0.2,\rh-0.18) {III \;generative cross-decoding};
  \node[font=\svsmall, align=left, anchor=north west, text width=\cw cm-0.5cm] at (\xo+\cw+0.2,\rh-0.62)
     {frozen diffusion / AR prior, Eq.~\eqref{eq:famIII};
      spurious-reconstruction risk peaks};
  \node[anchor=south east] at (\xo+2*\cw-0.18,0.18) {\chip{svfMRI}{fMRI}};
  \fill[svFamIV!10, rounded corners=3pt] (0,0) rectangle (0.95,3*\rh);
  \draw[svFamIV, line width=.9pt, svdash, rounded corners=3pt] (0,0) rectangle (0.95,3*\rh);
  \node[rotate=90, align=center, font=\svfont, text=svFamIV!85!black] at (0.48,1.5*\rh)
       {\textbf{IV \;alignment} $\mathbf{x}\mapsto T(\mathbf{x})$\\
        {\svsmall\color{svGrey}pre-composes with any cell, Eq.~\eqref{eq:compose}}};
  \foreach \r in {0.5,1.5,2.5}
     \draw[svarrow, draw=svFamIV] (0.98,\r*\rh) -- (1.5,\r*\rh);
\end{tikzpicture}
\endgroup
\caption{The two design axes of Sec.~\ref{subsec:pipeline}. Families~Ia, Ib, II
and~III are placed by their direction of inference (columns) and the kind of
generator (rows); hatched cells are occupied by no family. Family~IV, a
transform on the input, is orthogonal to both axes and pre-composes with any
cell under the conditions of Sec.~\ref{subsec:famIV}. Modality chips list the
modalities supplying anchors in Table~\ref{tab:families}.}
\label{fig:designaxes}
\end{figure*}
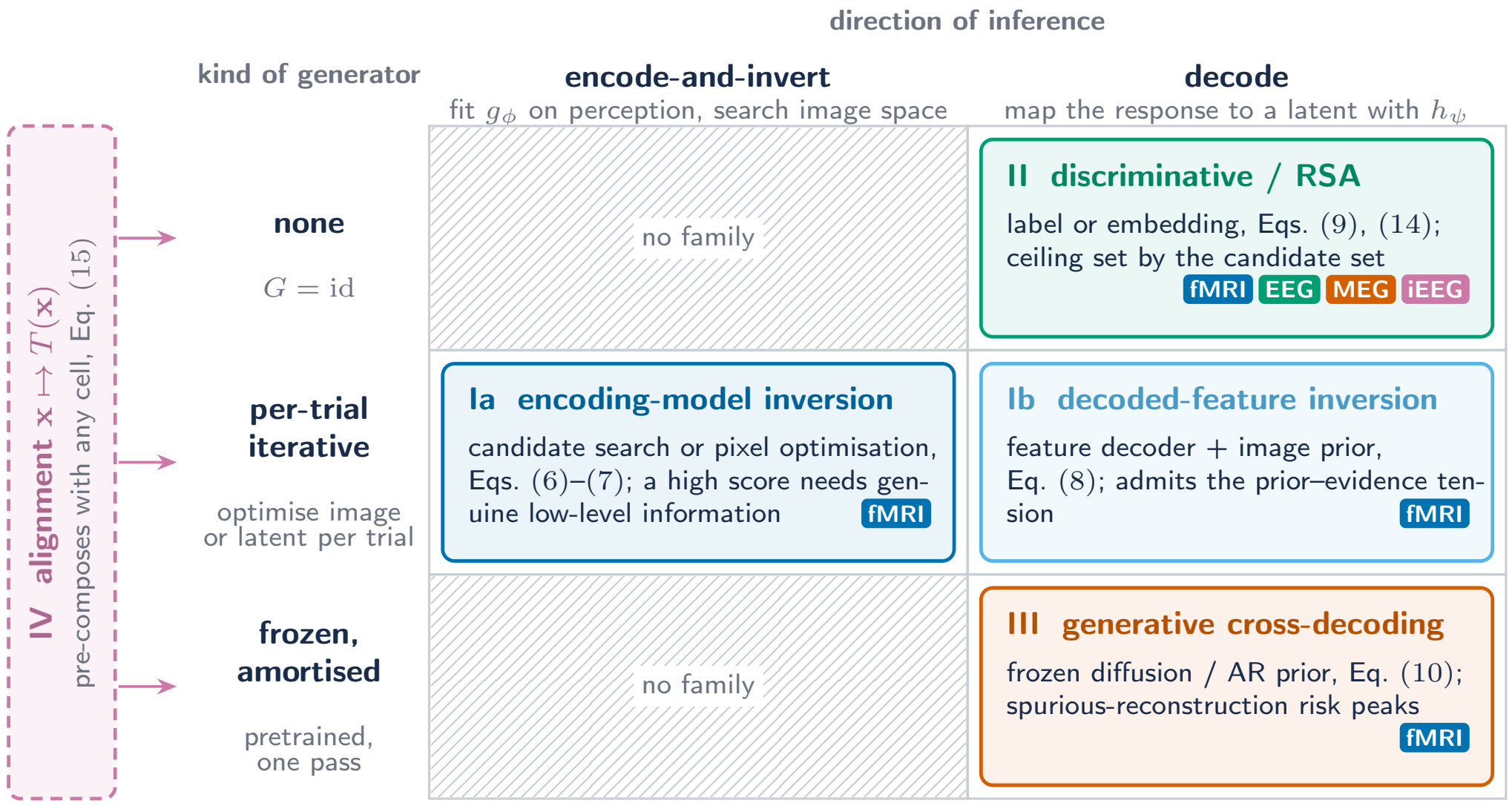

Fig.~\ref{fig:designaxes} places the families by their direction of
inference and kind of generator (Sec.~\ref{subsec:pipeline}). The caption of
Fig.~\ref{fig:pipeline}, which describes Family~I as inverting $\encmodel$,
applies strictly to Ia, and Table~\ref{tab:families} keeps the historical
label ``encoding-model inversion through a feature space'' for Ib; both are
to be read under these axes.

\subsection{The Causal Chain of Spurious Reconstruction}
\label{app:causalchain}

\begin{figure*}[!htbp]
\centering
\begin{tikzpicture}[sv, x=1cm, y=1cm,
   stage/.style={svbox, text width=2.05cm, minimum height=2.3cm, anchor=north, draw=svRed!70!black, fill=svRed!4},
   where/.style={font=\svsmall\bfseries, text=svGrey},
   std/.style={svtag, fill=svInk},
   lnk/.style={svarrow, draw=svRed!70!black}]
  \def\dx{2.72}
  \foreach \i/\w/\t/\e in {
     0/corpus/{Clustered training targets}/{$\approx$40 clusters\\[-1pt]in NSD},
     1/translator/{Output dimension collapse}/{Eq.~\eqref{eq:collapse}},
     2/generator/{Fluent class-typical image}/{Eq.~\eqref{eq:spurious}},
     3/metric/{Inflated scores}/{Eq.~\eqref{eq:pairwise-inflation}; fluency-\\[-1pt]rewarding raters},
     4/inference/{Score read as fidelity}/{$\Pr[T\,|\,R]$\\[-1pt]invalidated}}{
    \node[stage] (n\i) at (\i*\dx,0) {\textbf{\t}\\[2pt]{\svsmall\color{svGrey}\e}};
    \node[where, anchor=base] at (\i*\dx,0.15) {\w};
  }
  \foreach \a/\b in {0/1,1/2,2/3,3/4} \draw[lnk] (n\a) -- (n\b);
  \node[std, anchor=north] at ($(n0.south)+(0,-0.12)$) {S6};
  \node[std, anchor=north] at ($(n1.south)+(0,-0.12)$) {S2};
  \node[std, anchor=north] at ($(n2.south)+(0,-0.12)$) {S5};
  \node[std, anchor=north] at ($(n3.south)+(-0.25,-0.12)$) {S3};
  \node[std, anchor=north] at ($(n3.south)+(0.25,-0.12)$) {S7};
  \node[std, anchor=north, fill=svBlue] at ($(n4.south)+(0,-0.12)$) {S1};
  \node[svnote, anchor=north west, align=left] at ($(n0.south west)+(-0.05,-0.62)$)
       {\textcolor{svInk}{\textbf{S$k$}}: reporting standard of Table~\ref{tab:standards} that closes the link;\\
        \textcolor{svBlue}{\textbf{S1}} ($\Lambda$, Eq.~\eqref{eq:lambda}) tests the inference itself};
  \node[svnote, anchor=north east, align=right] (dep) at ($(n4.south east)+(0,-1.05)$)
       {at deployment: misleads a closed-loop user (Sec.~\ref{subsec:gap4})\\ or is misread as evidence (Sec.~\ref{subsec:forensic})};
  \draw[svarrow, draw=svGrey, svdash] ($(n4.south)+(0,-0.52)$) -- (dep.north -| n4.south);
\end{tikzpicture}
\caption{The causal chain of spurious reconstruction
(Sec.~\ref{subsec:spurious}). Each link carries the equation that formalises
it and, below, the standard of Table~\ref{tab:standards} that closes it; S1
tests the inference itself.}
\label{fig:causalchain}
\end{figure*}
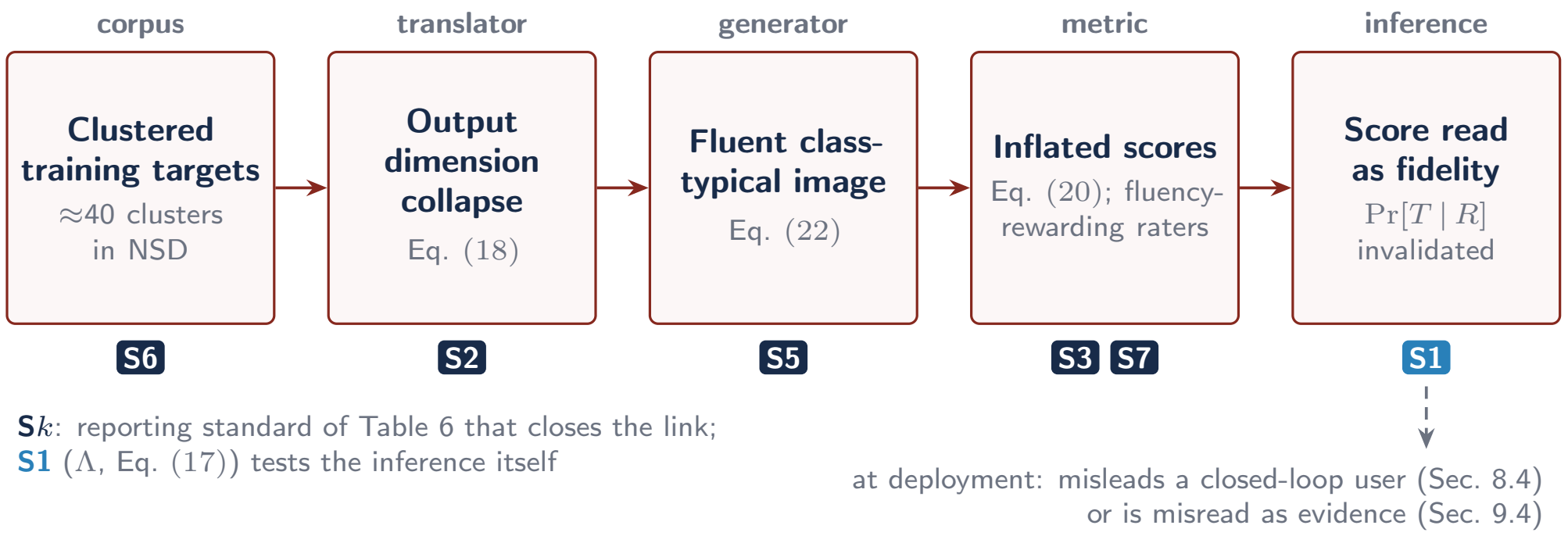

Fig.~\ref{fig:causalchain} summarises the chain described in
Sec.~\ref{subsec:spurious}: a clustered corpus induces output dimension
collapse, collapsed latents are rendered as fluent class-typical images,
categorical test structure and fluency-rewarding raters inflate their
scores, and the score is read as fidelity. The same chain applies at
deployment, where a hallucinated reconstruction could mislead a closed-loop
user (Sec.~\ref{subsec:gap4}) or be misread as evidence
(Sec.~\ref{subsec:forensic}).

\subsection{A Within-Subject Test of the Nature of the Cross-State Shift}
\label{app:h0h1}

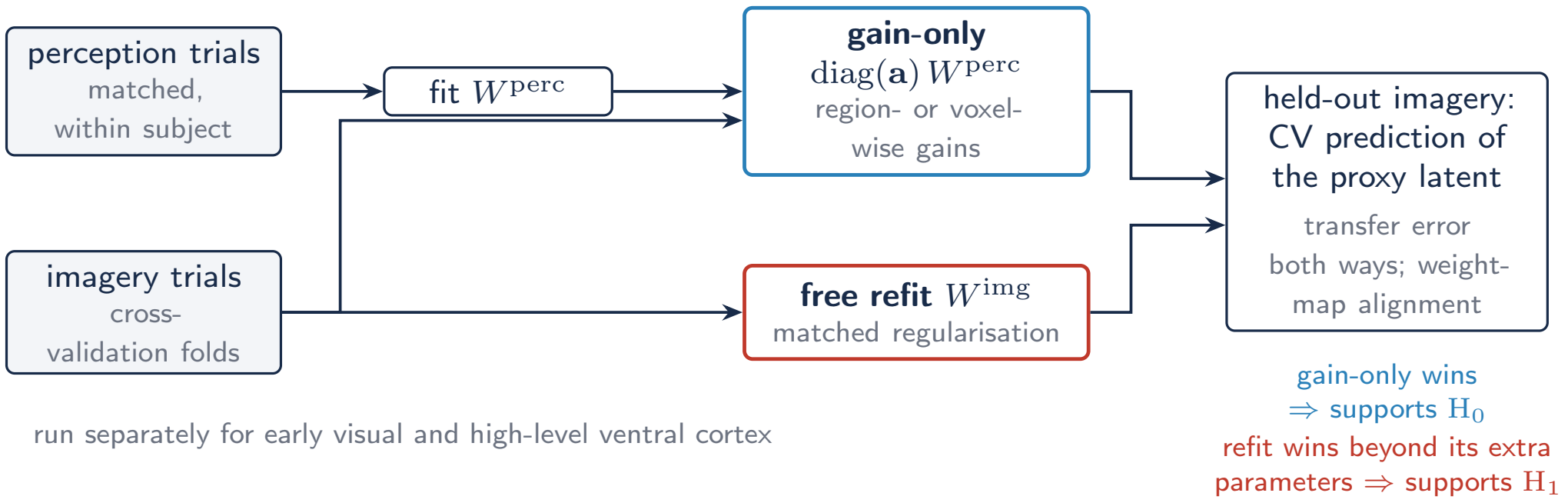
\begin{figure*}[!htbp]
\centering
\begin{tikzpicture}[sv, x=1cm, y=1cm, node distance=4mm,
   data/.style={svbox, fill=svPanel, text width=2.15cm},
   model/.style={svbox, text width=2.75cm, line width=.9pt},
   res/.style={svbox, text width=2.55cm, fill=white}]
  \node[data] (perc) at (0,0.95) {perception trials\\[-1pt]{\svsmall\color{svGrey}matched, within subject}};
  \node[data] (img)  at (0,-0.95) {imagery trials\\[-1pt]{\svsmall\color{svGrey}cross-validation folds}};
  \node[svbox, text width=1.75cm] (W) at (3.05,0.95) {fit $W^{\mathrm{perc}}$};
  \node[model, draw=svBlue] (H0) at (6.65,0.95)
       {\textbf{gain-only} $\operatorname{diag}(\mathbf{a})\,W^{\mathrm{perc}}$\\[-1pt]{\svsmall\color{svGrey}region- or voxel-wise gains}};
  \node[model, draw=svRed] (H1) at (6.65,-0.95)
       {\textbf{free refit} $W^{\mathrm{img}}$\\[-1pt]{\svsmall\color{svGrey}matched regularisation}};
  \node[res] (cmp) at (10.7,0)
       {held-out imagery: CV prediction of the proxy latent\\[2pt]{\svsmall\color{svGrey}transfer error both ways; weight-map alignment}};
  \draw[svarrow] (perc) -- (W);
  \draw[svarrow] (W) -- (H0);
  \draw[svarrow] (img.east) -- ++(0.5,0) |- ($(H0.west)+(0,-0.25)$);
  \draw[svarrow] (img) -- (H1);
  \draw[svarrow] (H0.east) -- ++(0.35,0) |- ($(cmp.west)+(0,0.2)$);
  \draw[svarrow] (H1.east) -- ++(0.35,0) |- ($(cmp.west)+(0,-0.2)$);
  \node[svnote, anchor=north, text width=3.1cm, align=center] at ($(cmp.south)+(0,-0.12)$)
       {\textcolor{svBlue}{gain-only wins $\Rightarrow$ supports $\mathrm{H}_0$}\\[1pt]
        \textcolor{svRed}{refit wins beyond its extra parameters $\Rightarrow$ supports $\mathrm{H}_1$}};
  \node[svnote, anchor=north west, align=left] at (-1.1,-1.75)
       {run separately for early visual and high-level ventral cortex};
\end{tikzpicture}
\caption{The within-subject analysis of Sec.~\ref{subsec:gap2} that would decide
between a shared code with state-specific gain ($\mathrm{H}_0$) and an
altered code ($\mathrm{H}_1$), using the matched perception and imagery
trials that NSD-Imagery and Deeprecon already contain.}
\label{fig:h0h1}
\end{figure*}

Fig.~\ref{fig:h0h1} outlines the analysis proposed in
Sec.~\ref{subsec:gap2}. Reporting transfer error in both directions and the
alignment of the weight maps, in addition to cross-validated prediction of
the proxy latent, would make the comparison between the gain-only and
refitted models interpretable region by region.

\subsection{The Imagery Corpus by Year and Modality}
\label{app:timeline}

Fig.~\ref{fig:timeline} arranges the corpus of Table~\ref{tab:corpus} and
the closed-loop study of Table~\ref{tab:ephys} by year and modality. Every
image-scored reconstruction lies in the fMRI column
(Sec.~\ref{subsec:eegmeg}).

\begin{figure*}[!htbp]
\centering
\begin{tikzpicture}[sv, x=1cm, y=1cm]
  \node[font=\svfont\bfseries, anchor=south west] at (1.05,0.08) {fMRI};
  \node[font=\svfont\bfseries, anchor=south west] at (5.35,0.08) {EEG};
  \node[font=\svfont\bfseries, anchor=south west] at (8.95,0.08) {MEG};
  \node[font=\svfont\bfseries, anchor=south west] at (11.1,0.08) {iEEG};
  \draw[svRule, line width=.3pt] (0,0.0) -- (13.4,0.0);
  \node[font=\svsmall\bfseries, text=svGrey, anchor=north west] at (0,-0.05) {2026};
  \node[font=\svsmall, anchor=north west, inner sep=1pt] at (1.05,-0.05) {\tikz[baseline=-.6ex]{\fill[svFamIII] (0,0) circle (2.1pt);\draw[svInk,line width=.6pt] (0,0) circle (3.3pt);}\ MIRAGE \cite{Kneeland2026mirage}};
  \node[font=\svsmall, anchor=north west, inner sep=1pt] at (5.35,-0.05) {\tikz[baseline=-.6ex]\node[rectangle,draw=svGrey,fill=white,line width=.7pt,inner sep=1.5pt]{};\ Gao (dataset) \cite{Gao2026eegvi}};
  \draw[svRule, line width=.3pt] (0,-0.47) -- (13.4,-0.47);
  \node[font=\svsmall\bfseries, text=svGrey, anchor=north west] at (0,-0.52) {2025};
  \node[font=\svsmall, anchor=north west, inner sep=1pt] at (1.05,-0.52) {\tikz[baseline=-.6ex]\fill[svFamII] (0,0) circle (2.1pt);\ Chang \cite{Chang2025imageless}};
  \node[font=\svsmall, anchor=north west, inner sep=1pt] at (1.05,-0.89) {\tikz[baseline=-.6ex]{\fill[svFamIII] (0,0) circle (2.1pt);\draw[svInk,line width=.6pt] (0,0) circle (3.3pt);}\ NSD-Imagery \cite{Kneeland2025nsdimagery}};
  \node[font=\svsmall, anchor=north west, inner sep=1pt] at (5.35,-0.52) {\tikz[baseline=-.6ex]\fill[svFamII] (0,0) circle (2.1pt);\ Kwon \cite{Kwon2025functional}};
  \draw[svRule, line width=.3pt] (0,-1.31) -- (13.4,-1.31);
  \node[font=\svsmall\bfseries, text=svGrey, anchor=north west] at (0,-1.36) {2024};
  \node[font=\svsmall, anchor=north west, inner sep=1pt] at (1.05,-1.36) {\tikz[baseline=-.6ex]{\fill[svFamIb] (0,0) circle (2.1pt);\draw[svInk,line width=.6pt] (0,0) circle (3.3pt);}\ Koide-Majima \cite{KoideMajima2024mental}};
  \node[font=\svsmall, anchor=north west, inner sep=1pt] at (5.35,-1.36) {\tikz[baseline=-.6ex]\fill[svFamII] (0,0) circle (2.1pt);\ Stecher \& Kaiser \cite{StecherKaiser2024alpha}};
  \draw[svRule, line width=.3pt] (0,-1.78) -- (13.4,-1.78);
  \node[font=\svsmall\bfseries, text=svGrey, anchor=north west] at (0,-1.83) {2022};
  \node[font=\svsmall, anchor=north west, inner sep=1pt] at (5.35,-1.83) {\tikz[baseline=-.6ex]\fill[svFamIII] (0,0) circle (2.1pt);\ Shimizu \cite{Shimizu2022improving}};
  \draw[svRule, line width=.3pt] (0,-2.25) -- (13.4,-2.25);
  \node[font=\svsmall\bfseries, text=svGrey, anchor=north west] at (0,-2.3) {2020};
  \node[font=\svsmall, anchor=north west, inner sep=1pt] at (5.35,-2.3) {\tikz[baseline=-.6ex]\fill[svFamII] (0,0) circle (2.1pt);\ Xie \cite{Xie2020alpha}};
  \draw[svRule, line width=.3pt] (0,-2.7199999999999998) -- (13.4,-2.7199999999999998);
  \node[font=\svsmall\bfseries, text=svGrey, anchor=north west] at (0,-2.7699999999999996) {2019};
  \node[font=\svsmall, anchor=north west, inner sep=1pt] at (1.05,-2.7699999999999996) {\tikz[baseline=-.6ex]{\fill[svFamIb] (0,0) circle (2.1pt);\draw[svInk,line width=.6pt] (0,0) circle (3.3pt);}\ Shen (Deeprecon) \cite{Shen2019deep}};
  \node[font=\svsmall, anchor=north west, inner sep=1pt] at (1.05,-3.1399999999999997) {\tikz[baseline=-.6ex]{\fill[svFamI] (0,0) circle (2.1pt);\draw[svInk,line width=.6pt] (0,0) circle (3.3pt);}\ Senden \cite{Senden2019imagery}};
  \node[font=\svsmall, anchor=north west, inner sep=1pt] at (5.35,-2.7699999999999996) {\tikz[baseline=-.6ex]\fill[svFamII] (0,0) circle (2.1pt);\ Shatek \cite{Shatek2019decoding}};
  \draw[svRule, line width=.3pt] (0,-3.5599999999999996) -- (13.4,-3.5599999999999996);
  \node[font=\svsmall\bfseries, text=svGrey, anchor=north west] at (0,-3.6099999999999994) {2018};
  \node[font=\svsmall, anchor=north west, inner sep=1pt] at (8.95,-3.6099999999999994) {\tikz[baseline=-.6ex]\fill[svFamII] (0,0) circle (2.1pt);\ Dijkstra \cite{Dijkstra2018differential}};
  \draw[svRule, line width=.3pt] (0,-4.029999999999999) -- (13.4,-4.029999999999999);
  \node[font=\svsmall\bfseries, text=svGrey, anchor=north west] at (0,-4.079999999999999) {2017};
  \node[font=\svsmall, anchor=north west, inner sep=1pt] at (1.05,-4.079999999999999) {\tikz[baseline=-.6ex]\fill[svFamII] (0,0) circle (2.1pt);\ Horikawa \& Kamitani \cite{Horikawa2017generic}};
  \draw[svRule, line width=.3pt] (0,-4.499999999999999) -- (13.4,-4.499999999999999);
  \node[font=\svsmall\bfseries, text=svGrey, anchor=north west] at (0,-4.549999999999999) {2016};
  \node[font=\svsmall, anchor=north west, inner sep=1pt] at (1.05,-4.549999999999999) {\tikz[baseline=-.6ex]\fill[svFamIb] (0,0) circle (2.1pt);\ Lee \& Kuhl \cite{LeeKuhl2016reconstructing}};
  \draw[svRule, line width=.3pt] (0,-4.969999999999999) -- (13.4,-4.969999999999999);
  \node[font=\svsmall\bfseries, text=svGrey, anchor=north west] at (0,-5.019999999999999) {2015};
  \node[font=\svsmall, anchor=north west, inner sep=1pt] at (1.05,-5.019999999999999) {\tikz[baseline=-.6ex]{\fill[svFamI] (0,0) circle (2.1pt);\draw[svInk,line width=.6pt] (0,0) circle (3.3pt);}\ Naselaris \cite{Naselaris2015voxelwise}};
  \draw[svRule, line width=.3pt] (0,-5.439999999999999) -- (13.4,-5.439999999999999);
  \node[font=\svsmall\bfseries, text=svGrey, anchor=north west] at (0,-5.489999999999998) {2010};
  \node[font=\svsmall, anchor=north west, inner sep=1pt] at (1.05,-5.489999999999998) {\tikz[baseline=-.6ex]\fill[svFamII] (0,0) circle (2.1pt);\ Reddy \cite{Reddy2010reading}};
  \node[font=\svsmall, anchor=north west, inner sep=1pt] at (11.1,-5.489999999999998) {\tikz[baseline=-.6ex]\fill[svFamII] (0,0) circle (2.1pt);\ Cerf \cite{Cerf2010online}};
  \draw[svRule, line width=.3pt] (0,-5.909999999999998) -- (13.4,-5.909999999999998);
  \node[font=\svsmall\bfseries, text=svGrey, anchor=north west] at (0,-5.959999999999998) {2009};
  \node[font=\svsmall, anchor=north west, inner sep=1pt] at (1.05,-5.959999999999998) {\tikz[baseline=-.6ex]\fill[svFamII] (0,0) circle (2.1pt);\ Stokes \cite{Stokes2009topdown}};
  \node[font=\svsmall, anchor=north west, inner sep=1pt] at (1.05,-6.329999999999998) {\tikz[baseline=-.6ex]\fill[svFamII] (0,0) circle (2.1pt);\ Harrison \& Tong \cite{HarrisonTong2009decoding}};
  \draw[svRule, line width=.3pt] (0,-6.749999999999998) -- (13.4,-6.749999999999998);
  \node[font=\svsmall\bfseries, text=svGrey, anchor=north west] at (0,-6.799999999999998) {2006};
  \node[font=\svsmall, anchor=north west, inner sep=1pt] at (1.05,-6.799999999999998) {\tikz[baseline=-.6ex]\fill[svFamI] (0,0) circle (2.1pt);\ Thirion \cite{Thirion2006inverse}};
  \draw[svRule, line width=.3pt] (0,-7.219999999999998) -- (13.4,-7.219999999999998);
  \node[font=\svsmall\bfseries, text=svGrey, anchor=north west] at (0,-7.269999999999998) {2000};
  \node[font=\svsmall, anchor=north west, inner sep=1pt] at (11.1,-7.269999999999998) {\tikz[baseline=-.6ex]\fill[svFamII] (0,0) circle (2.1pt);\ Kreiman \cite{Kreiman2000imagery}};
  \draw[svRule, line width=.3pt] (0,-7.689999999999998) -- (13.4,-7.689999999999998);
  \draw[svRule, line width=.3pt] (5.2299999999999995,0.45) -- (5.2299999999999995,-7.689999999999998);
  \draw[svRule, line width=.3pt] (8.83,0.45) -- (8.83,-7.689999999999998);
  \draw[svRule, line width=.3pt] (10.98,0.45) -- (10.98,-7.689999999999998);
  \node[anchor=north west, font=\svsmall] at (0,-7.809999999999998)
    {family: \tikz[baseline=-.6ex]\fill[svFamI] (0,0) circle (2.1pt);\,Ia\quad \tikz[baseline=-.6ex]\fill[svFamIb] (0,0) circle (2.1pt);\,Ib\quad \tikz[baseline=-.6ex]\fill[svFamII] (0,0) circle (2.1pt);\,II\quad \tikz[baseline=-.6ex]\fill[svFamIII] (0,0) circle (2.1pt);\,III\quad \tikz[baseline=-.6ex]\node[rectangle,draw=svGrey,fill=white,line width=.7pt,inner sep=1.5pt]{};\,dataset only\qquad \tikz[baseline=-.6ex]{\fill[svFamII] (0,0) circle (2.1pt);\draw[svInk,line width=.6pt] (0,0) circle (3.3pt);}\,image-scored reconstruction (Sec.~\ref{subsec:datasetstable})};
\end{tikzpicture}
\caption{The imagery corpus of Table~\ref{tab:corpus}, with the closed-loop study
of Table~\ref{tab:ephys}, by year and modality. Colour gives the method family
of Table~\ref{tab:families} and a ring an image-scored reconstruction
(Sec.~\ref{subsec:datasetstable}); every ring lies in the fMRI column
(Sec.~\ref{subsec:eegmeg}).}
\label{fig:timeline}
\end{figure*}

\FloatBarrier

\end{document}